\documentclass{aa}
\usepackage{graphicx}
\usepackage{txfonts,textcomp}
\usepackage{natbib}
\usepackage{upgreek}

\newcommand{\twCN}{$^{12}$CN}
\newcommand{\twCO}{$^{12}$CO}
\newcommand{\thCO}{$^{13}$CO}
\newcommand{\CeiO}{C$^{18}$O}
\newcommand{\HCOp}{HCO$^{+}$}
\newcommand{\Jone}{\mbox{(1--0)}}
\newcommand{\Jtwo}{\mbox{(2--1)}}

\usepackage{hyperref}
\hypersetup{colorlinks,allcolors=magenta}

\usepackage{orcidlink}
\newcommand{\orcid}[1]{\orcidlink{#1}}

\makeatletter
\DeclareRobustCommand{\HII}{%
  \mbox{H\check@mathfonts\fontsize\sf@size\z@\selectfont II}%
}
\makeatother

\makeatletter
\DeclareRobustCommand{\HI}{%
  \mbox{H\check@mathfonts\fontsize\sf@size\z@\selectfont I}%
}
\makeatother
\begin{document} 

   \title{The magnetic field in the Flame nebula}

 %  \subtitle{The Magnetic field in NGC~2024}

% \titlerunning{The Magnetic field in NGC~2024}
   \authorrunning{Be\v{s}li\'c}

   \author{I.~Be\v{s}li\'c \inst{\ref{lerma}} \orcid{0000-0003-0583-7363}
   \and 
        S.~Coud\'e  \inst{\ref{worc}, \ref{cfa}} \orcid{0000-0002-0859-0805}
   \and 
        D.~C.~Lis  \inst{\ref{jpl}} \orcid{0000-0002-0500-4700}
   \and 
        M.~Gerin  \inst{\ref{lerma}} \orcid{0000-0002-2418-7952}
   \and 
        P.~F.~Goldsmith  \inst{\ref{jpl}}  \orcid{0000-0002-6622-8396}
   \and 
        J.~Pety  \inst{\ref{lerma}, \ref{iram}} \orcid{0000-0003-3061-6546}
   \and
        A.~Roueff \inst{\ref{toulon}}
   \and
        K.~Demyk \inst{\ref{toulouse}} \orcid{0000-0002-5019-8700}
    \and
        C.~D.~Dowell \inst{\ref{jpl}}
    \and 
        L.~Einig \inst{\ref{iram},\ref{grenoble}} \orcid{0000-0003-4250-7638}
    \and 
        J.~R.~Goicoechea \inst{\ref{csic}} \orcid{0000-0001-7046-4319}
    \and 
        F.~Levrier \inst{\ref{ens}} \orcid{0000-0002-3065-9944}
    \and 
        J.~Orkisz \inst{\ref{iram}} \orcid{0000-0002-3382-9208}
    \and 
        N.~Peretto \inst{\ref{cardiff}} \orcid{0000-0002-6893-602X}
    \and 
        M.~G.~Santa-Maria \inst{\ref{csic}} \orcid{0000-0002-3941-0360}
    \and 
        N.~Ysard \inst{\ref{toulouse}, \ref{saclay}} \orcid{0000-0003-1037-4121}
    \and 
        A.~Zakardjian \inst{\ref{toulouse}} \orcid{0000-0002-4240-6012}
          }

   \institute{LERMA, Observatoire de Paris, PSL Research University, CNRS, Sorbonne Universit\'es, 75014 Paris, France \label{lerma}
   \and
        Department of Earth, Environment, and Physics, Worcester State University, Worcester, MA 01602, USA \label{worc}
   \and 
        Center for Astrophysics $\vert$ Harvard \& Smithsonian, 60 Garden Street, Cambridge, MA 02138, USA \label{cfa}
   \and 
        Jet Propulsion Laboratory, California Institute of Technology, 4800 Oak Grove Drive, Pasadena, CA 91109, USA \label{jpl}
   \and 
        IRAM, 300 rue de la Piscine, 38406 Saint Martin d’Hères, France \label{iram}
   \and 
        Université de Toulon, Aix Marseille Univ, CNRS, IM2NP, Toulon, France \label{toulon}
    \and 
        IRAP, CNRS, Université de Toulouse, 9 avenue du Colonel Roche, 31028 Toulouse Cedex 4, France \label{toulouse}
    \and 
        Université Grenoble Alpes, CNRS, GIPSA-lab, Grenoble INP, Grenoble, 38000, France \label{grenoble}
    \and 
        Instituto de Física Fundamental (CSIC). Calle Serrano 121-123, 28006, Madrid, Spain \label{csic}
   \and 
        Laboratoire de Physique de l’ENS, ENS, Université PSL, CNRS, Sorbonne Université, Université Paris Cité, Observatoire de Paris, F-75005 Paris, France \label{ens}
    \and 
        Cardiff Hub for Astrophysics Research $\&$ Technology, School of Physics $\&$ Astronomy, Cardiff University, Queens Buildings, The parade, Cardiff CF24 3AA, UK \label{cardiff}
    \and 
         Université Paris-Saclay, CNRS, Institut d’Astrophysique Spatiale, 91405, Orsay, France \label{saclay}
    %          \email{ivana.beslic@obspm.fr}
             }

   \date{Received 24 October 2023; accepted 20 January 2024}

  \abstract
{Star formation drives the evolution of galaxies and the cycling of matter between different phases of the interstellar medium and stars. The support of interstellar clouds against gravitational collapse by magnetic fields has been proposed as a possible explanation for the low observed star formation efficiency in galaxies and the Milky Way. The Planck satellite provided the first all-sky map of the magnetic field geometry in the diffuse interstellar medium on angular scales of 5-15$'$. However, higher spatial resolution observations are required to understand the transition from diffuse, subcritical gas to dense, gravitationally unstable filaments.}
{NGC 2024, also known as the Flame nebula, is located in the nearby Orion B molecular cloud. It contains a young, expanding \HII\, region and a dense supercritical filament. This filament harbors embedded protostellar objects and is likely not supported by the magnetic field against gravitational collapse. Therefore, NGC 2024 provides an excellent opportunity to study the role of magnetic fields in the formation, evolution, and collapse of dense filaments, the dynamics of young \HII\, regions, and the effects of mechanical and radiative feedback from massive stars on the surrounding molecular gas.}
{We combined new 154 and 216 \textmu m dust polarization measurements carried out using the HAWC+ instrument aboard SOFIA with molecular line observations of \twCN\Jone\, and \HCOp\Jone\, from the IRAM 30-meter telescope to determine the magnetic field geometry, and to estimate the plane of the sky magnetic field strength across the NGC~2024 \HII\, region and the surrounding molecular cloud.}
{The HAWC+ observations show an ordered magnetic field geometry in NGC~2024 that follows the morphology of the expanding \HII\, region and the direction of the main dense filament. The derived plane of the sky magnetic field strength is moderate, ranging from 30 to 80 \textmu G. The strongest magnetic field is found at the east edge of the \HII\ region, characterized by the highest gas densities and molecular line widths. In contrast, the weakest field is found toward the main, dense filament in NGC~2024.}
{We find that the magnetic field has a non-negligible influence on the gas stability at the edges of the expanding \HII\, shell (gas impacted by stellar feedback) and the filament (site of current star formation).}
  %{We find that the UV-illuminated gas at the edge of the \HII\, region, which is impacted by the stellar feedback, is supported by the magnetic field against gravitational collapse. In contrast, the dense filament is not magnetically supported and thus represents a location where star formation can take place.}

   \keywords{ISM: magnetic fields --
                ISM: molecules --
                Galaxies: star formation
               }

   \maketitle
%-------------------------------------------------------------------

% \input{Sections_referee/01-Introduction}

\section{Introduction}
\label{sec:introduction}
The question of what controls the star formation efficiency in molecular clouds has long been at the center of star formation research. Early studies \citep{zuckerman_evans_1974} showed that if all the gas within dense interstellar clouds were to collapse freely under self-gravity, the star formation rate in the Milky Way would be two orders of magnitude higher than the observed rate of 2 M$_\odot$/year \citep{robitaille_2010}. Theories proposed to explain such a low star formation rate invoke the presence of turbulence or magnetic fields supporting interstellar clouds against gravitational collapse. In some models \citep{tan_2006}, turbulent or magnetic pressure gradients are strong enough to maintain clouds in approximate hydrostatic equilibrium on all spatial scales for tens of free-fall times. In other models, high-density regions are rapidly contracting, converting a large fraction \citep[as high as 40\%;][]{bonnell_2011} of their mass into stars in only a few free-fall times. However, the low-density parts of the cloud, which contain up to 90\% of the cloud mass \citep{battisti_2014}, disperse over the scales of free fall time as a result of turbulence generated by galactic shear or by energy input from supernova explosions \citep{dobbs_2011,walch_2015}, or due to radiative and mechanical stellar feedback from high-mass stars that formed early on during the evolution of the cloud \citep{murray_2011, colin_2013, pabst_2020, chevance_2022, suin_2022}.

In addition to preventing the gas from collapsing, the turbulence and the magnetic field can also bolster the star formation processes. For example, the coupling between the magnetic field and the neutral gas can allow parts of low-density clouds to fragment and initiate star formation \citep{fiedler_1993}. In this scenario, while neutral gas is collapsing, a part of the magnetic flux is removed, which further contributes to the gravitational instability \citep{mouschovias_1999, lazarian_2012, priestley_2022b, tritsis_2023}.

However, it has been difficult to test these star formation models observationally. Despite all of the effort made over the past 40 years to characterize interstellar turbulence, we still struggle to understand the different energy sources that contribute to the observed line-of-sight gas velocity dispersion. The only available way to measure turbulent motions is by observing the gas velocity along the line of sight \citep{larson_1981, hennebelle_2012}, which does not provide a complete picture. As noted by \cite{ballesteros_paredes_2011}, a collapsing cloud has energetic properties (at least in terms of the total kinetic energy) similar to those of an identical cloud supported against gravity by turbulence. In addition, measurements of magnetic fields are observationally extremely challenging, as they typically involve relatively weak signals, as is the case for Zeeman gas line splitting \citep{crutcher_2012} or (sub)millimeter/far-infrared (FIR) dust polarization measurements \citep[see,][for a recent review]{pattle_2019}.

\begin{figure*}[t!]
    %%\centering
    \includegraphics[width=\textwidth]{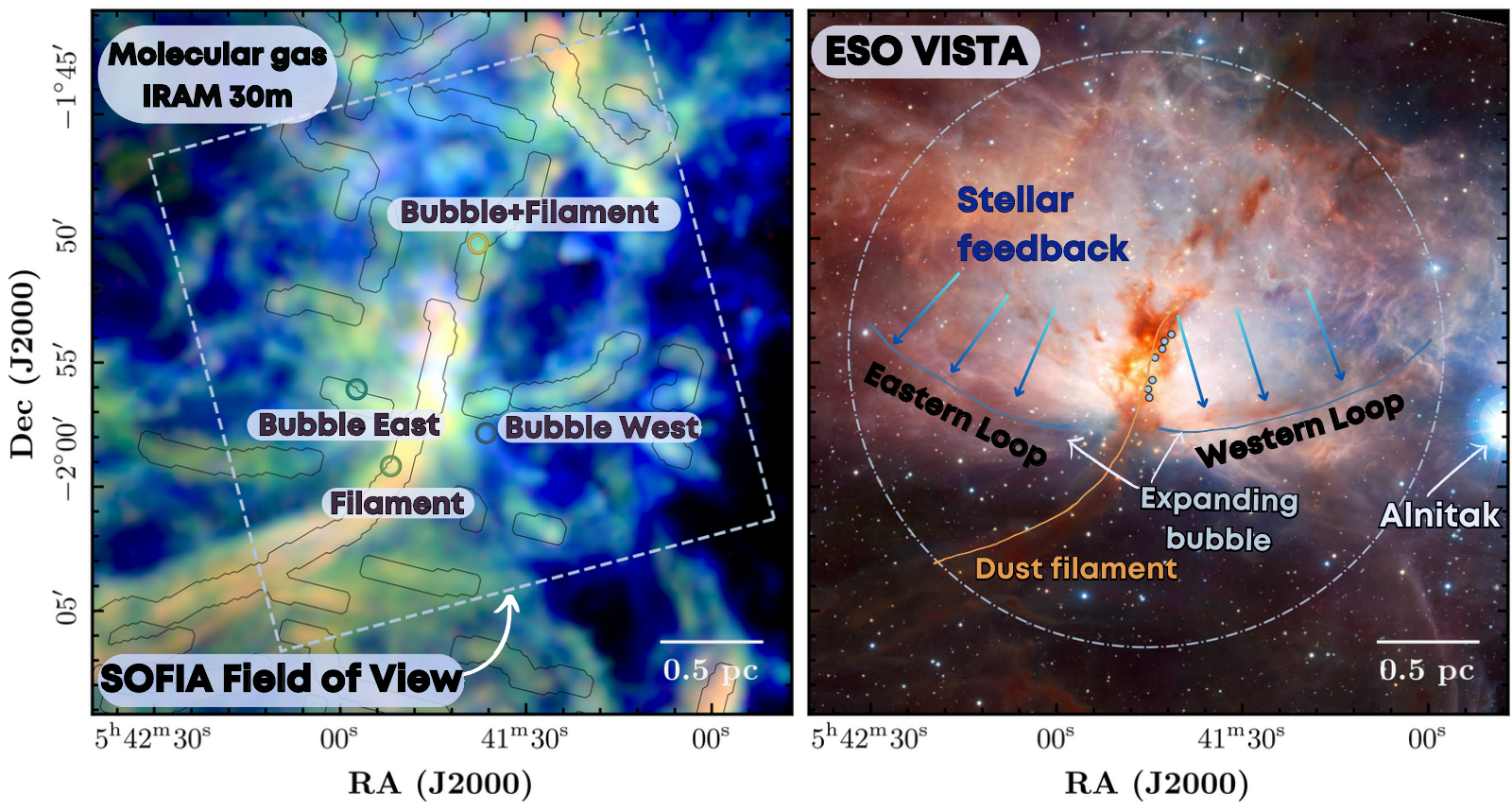}
    \caption{Multiwavelength image of NGC~2024. (Left panel) Color-composite image of the Flame nebula (NGC~2024) showing peak intensities of $^{12}$CO\,$(1-0)$ (blue) emission and isotopologues, $^{13}$CO\,$(1-0)$ (green) and C$^{18}$O\,$(1-0)$ (red) obtained by the IRAM 30-meter telescope \citep[image credits:][]{pety_2017}. We overlaid the SOFIA HAWC+ field of view as a white dashed rectangle. Gray contours show a network of filaments presented in \cite{orkisz_2019}. We label the regions we investigate in this work: Bubble East, Bubble West, Filament, and Bubble+Filament. (Right panel) ESO-VISTA image (ESO/J. Emerson/VISTA, Cambridge Astronomical Survey Unit). We sketched the environments seen across NGC~2024: the \HII\ region \citep[white dashed circle, see Tab.\,2 in][]{gaudel_2023} and the filament \citep[orange line,][]{orkisz_2019}. Light blue points are the positions of far-infrared sources in the background \citep{mezger_1988, mezger_1992}. In addition, we labeled the edges of the expanding \HII\ region and the direction of the stellar feedback driven by the radiation produced by recent star formation.}
    \label{fig:fig1}
\end{figure*}

%\subsection{Dust Polarimetry}
%\label{sec:intro_dustpol}
%%%%%%%%%%%%%%%%%%%%%%%%%%%%%%%%%%%%%%%%%%%%%%%%%%%%%%%%%%%%%%%%%%%%%%%%%%%%%%%%%%%%%%%%%%%%%%%%%%%%%%%%%%%%%%%%%%%%%%%%%%%
Interstellar dust thermal emission is polarized \citep{hall_1949,hiltner_1949} due to the presence of $B$ fields and non-spherical dust grains in the interstellar medium (ISM). The explanation for this phenomenon was firstly proposed by \cite{davis_1951} as a paramagnetic alignment with the magnetic field. Later on, this process was described by the radiative alignment torques (RAT) theory \citep[][and references therein]{HoangLazarian2014,andersson_2015}, in which the minor axis of dust grains is aligned parallel to the direction of the $B$ field. Consequently, the (sub)millimeter dust continuum emission is polarized perpendicular to the direction of the component of the $B$ field in the plane of the sky (POS).

Regardless of the observational challenges, technological advances in the past decade are now allowing for polarized dust emission to be measured over increasingly extended regions. The Planck satellite provided an all-sky map of the magnetic field geometry in the diffuse ISM, albeit at a low angular resolution of 5–15$^\prime$. These observations revealed that the galactic $B$ field is intertwined with the filamentary structure of the ISM. In particular, the POS orientation of the $B$ field and the filamentary structures is correlated with the observed column density, with the field largely parallel to diffuse structures (striations) while perpendicular to dense filaments \citep{planck_2016a, planck_2016b}. While supported by independent studies \citep{mcclure-griffiths_2006,goldsmith_2008,peretto_2012,clark_2019}, this picture is complicated by potential projection effects \citep{panopoulou_2016} or the presence of stellar feedback \citep[\HII\, bubble shells,][]{chapman_2011}.

The Balloon-borne Large Aperture Submillimeter Telescope for Polarimetry (BLASTPol) has also observed a similar $B$-field morphology at a higher angular resolution \citep[2.5$^\prime$,][]{fissel_2019}, which has been interpreted as evidence of the magnetic field dominating the energy balance in the diffuse gas. In contrast, gravity is dominant in the dense, star-forming regions. This interpretation has recently been reinforced by high angular resolution ($14''$) ground-based observations using the POL-2 instrument on the James Clerk Maxwell Telescope \citep[JCMT;][]{pattle_2017, liu_2019, wang_2019}, which have shown that field lines bend along dense star-forming filaments, possibly as a result of gravitationally driven flows \citep{goldsmith_2008, chapman_2011, hwang_2023}. % {\bf see suggestions for additional references}

A transition from magnetically dominated to gravitationally dominated gas requires a redistribution of magnetic flux \citep[][]{tritsis_2022}. In principle, such a transition occurs before the cores form \citep{ching_2022}, which implies that cores should be generally super-critical. Nevertheless, a change in the gravitational stability of the gas should be accompanied by a corresponding change in its kinematic properties, arguing for the necessity of combining high angular resolution dust polarization and velocity-resolved molecular emission imaging. Such studies have only recently started and cover relatively small isolated fields. A recent study \citep{tang_2019} has indeed investigated the relation between dense gas velocity gradients (as imaged in N$_2$H$^+$) and the magnetic field direction in massive star-forming infrared dark clouds and concluded that the two are strongly correlated as a result of gravity dragging matter toward the center of the massive ridge. Extending such studies to larger fields is of crucial importance and the High-resolution Airborne Wide-band Camera-Plus \citep[HAWC+,][]{harper_2018} instrument on the Stratospheric Observatory for Infrared Astronomy (SOFIA) offered exceptional capabilities for mapping the magnetic field geometry, as demonstrated, for example, by the observations of the OMC-1 region at 53, 89, 154, and 214 \textmu m \citep{chuss_2019}. The capabilities of HAWC+ for large-area magnetic field mapping have further improved with the commissioning of the on-the-fly map (OTFMAP) polarimetric mode for fast wide-field polarimetric imaging.

When estimating the strength of the POS $B$ field, it is critical to know the level of turbulence in the gas because the small-scale turbulence causes local deviations from the mean direction of the magnetic field. The Davis-Chandrasekhar-Fermi (DCF) method \citep[][]{davis_1951,chandrasekhar_fermi_1953} provides a way of calculating the POS $B$-field strength using information about gas density, turbulence, and changes in the direction of the magnetic field. This method assumes linear geometry and sub-Alfvénic (magnetically dominated) turbulence. Therefore, it is necessary to obtain information about the turbulence of the gas and dust polarization to use this method.
Several variations of the DCF have been developed over the last few decades. The main modification lies in taking into account the number of turbulent cells present along the line of sight and those captured within the beam by including a correction factor Q, which takes a value between zero and one \citep[see Eq.\,16 in][]{ostriker_2001}. In another variation, the deviation in the mean direction of the $B$ field is substituted by the ratio between the ordered and turbulent magnetic field component \citep[][and references therein]{houde_2009}. 
The DCF method assumes that imcompressible motions cause the dispersions of the observed polarization angles, which is not always applicable within the ISM. \cite{skalidis_2021a} provided an alternative method for deriving the magnetic field strength from dust polarization measurements -- the Skalidis-Tassis (ST) method. In addition to Alfvénic modes, this study accounts for the presence of compressible motions in the gas without discarding the anisotropic nature of the turbulence, providing a physically motivated approach to estimating the magnetic field strength \citep{skalidis_2021b}. For this work, we used the ST method to investigate highly structured regions in the Flame nebula in the Orion~B complex, while also presenting the results from the modified DCF method \citep[for example,][]{ostriker_2001,crutcher_2004,lyo_2021} for comparison.

%\subsection{The Orion-B GMC}

Although OMC-1 is the closest and best-studied high-mass star-forming region, it is not a typical cloud to study the role of the magnetic field in star formation. The reason is that OMC-1 is particularly dense (10$^5$ cm$^{-3}$) and very active in terms of star formation, affected by strong shocks \citep[including an explosive outflow resulting from the merger of three massive protostars, see,][]{bally_2008}. These shocks exhibit exceptionally intense UV illumination, with enhancement factors of $G_0$ $\sim$10$^4$, up to 10$^5$ relative to the standard interstellar radiation field (ISRF) \citep[in the vicinity of the Trapezium cluster,][]{goicoechea_2015, goicoechea_2019}. The presence of a strong radiation field implies that the thermal pressure is high, and the magnetic field plays a limited role in the photon-dominated region (PDR) gas dynamics.

Fortunately, the vast quantity of molecular data already obtained in Orion~B makes this region a particularly well-suited location for studying the correlation between changes in the magnetic field geometry and associated gas velocity gradients. The Orion~B giant molecular cloud (GMC) hosts NGC~2024, one of the closest high-mass star-forming regions \citep[at a distance of $d=410\,$pc;][]{cao_2023}, and it is overall more representative of a standard GMC in our galaxy and normal galaxies, with the far-ultraviolet (FUV) radiation field $G_0$ in the range of $\sim10^3-10^4$ \citep{santa-maria_2023}.

In a recent study, \cite{orkisz_2017} used \twCO\Jone\, and \thCO\Jone\, lines to characterize observationally the ratio of compressive versus solenoidal motions in the turbulent flow and to relate this to the star formation efficiency in various regions of Orion~B. \cite{orkisz_2019} accurately analyzed the dynamics of the filamentary network using \CeiO\Jone. Most identified filaments in Orion~B are low-density, thermally subcritical structures, not collapsing to form stars. 
Only about 1\% of the Orion~B cloud mass can be found in super-critical, star-forming filaments, consistent with the low overall star formation efficiency of the region \citep{orkisz_2019}. 

%\subsubsection{NGC~2024}

NGC~2024 (Fig.\,\ref{fig:fig1}) is located east of Alnitak ($\zeta$ Ori) in the Orion~B complex \citep[][]{meyer_2008}. This region contains a massive and young \HII\, region \citep[age $2\cdot10^5$\,yr, ][and private communication]{tremblin_2014}, deeply embedded in dust, located in the foreground, and extending to the south \citep[][see gray contrours showing the network of filaments from \cite{orkisz_2019} in the left panel and labels in the right panel in Fig.\,\ref{fig:fig1}]{barnes_1989}. The north-south filament in NGC~2024 is super-critical \citep{orkisz_2019} and the site of ongoing star formation as recently confirmed in the southern part of this filament observed by NOEMA \citep{schimajiri_2023}. %, whereas solenoidal motions dominate the bulk of the cloud \citep{orkisz_2017}.

The dust bar observed along the line of sight is visible in the ESO-Vista image (right panel in Fig.\,\ref{fig:fig1}) as the dark extinction pattern across the image, leading to an apparent cooler dust temperature derived from Herschel observations \citep{lombardi_2014}. The central part of NGC~2024 contains warm dust and gas, heated by the \HII\, region, as well as embedded protostellar objects \citep{mezger_1988,mezger_1992,lis_1991} located in the background (light blue points in the right panel in Fig.\,\ref{fig:fig1}). The young \HII\, region is expanding, strongly impacting its parental cloud, and creating sharp ionization fronts toward the south, which makes it a good example demonstrating how such systems can efficiently exert stellar feedback. The edges of the bubble are seen toward the west and east of the center of NGC~2024 (labeled as Eastern and Western Loop in Fig.\,\ref{fig:fig1}).

\section{Observations}
\label{sec:obs}

\subsection{Dust polarization measurements using SOFIA HAWC+}
\label{subsec:hawc+}

Our work employs the dust polarization measurements acquired using HAWC+ on SOFIA in September~2021 for program 09\_0015 (PI: D.~Lis). Specifically, we observed NGC~2024 at 154~\textmu m (Band~D) and 214~\textmu m (Band~E) with HAWC+ in polarization mode on flights F779 (8 September 2021), F780 (9 September 2021), F782 (11 September 2021), F783 (14 September 2021), and F784 (15 September 2021). The map obtained at each wavelength comprises ten $20' \times 7'$ strips in a weave pattern: five vertical and five horizontal. These strips were designed using the On-The-Fly scan mode of the HAWC+ camera, which covers the requested area using Lissajous patterns on the sky \citep[][]{harper_2018}. Each strip was repeated at least twice.

%%% data reduction
The level~0 data for each flight were downloaded from the NASA/IPAC Infrared Science Archive (IRSA) and reduced in July~2023 using the SOFIA Data Reduction software \citep[SOFIA Redux version 1.3.0;][]{melanie_clarke_2023}. The HAWC+ scan mode reduction package from SOFIA Redux was initially built in Python from the Java-based \texttt{CRUSH} data reduction software \citep{Kovacs_2008}. All available files at a given wavelength (104 each for Band~D and Band~E) were loaded in SOFIA Redux to be reduced using the default parameters of the software except for two options at the \texttt{Compute Scan Map} step of the pipeline. Specifically, the \texttt{fixjumps} option was set to \texttt{True} to filter out flux jumps in individual detectors during observations, and the \texttt{rounds} option was set to 40 to improve the recovery of astronomical large scale flux from the background subtraction. Appendix\,\ref{sec:sofia_flux} shows the comparison of the Stokes~$I$ values from the resulting data to archival Herschel fluxes and the improvement relative to the Level~4 data available on IRSA at the time of writing.

The pixel scale of the HAWC+ polarization maps and the effective beam size is 3.4\,$''$ and $14.0''$, respectively, in Band~D, and $4.6''$ and $18.7''$ in Band~E. The final data products for both Band~D and E contain the $I$, $Q$ and $U$ Stokes parameters, the polarization fraction $P$ and angle $\theta$, and their uncertainties \citep{Gordon2018}. \citet{melanie_clarke_2023} gives the full calculation for each quantity, which we summarize here, assuming that the cross-terms of the error covariance matrix are negligible.  

Stokes $I$ describes the total dust thermal emission, and its polarized component $I_p'$ is calculated using the following equation:
\begin{equation}
    I_p' = \sqrt{Q^2 + U^2}.
\label{eq:pi'}
\end{equation}
The uncertainty $\delta I_p'$ on the polarized intensity $I_p'$ is then 
\begin{equation}
    \delta I_p' = \dfrac{\sqrt{ (Q \, \delta Q)^2 + (U \, \delta U)^2 }}{I_p'},
\label{eq:delta_pi'}
\end{equation}
where $\delta Q$ and $\delta U$ are the uncertainties on Stokes~$Q$ and $U$, respectively. 

The polarized intensity $I_p'$ has to be corrected for the bias created by the quadratic addition of the noise in the Stokes $Q$ and $U$ maps \citep{Wardle1974,Naghizadeh1993}. This de-biased polarized intensity $I_p$ \citep[although this is not the only way to do so,][]{montier_2015}, used in this work, is calculated using 
%the following equation%:
\begin{equation}
    I_p = \sqrt{I_p'^2 - \delta I_p'^2},
\label{eq:pi}
\end{equation}
with uncertainty $\delta I_p = \delta I_p'$.

The polarization fraction $P$ is then obtained from
\begin{equation}
    P = 100 \, \frac{I_p}{I},
\label{eq:pfrac}
\end{equation}
with the uncertainty $\delta P$ given by
\begin{equation}
    \delta P = P \, \sqrt{\left( \frac{\delta I_p'}{I_p'} \right) ^2 + \left( \frac{\delta I}{I} \right) ^2}.
\label{eq:dpfrac}
\end{equation}

The polarization angle $\theta_p$ is defined using the Stokes parameters $Q$ and $U$ as
\begin{equation}
    \theta_p = \dfrac{1}{2} \, \arctan{\dfrac{U}{Q}},
\label{eq:theta}
\end{equation}
with the uncertainty $\delta \theta_p$ given by
\begin{equation}
    \delta \theta_p = \dfrac{1}{2} \, \dfrac{\sqrt{ (Q \, \delta U)^2 + (U \, \delta Q)^2 }}{I_p'^2}.
\label{eq:deltatheta}
\end{equation}
Since the thermal emission from interstellar dust grains is preferentially polarized perpendicular to the plane of the sky magnetic field lines \citep[][and references therein]{HoangLazarian2014,andersson_2015}, the direction of the magnetic field in the plane of the sky can be obtained by adding $\pi/2$ to Eq.\,\ref{eq:theta}.

Prior to the data analysis, we filter out low signal-to-noise ratio (S/N) data points. After masking the data, we keep pixels that show an S/N $\geq50$ in total intensity, $\leq30^o$ in polarization angle uncertainty, and $\leq30\%$ in polarization fraction. 
Our HAWC+ observations filter out a non-negligible fraction of low-level extended emission. We estimate the amount of missing flux in the SOFIA observations by comparing our data with PACS measurements in Appendix\,\ref{sec:sofia_flux}. The data set used in this study was reduced using a larger number (40) of iterations, or rounds, than the default (15) used for the original Level~4 data available on IRSA. A larger number of iterations during data reduction typically improves the recovery of diffuse large-scale emission for scan mode maps. We compare our reduced maps of each Stokes parameter and the polarization angle with those derived from the standard pipeline setup and show them in App.\,\ref{sec:sofia_flux} in Figs.\,\ref{fig:data_comp} and \ref{fig:data_comp2}. We find an overall good agreement between these two data sets. 

\subsection{IRAM 30-meter observations of Orion~B}
\label{subsec:iram}

\begin{table*}[t!]
\centering
\caption{Components of the hyperfine structure of the $^{12}$CN\Jone\ transition, their relative offsets, and intensities relative to the sum of all components \citep{milam_2009}.}
%%%%%%%%%%%%%%%%%%%%%%%%%%%%%%%%%%%
\begingroup
\setlength{\tabcolsep}{8pt} % Default value: 6pt
\renewcommand{\arraystretch}{1.2} % Default value: 1
%%%%%%%%%%%%%%%%%%%%%%%%%%%%%%%%%%%
\begin{tabular}{cccc}
\hline
Transition F        & Rest frequency {[}GHz{]} & Relative offset {[}km/s{]} & Relative intensity \\ \hline \hline
3/2 $-$ 1/2 & 113.488142               & 7.51                       & 0.1235       \\
5/2 $-$ 3/2 & 113.490985               & 0                          & 0.3333       \\
1/2 $-$ 1/2 & 113.499643               & -22.88                     & 0.0988       \\
3/2 $-$ 3/2 & 113.508934               & -47.45                     & 0.0988       \\
1/2 $-$ 3/2 & 113.520414               & -77.8                      & 0.0123       \\    \hline  
\end{tabular}
\endgroup
\label{tab:hfs}
\end{table*}

In our work, we make use of information of the \twCN\Jone\, and \HCOp\Jone\, emission from the ongoing IRAM-30m ORION-B Large Program \citep[PIs: M.~Gerin \& J.~Pety, see the left panel in Fig.\,\ref{fig:fig1} taken from][]{pety_2017}. ORION~B images a 5 square degree field ($\sim$20 pc across) in the Orion~B molecular cloud, at an angular resolution of 26$^{\prime\prime}$ ($10^4$ AU, $0.05$\,pc) in at least 30 molecular lines in the (71--79) and (84--116)\,GHz range with a spectral resolution $\sim$0.6 km s$^{-1}$. These include common molecular tracers such as \twCO\Jone, \HCOp\Jone, HCN\Jone, and CS\Jtwo, as well as their optically thin isotopologues, which have narrow line widths and are most sensitive to kinematic variations. The resulting wide-field hyper-spectral data cube is genuinely unique in terms of its massive information content ($\sim$820,000 pixels, $\sim$240,000 spectral channels per pixel), enabling an unprecedented characterization of the physical structure, chemistry, and dynamics of a GMC, and their connection to its star formation activity. 

The $J=1 - 0$ lines were observed in 2013--2020 in the context of the ORION-B Large Program, using a combination of the EMIR receivers and FTS spectrometers. The data reduction is described in~\citet{pety_2017}. It uses the standard methods provided by the \texttt{GILDAS}\footnote{See \url{http://www.iram.fr/IRAMFR/GILDAS} for more information about the GILDAS software~\citep{pety_2005}.}\texttt{/CLASS} software. The data cubes were reprojected on the same astrometric grid as the SOFIA data. We use the data cubes at their original spatial resolution of $30''$ and $23''$ for the \HCOp\, and \twCN\, lines, respectively. The velocity spacing is 0.5\,km/s. The data are calibrated on the main beam temperature scale. The achieved noise levels are 0.26 and 0.58\,K \citep[see Tab.\,2 in][]{gratier_2017}, respectively, over the studied field of view.

We use spectroscopic data of cyanide (\twCN) and formyl cation (\HCOp) ($J= 1 - 0$ transition line) to trace the UV-irradiated gas \citep{bron_2018}. CN is present in UV-illuminated edges as a photodissociation product of HCN and HNC, resulting from UV-dominated chemistry, and can be collisionally excited by electrons and neutrals \citep{santa-maria_2023}. Therefore, the CN rotational lines remain bright in dense UV-illuminated edges, which makes them excellent tracers of the UV-dominated regions, like the one seen in NGC~2024. In addition, CN shows the multiple hyperfine components \citep{penzias_1974}, spectrally resolved by our observations. We report the properties of each hyperfine component used in this work in Tab.\,\ref{tab:hfs}, which allow the opacity \citep{milam_2009} and excitation temperature of the $\Jone$ transition to be derived, enabling its excitation to be constrained accurately and for limits on the gas density to be obtained.

\HCOp\,, similarly to CN, traces photon-dominated regions \citep{youngowl_2000,lis_2003}. \HCOp\, has different production pathways dominant at different physical conditions of the gas. At the edges of the dissociation region, \HCOp\, is a precursor of CO, and can be produced through the CO$^+$ \citep{hogerheijde_1995, young_2000, van-der-werf_1996} or CH$^+$ molecule \citep[][]{goicoechea_2016}. Recent studies detected \HCOp\, emission the edge of the PDR in the Orion bar \citep{goicoechea_2016}, and the Horsehead nebula \citep{hernandez-vera_2023}.
\cite{pety_2017, gratier_2017} found that CN and \HCOp\, are sensitive to the UV radiation. Moreover, by applying a clustering algorithm, \cite{bron_2018} found that CN and \HCOp\ trace the UV-radiated gas, contrary to the \CeiO\ molecule, which gets photodissociated. Therefore, using CN and \HCOp\ emission in our work, together with the FUV heated dust traced by SOFIA HAWC+ observations, we can directly assess the gas impacted by the radiative stellar feedback in NGC~2024. 

%%%%%%%%%%%%%%%%%%%%%%%%%%%%%%%%%%%%%%%%%%%%%%%%%%%%%%%%%%%%%%%%%%%%%%%%%%%%%%%%%%%%%%%%%%%%%%%%%%%%%%%%%%%%%%%%%%%%%%%%%%%%%%%

\begin{figure*} %[t!]
    \centering
    \includegraphics[height=0.92\textheight]{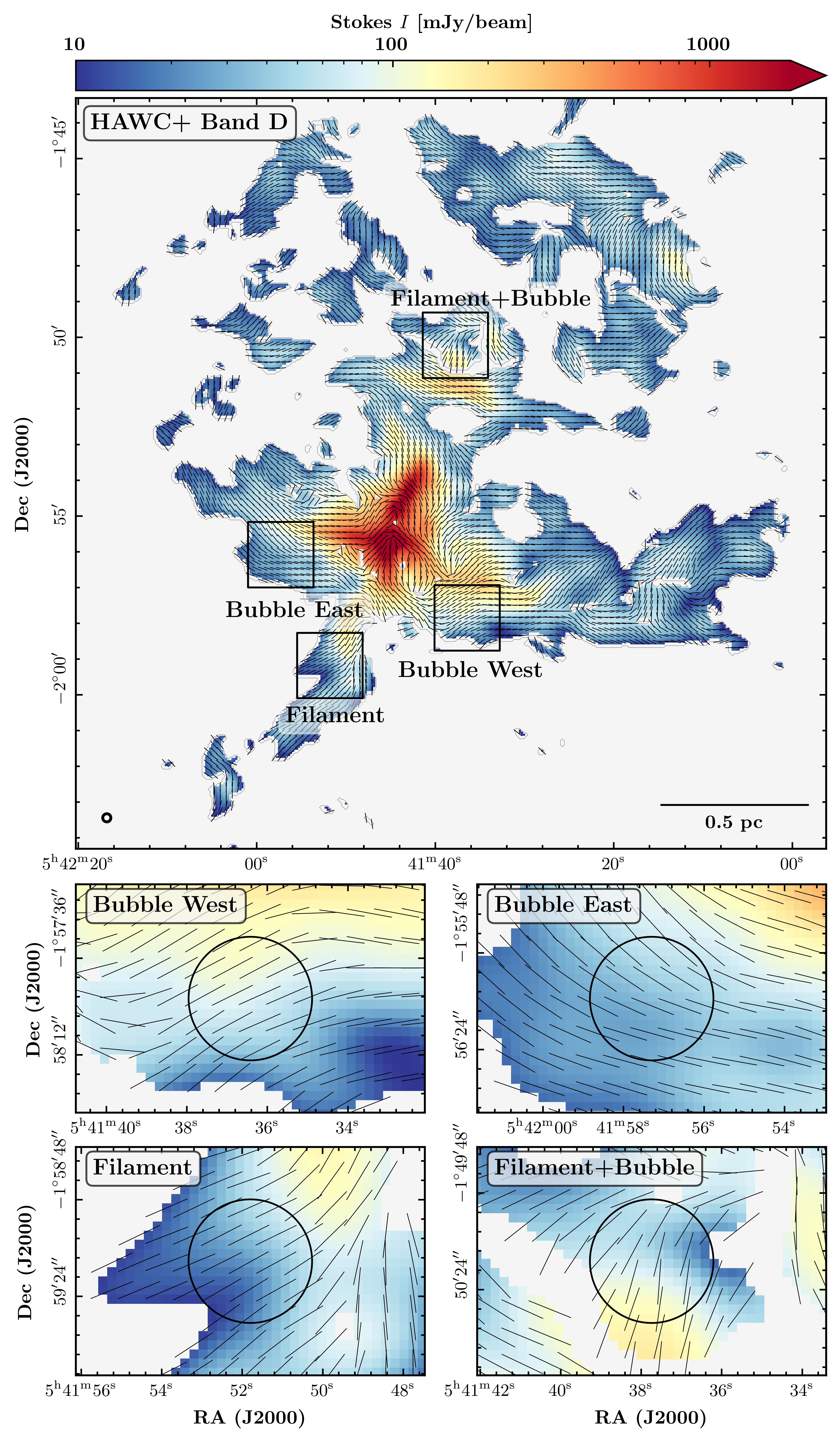}
    \caption{SOFIA HAWC+ $154\,$\textmu m (Band~D, all panels) dust continuum maps at $13.6^{\prime\prime}$ angular resolution. The maps are masked based on the measured S/N in measured Stokes intensity and polarization angle. Black lines in all panels show the orientation of the magnetic field for every fifth pixel. We show Band~D dust polarization map across NGC~2024 (top panel). Bottom rows show the zoomed-in regions we analyse in this work: egde of the bubble on the western (middle left panel), eastern (middle right panel), in the filament (bottom left panel), and in the overlap region of the filament and bubble (bottom right panel). A circle in each of these zoomed-in panels shows the region we use to compute the dispersion in the mean angle of magnetic field, as well as number density and turbulent velocity dispersion.}
    \label{fig:dust_pol}
\end{figure*} 

\section{Dust polarization}
\label{sec:mag_fields}

In this section, we discuss our SOFIA HAWC+ dust polarization measurements introduced in Sec.\,\ref{subsec:hawc+}, and present the magnetic field morphology across NGC~2024. 

The SOFIA HAWC+ Band~D measurement is shown in Fig.\,\ref{fig:dust_pol}. The map is masked, as explained above, and is presented at its native angular resolution of 13.6$^{\prime\prime}$, which corresponds to linear scales of $\sim 0.027$\,pc. The background of this figure (all panels) shows total Stokes intensity, and the black lines indicate the orientation of the magnetic field. In our work, we mainly focus on four specific regions across NGC~2024: the edges of the bubble (middle panels in Fig.\,\ref{fig:dust_pol}), the filament (bottom left panel in Fig.\,\ref{fig:dust_pol}), and the intersection of filament and bubble (bottom right panel in Fig.\,\ref{fig:dust_pol}). Each of these panels shows a circular-shaped, beam-sized region within which we computed the magnetic field strength. We chose a beam-sized region for data analysis rather than taking the information from a single pixel to get suitable properties of each region.

We show SOFIA HAWC+ Band~E map in Fig.\,\ref{fig:hawc+_e} in App.\,\ref{app:nir_pol}. The overall agreement in magnetic field direction traced by dust polarization at 154 and 214\,\textmu m is observed within NGC~2024. We show the difference in measured polarization angles and their rms in Fig.\,\ref{fig:hawc+_comparison} in App.\,\ref{app:nir_pol}. While the rms is only a few degrees ($\sim2\,\deg$) across a large part of NGC~2024, we observe higher values ($>10\,\deg$) toward the center of NGC~2024. Nevertheless, we consider only Band~D data in this work because of an overall good correspondence (measured as a low rms) between dust polarization angles measured from Band~D and E across the regions we further analyze in this work. %In the following, we analyze only Band~D.}

%%%%% comparison of these two maps 
The dust continuum emission is strongest toward the center of NGC~2024. A significant emission is observed along the filament that is in the front of the ionization region caused by the young massive star IRS~2b \citep[the central part of our map,][]{bik_2003}, and along the ionization fronts on the southeastern and southwestern part of NGC~2024 (labeled as Bubble West and Bubble East, and shown in the middle row in Fig.\,\ref{fig:dust_pol}). 
%Next, we investigate the orientation of the magnetic field across NGC~2024. 

In addition, we compare our magnetic field directions to the direction inferred from the near-infrared (NIR) polarization of young stars \citep{kandori_2007}. The NIR polarization directly traces the direction of the magnetic field (that is, there is no $\pi/2$ difference as for the FIR dust polarization - Sec.\,\ref{subsec:hawc+}). We show this comparison in the top panels of Fig.\,\ref{fig:hawc+_sources} in App.\,\ref{app:nir_pol}. As seen from the figure, we find that magnetic field vectors derived from NIR polarization and our work are parallel to each other, indicating an agreement between these two datasets in NGC~2024 (see Fig.\,\ref{fig:hawc+_sources} in App.\,\ref{app:nir_pol}). In addition, we compare the magnetic field direction with the 100\,\textmu m \citep{dotson_2000} and submillimeter (850\,\textmu m) polarization measurements \citep{matthews_2002} in the central part of NGC~2024 (labeled with the dashed black rectangle in top panels of Fig.\,\ref{fig:hawc+_sources}) and show zoomed-in panels in the bottom row of Fig.\,\ref{fig:hawc+_sources}.

\begin{figure*}[t!]
    \centering
    \includegraphics[width=\textwidth]{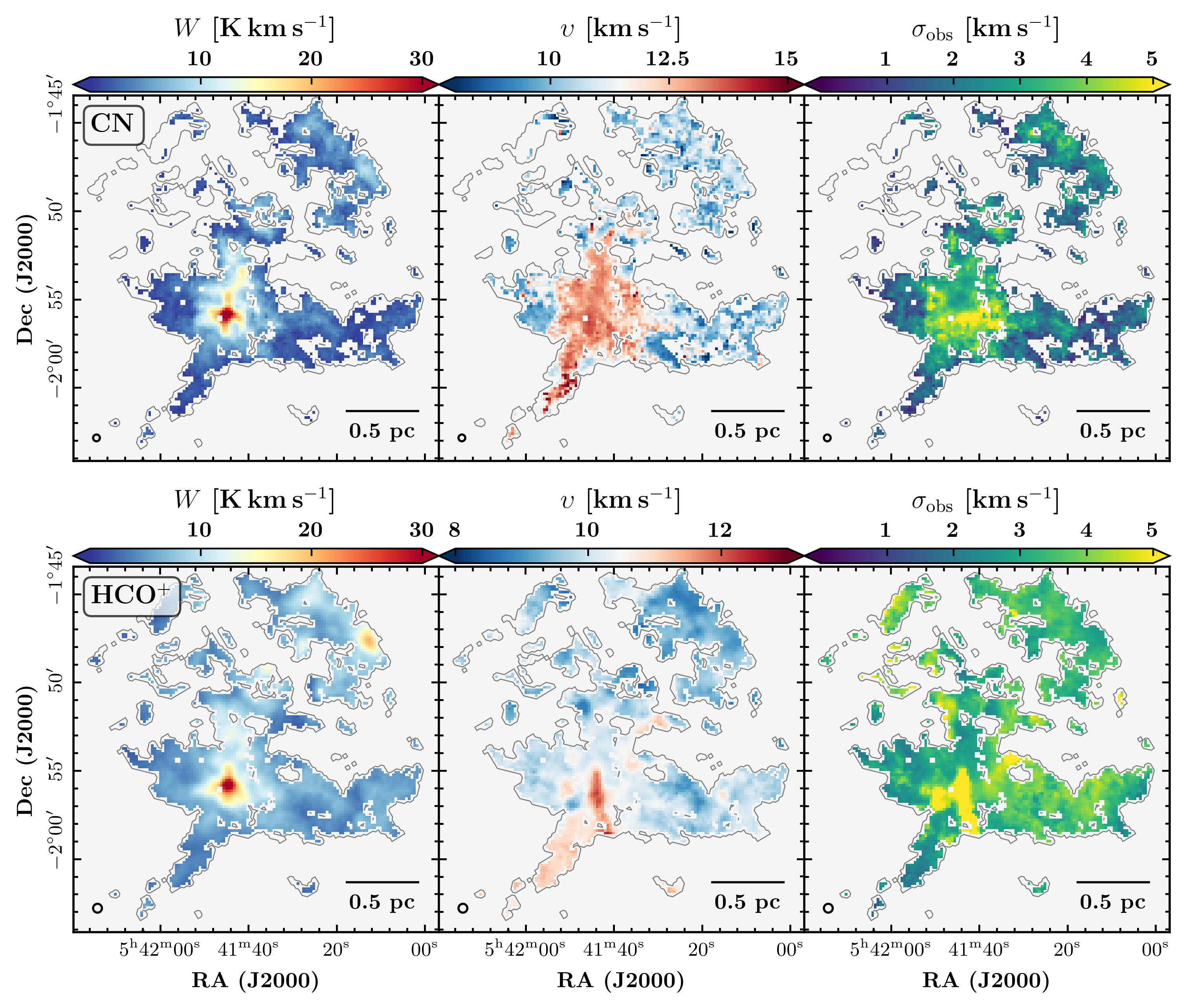}
    \caption{Moment maps of \twCN\Jone\, and \HCOp\Jone\, emission across NGC~2024. Top: Maps of the CN emission: integrated intensity (the zeroth moment, left panel), the centroid velocity (the first moment, middle panel), and the FWHM (the second moment, right). Bottom: The same as in the top row, but for \HCOp. We show the beam size for both molecular lines at the bottom left corner of each panel and the $0.5\,$pc scalebar at the bottom right corner. All pixels shown in these maps result from the masking technique presented in \cite{einig_2023}. We refer the reader to the description of the production of these moment maps in Sec.\,\ref{sec:mom_maps}.} % The light black contour corresponds to $3\sigma$ limit on determination of the relevant quantity. }
    \label{fig:moments}
\end{figure*}

%%%% the part with the outflow
The area labeled with dark blue contours in the top panels in Fig.\,\ref{fig:hawc+_sources} is particularly interesting because we observe an outflowing feature in the \HCOp\ emission at velocities of $\sim14$\,km\,s$^{-1}$. The outflow originates from the FIR\,5 source located in the dense molecular cloud behind the expanding \HII\, region \citep[][]{richer_1992, greaves_2001, choi_2015}. Dust continuum emission at 154\,\textmu m possibly traces the outflow in this region (see also the bottom middle panel in Fig.\,\ref{fig:moments}), as we note that magnetic field lines follow the direction of the outflow. We have not observed this behavior at 214\,\textmu m.

The magnetic flux is frozen within the molecular gas, and as a consequence, the $B$ field will trace its morphology. Therefore, the local environmental conditions that shape the distribution of molecular gas will also impact the magnetic field. On the one hand, the magnetic field is highly ordered in some regions of NGC~2024; for instance, $B$ field follows the dusty filament to the south of NGC~2024 (see, for instance, the bottom left panel of Fig.\,\ref{fig:dust_pol}), except for the very southern part of the filament (close to the bottom left corner of the top panel in Fig.\,\ref{fig:dust_pol}), where the magnetic field is perpendicular to the filament. At the northern part of NGC~2024, the magnetic field direction varies from following the filament to being perpendicular.

We observe the nearly horizontal direction of the $B$-field at the edges of the expanding \HII\ region (middle panels in Fig.\,\ref{fig:dust_pol}). Gas affected by the stellar feedback (for example, UV radiation, stellar winds) is pushed outward from the \HII\, region. Consequently, magnetic field lines become parallel to the edges of such an expanding shell \citep[for example,][]{tahani_2023}.
On the other hand, the magnetic field appears chaotic in the central and densest parts of NGC~2024. This area shows great complexity, as it results from the mixture of the filament located in the front, the \HII\, region and the dense molecular gas in the background (see Fig.\,8 in \citealt{matthews_2002}, or Fig.\,5 in \citealt{roshi_2014}). One possible explanation for this observed morphology of the magnetic field is that the magnetic field changes its orientation relative to the filament depending on the column density contrast between the filament and the background emission, from being parallel to perpendicular to the molecular gas distribution \citep[$\Delta N_\mathrm{H_2}>10^{20}$\,cm$^{-2}$, for example,][]{planck_2016c, alina_2019}. Additionally, the magnetic field could be ''pinched" due to the gravitational collapse of the gas \citep{basu_2000, lai_2002, doi_2020, doi_2021}, which can also explain the magnetic field direction at the northern part of NGC~2024 and the very southern part of the filament.

\section{Characterizing the turbulence and density structures in NGC~2024}
\label{sec:cn_analysis}

The DCF method (Sec.\,\ref{sec:introduction}) requires the characterization of the turbulent velocity dispersion and the gas density in the same material traced by the dust emission. Therefore, in this section, we briefly describe our analysis of molecular data from the ORION~B large program and present moment maps of \twCN\Jone\, and \HCOp\Jone\, and then provide results on measured turbulent line widths and gas volume densities of each region shown on the left panel of Fig.\,\ref{fig:fig1}. We derive these quantities using non-LTE radiative transfer models \citep[\texttt{RADEX},][]{vandertak_2007}.

\subsection{Moment maps}
\label{sec:mom_maps}

Fig.\,\ref{fig:moments} shows moment maps for the CN\Jone\, (top row) and \HCOp\Jone\, (bottom row) emission. We create these moment maps using \texttt{CUBE} in \texttt{GILDAS}. Before creating moment maps, we mask CN and \HCOp\ cubes. The mask is created using the segmentation technique in \texttt{CUBE}, where we identify neighboring voxels with continuous S/N; we have selected all voxels with S/N above 2 \citep[see][for a complete description]{einig_2023}. In this case, we integrate all molecular emission along each unmasked line of sight, without considering the possible presence of multiple velocity components. Finally, we apply to the moment maps the spatial mask where the dust polarization measurements are reliable as described in Sec.\,\ref{sec:mag_fields}. Similarly, as for the intensity map, the line width map shown in this figure corresponds to the measured second moment, which does not consider possible spectral complexity (multiple velocity components or the presence of an outflow) or a correction for the opacity broadening. These corrections could bias our measurement, particularly in the central part of NGC~2024, as shown by the spectral decomposition results of CN and \HCOp\, in Figs.\,\ref{fig:cn_fit} and \ref{fig:hcop_fit} in App.\,\ref{sec:fit_cn} and \ref{sec:fit_hcop} respectively).

The integrated line intensity (moment 0) is shown in the left panels of Fig.\,\ref{fig:moments}. Both CN and \HCOp\ show the brightest emission toward the central part of NGC~2024, at the heavily dust-obscured region, as seen in Fig.\,\ref{fig:fig1}. 

The first moment, also known as the centroid velocity map, is shown in the central panels of Fig.\,\ref{fig:moments}. The central velocity is at 10\,km/s. However, we note the presence of multiple velocity components along the line of sight in both CN and \HCOp\,, as previously identified in the \thCO\Jone\ and \CeiO\Jone\ emission \citep{gaudel_2023}. We discuss velocity components in App.\,\ref{sec:fit_cn} and \ref{sec:fit_hcop}. 

The second moment map (the observed line width ($\sigma_\mathrm{obs}$) map) is shown in the right panels of Fig.\,\ref{fig:moments}. We note that to first order the \HCOp\, line is broader than CN, which has several possible reasons. Firstly, \HCOp\ is brighter and more spatially distributed than CN and, therefore, could result in having a broader line. Second, \HCOp\, emission can be optically thick \citep[see,][]{barnes_1990}, which additionally broadens the line. Third, it is possible that \HCOp, similarly to \thCO\, and \CeiO\, has multiple velocity layers \citep{gaudel_2023}. Nevertheless, we attempt to correct these effects in our analysis described in the following sections. The regions we analyze in this work do not contain multiple velocity features in the CN emission (see Fig.\,\ref{fig:cn_fit}). However, in the case of \HCOp\, emission, we do not spectrally resolve the multiple-component \HCOp\, emission. However, we note the possibility of \HCOp\, showing multiple velocity components (Fig.\,\ref{fig:hcop_fit}) in NGC~2024. The more thorough analysis of the \HCOp\, velocity field is, therefore, beyond the scope of this paper, and will be an aim of the upcoming studies.

\subsection{Measuring turbulent velocity dispersion}
\label{sec:fwhm_etc}

Here, we briefly describe steps in order to constrain the turbulent velocity dispersion. Prior to this, we note the different natures of the \twCN\ and \HCOp\ line emission profiles. For instance, the \twCN\Jone\, has a hyperfine structure (see Fig.\,\ref{fig:cn_hcop} in App.\,\ref{sec:fit_cn}). We do not observe any anomalous hyperfine structure emission in the CN emission, as reported, for example, for the HCN emission in Orion~B \citep{santa-maria_2023}. Therefore, all components of the multiplet have the same line width. We provide more information on the properties of the hyperfine structure of the \twCN\Jone\, emission in App.\,\ref{sec:fit_cn}. The profile of the \HCOp\, can be described by a Gaussian function (see Fig.\,\ref{fig:cn_hcop} in App.\,\ref{sec:fit_hcop}) assuming its optically thin emission. In the case of the optically thick emission, the line profile will be changed and depend on the observed optical depth.

To derive the turbulent line width for both CN and \HCOp, we correct their measured line widths (that we will infer from the RADEX modeling) as follows. First, we correct our measurements for the contribution of thermal broadening:

\begin{equation}
    \sigma_\mathrm{NT, mol} = \sqrt{\sigma_\mathrm{obs, mol}^2 - \sigma_\mathrm{TH, mol}^2},
\label{eq:sigma_nt}
\end{equation}

\noindent where $\sigma_\mathrm{obs, mol}$ is the measured FWHM of a molecular line, and $\sigma_\mathrm{TH, mol}$ is the line width of the thermal component. We calculate the thermal broadening using the following equation:

\begin{equation}
    \sigma_\mathrm{TH, mol} = \sqrt{\dfrac{kT_\mathrm{k}}{m_\mathrm{mol}}},
\label{eq:sigma_th}
\end{equation}
%Next, we correct for the instrumental broadening caused by limited channel width. 

\noindent where $k$ is Boltzman's constant, $T_\mathrm{k}$ is the kinetic temperature obtained using the CO\Jone\, measurements \citep{orkisz_2017}, and $m_\mathrm{mol}$ is the mass of a molecule. A final correction that we apply in our analysis is for opacity broadening of molecular lines, using the following equation:

\begin{equation}
    \Delta\upsilon = \dfrac{\sigma_\mathrm{NT, mol}}{\beta},
\label{eq:tau_broadening}
\end{equation}

\noindent where the factor $\beta$ is a function of the optical depth ($\tau_0$) at the line center, defined as \citep{philips_1979, hacar_2016,orkisz_2017}:

\begin{equation}
 \beta = \dfrac{1}{\sqrt{\ln{2}}} \cdot \left [ \ln{\left \{  \dfrac{\tau_0}{ \ln{\left ( \dfrac{2}{\exp{(- \tau_0)} + 1} \right ) } } \right \}  } \right ]^{\dfrac{1}{2}} .
\end{equation}

The equations presented above are crucial at subparsec scales since these effects significantly contribute to the measured line width at these scales. For example, the optical depth effect could broaden the line by a few tens of percent, which is the case of CN and \HCOp. This is discussed in the following section.

The FWHM of a molecular line, corrected for the contribution mentioned above of the thermal and opacity broadening, can be derived in multiple ways. For example, the FWHM can be derived directly from measuring the second order moment map (right panel in Fig.\,\ref{fig:moments}, Sec.\,\ref{sec:mom_maps}), or from the line fitting by using, for example, the spectral decomposition (see Apps.\,\ref{sec:fit_cn} and \ref{sec:fit_hcop}), or directly from the non-LTE modeling of the emission spectrum for a set of input parameters, that describe the physical conditions of the gas within a selected region. In this work, we use the latter method, as we aim to find a set of physical parameters that describe the edges of the bubble exposed to the FUV emission and the gas coming from the filamentary structure. Therefore, we derive the FWHM from the non-LTE modeling and RADEX analysis, as well as the gas number density described in the following section.

\subsection{Radiative transfer results}
\label{subsec:radex}

To derive gas volume densities and measure the turbulence in NGC~2024, we have employed the non-LTE radiative transfer code \texttt{RADEX} using Python wrapper \citep[\texttt{SpectralRadex,}][]{holdship_inprep} because it allows the user to compute the spectrum of a line and directly compare models to observations. Using the excitation of CN and \HCOp\, and some a priori information and assumptions, we derive gas number densities, $n_\mathrm{H_2}$ and line widths across four regions in NGC~2024: two edges of the expanding \HII\, bubble (located to the west and the east), the filament, and the region consisting of the edge of the bubble and the filament (located in the north of NGC~2024).

In the following, we provide information about our input parameters and assumptions. We take the value of 2.73\,K as the background temperature, and use the kinetic temperature ($T_\mathrm{kin}$) derived from the \twCO\Jone\, peak intensity \citep{orkisz_2017}, and shown in Tab.\,\ref{tab:radex}. We create a grid of column densities in the range of (1-4)$\cdot10^{14}$\,cm$^{-2}$ for CN and (1-4)$\cdot10^{13}$\,cm$^{-2}$ for \HCOp\,, following results from \cite{bron_2018}. The selected grid of line widths covers the range from or 1 to 2.5 km/s. Next, we create a grid in molecular hydrogen densities from $10^2$ to $10^5$\,cm$^{-3}$. The density of H$_2$ is comprised of para- and ortho-H$_2$, assuming the temperature dependence of the ortho-to-para ratio: $9\cdot e^{-170/T_\mathrm{kin}}$ \citep{mandy_1993}. Here, we assume a fixed electron fraction (electron-to-H$_2$ density ratios), $f_\mathrm{e} = 1.4\cdot10^{-4}$, which is a maximum value found in PDRs for the case when number densities of ionized and neutral carbon (including CO) are equal \citep{sofia_2004,graf_2012}. Our choice of $f_\mathrm{e}$ corresponds to values of the ionization fraction for translucent medium within Orion~B found in \cite{bron_2021}. With fixed ortho-to-para H$_2$ ratio and electron fraction, we run a grid in three independent variables: line width, column density, and H$_2$ volume density.

We consider all relevant collision partners for CN and \HCOp \, in our input files. For the CN emission, we take into account its hyperfine structure \citep{muller_2005}, and include collisions with ortho- and para-H$_2$ \citep{kalugina_2012}, as well as electrons \citep{harrison_2013, santa-maria_2023}. The input file with collision partners comes from the combination of two data files from the Leiden Atomic and Molecular Database \citep[\href{https://home.strw.leidenuniv.nl/~moldata/}{LAMDA};][]{lamda_2005} and Excitation of Molecules and Atoms for Astrophysics \citep[\href{https://emaa.osug.fr/}{EMAA};][]{emaa_2021} to include all relevant collisional partners and take into account the hyperfine structure of CN. In the case of \HCOp\,, we consider collisions with ortho-, para-H$_2$ \citep{denis_alpizar_2020} and electrons \citep{fuente_2008} as well. 

The output parameters from the RADEX modeling are the excitation temperature, $T_\mathrm{ex}$ and opacity, $\tau$. We additionally compute the peak temperature of the spectrum generated from each model using \texttt{SpectralRadex}. In the case of CN, we calculate excitation temperature and the opacity of each component of the multiplet, which is scaled using information about relative intensities (Tab.\,\ref{tab:hfs}).

We compute model CN\Jone\, and \HCOp\Jone\, spectra assuming the optical depth to be a Gaussian function of frequency, $\tau(\nu)$:

\begin{equation}
    T_{mb}(\nu) = (T_\mathrm{ex} - T_\mathrm{bg})\cdot(1 - e^{-\tau(\nu)}).
\label{eq:temp}    
\end{equation}

The best-fit model is found based on the modeled spectrum for each combination of $(n_\mathrm{H_2}, N_\mathrm{mol}, \Delta\upsilon)$. We perform a $\chi^2$ minimization for the modeled and observed peak temperature, $T_\mathrm{peak}$. Then, we calculate the turbulent line width following the prescription in Sec.\,\ref{sec:fwhm_etc}. We report our turbulent line width and gas number density results in Tab.\,\ref{tab:radex}. Model spectra inferred from the RADEX analysis of CN and \HCOp\, for a selection of physical parameters that best describe the observed spectra in each region we investigate in this work are shown in Fig.\,\ref{fig:cn_hcop} in App.\,\ref{sec:radex_all}. 

\begin{table*}
\centering
\caption{Derived physical conditions of molecular gas traced by CN and \HCOp\, emission for the four regions we investigate in NGC~2024, which are shown in Fig.\,\ref{fig:dust_pol}. We tabulate position of each region, and kinetic temperature inferred from the CO data. Next, we tabulate results from the non-LTE radiative transfer modeling using RADEX: volume number density, and measured turbulent line widths.}
%%%%%%%%%%%%%%%%%%%%%%%%%%%%%%%%%%%
\begingroup
\setlength{\tabcolsep}{8pt} % Default value: 6pt
\renewcommand{\arraystretch}{1.25} % Default value: 1
%%%%%%%%%%%%%%%%%%%%%%%%%%%%%%%%%%%
\resizebox{\textwidth}{!}{ \begin{tabular}{lccccccc}
\hline
Region & RA [deg] & Dec [deg] & $T_\mathrm{kin}$ [K] & $\Delta\upsilon_\mathrm{CN}$ [km/s] & $n_\mathrm{H_2}\mathrm{(CN)}$ [$10^3$cm$^{-3}$] & $\Delta\upsilon_\mathrm{HCO^+}$ [km/s] & $n_\mathrm{H_2}\mathrm{(HCO^+)}$ [$10^3$cm$^{-3}$] \\ \hline \hline
Bubble West & 85.4018 & -1.9642 & 53.04 & 0.56 & 2.27 & 1.10& 2.93 \\
Bubble East & 85.4768 & -1.9449 & 32.63 & 1.24 & 4.87 & 1.24& 4.93 \\
Filament & 85.4658 & -1.9864 & 27.18 & 0.62 & 4.72 & 0.96& 2.31 \\
Filament+Bubble & 85.3490 & -1.7587 & 30.80 & 0.94 & 13.37 & 1.07& 5.39 \\
\hline
\hline
\end{tabular} }
\endgroup
\label{tab:radex}
\end{table*}

\begin{figure*}[t!]
    \centering
    \includegraphics[width=\textwidth]{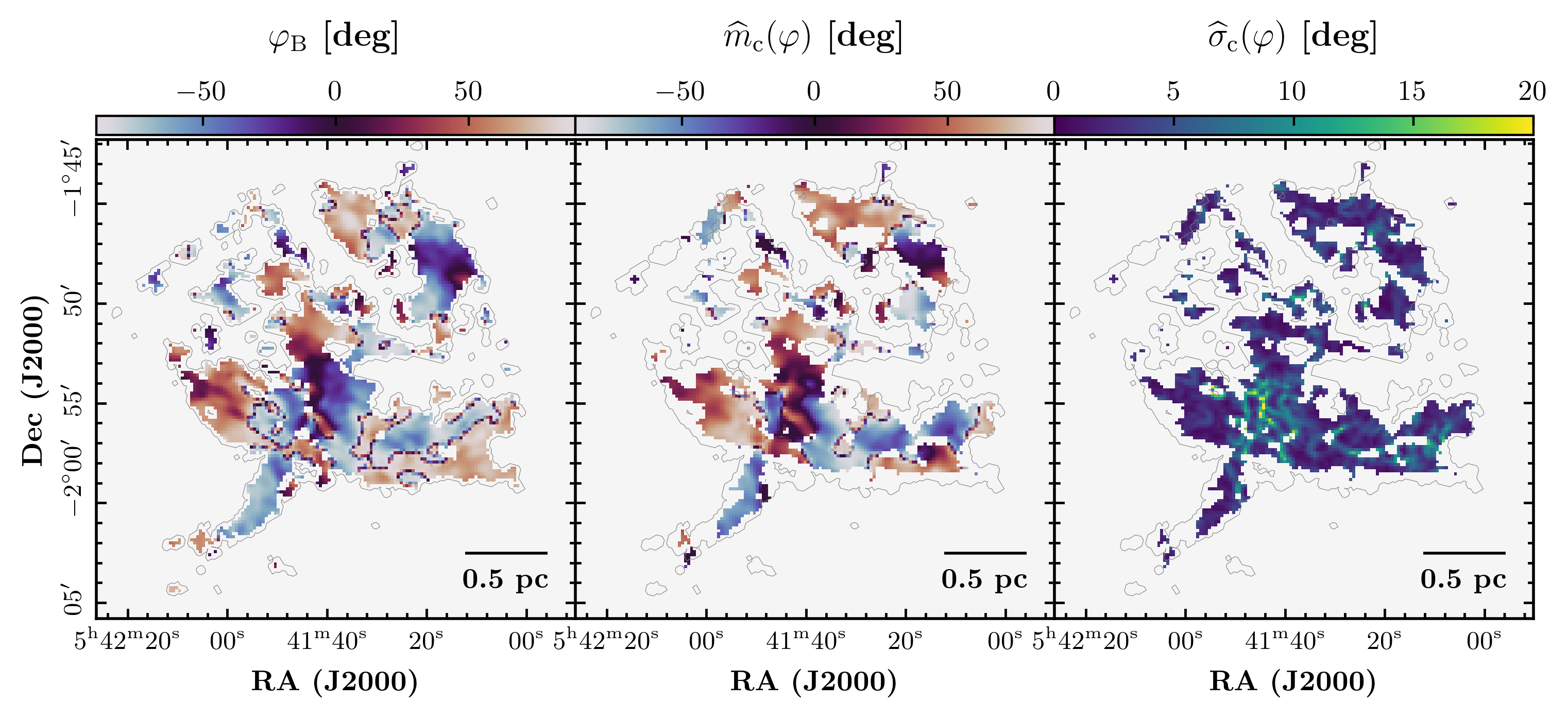}
    \caption{Direction of the magnetic field (left panel), the mean angle of the magnetic field (middle panel), and the rms of the magnetic field direction (right panel). Maps in the middle and right panels have been made using a $12\times 12$ pixels sliding window with weight inversely proportional to the measured uncertainty variance. All maps have been regridded to match the grid size of the CN and \HCOp\, data. The gray contour indicates the mask we defined for dust polarization measurements, described in Sec.\,\ref{subsec:hawc+}.}
    \label{fig:angle}
\end{figure*} 

\section{Magnetic field strength}
\label{sec:Bfield_results}

\subsection{Davis-Chandrasekhar-Fermi method}
\label{subsec:dcf}

To derive the strength of the magnetic field, we use the Davis-Chandrasekhar-Fermi method \citep{davis_1951, chandrasekhar_fermi_1953}, which provides a recipe for calculating the POS $B$-field strength as following:
\begin{equation}
    B_\mathrm{pos} = \sqrt{4\pi\rho}\dfrac{\sigma_\mathrm{NT}}{ \widehat{\sigma}_\mathrm{c}(\varphi) },
\label{eq:bpos}
\end{equation}
where $\rho$ is the gas volume density, $\sigma_\mathrm{NT}$ is the nonthermal velocity dispersion ($\sigma_\mathrm{NT} = \Delta\upsilon / \sqrt{8\ln{2}}$), where $\Delta\upsilon$ is the measured FWHM of the line, corrected for the line broadening and opacity effects - Eq.\,\ref{eq:sigma_nt} and \ref{eq:tau_broadening} - Sec.\,\ref{sec:fwhm_etc}, and $\widehat{\sigma}_\mathrm{c}(\varphi)$ is the spatial dispersion of magnetic field angle.

This equation is in CGS units, and the DCF method also assumes that any perturbation in the magnetic field originates from local, small-scale turbulence. The stronger the magnetic field, the smaller will be the perturbation caused by turbulence. By calculating all constants and keeping the units of number density, line width and the angle dispersion in cm$^{-3}$, km\,s$^{-1}$, and deg, respectively \citep[][]{lyo_2021}, Eq.\,\ref{eq:bpos} can be expressed as: 

\begin{equation}
    B_\mathrm{pos} \approx 9.3\cdot\sqrt{n_\mathrm{H_2}}\cdot\dfrac{\Delta\upsilon}{\widehat{\sigma}_\mathrm{c}(\varphi)}\,[\rm \upmu G].
\label{eq:bpos2}
\end{equation}

We have included in the above a factor of 0.5 for overestimation of the magnetic field strength due to line of sight integration effects \citep[see, for instance,][]{ostriker_2001}.

\subsection{Skalidis-Tassis method}
\label{subsec:skalidis}

The DCF method is widely used in the literature to derive the strength of the magnetic field \citep[see,][]{pattle_2019}. However, it assumes that isotropic turbulence and Alfvénic waves in an incompressible medium cause the observed dispersion in polarization angle.
Other mechanisms such as magneto-hydrodynamic waves \citep{heyvaerts-priest_1983} and entropy modes \citep{lithwick-goldreich_2001} also cause fluctuations in polarization angle. Therefore, to derive the magnetic field strength in NGC~2024, it is important to acknowledge the contribution of non-Alfvénic motions and the compressible nature of the ISM. In this work, we derive the $B_\mathrm{POS}$ using the prescription presented in \cite{skalidis_2021a, skalidis_2021b}:
\begin{equation}
   B_\mathrm{pos} = \sqrt{2\pi\rho}\dfrac{\sigma_\mathrm{NT}}{ \sqrt{\widehat{\sigma}_\mathrm{c}(\varphi)} },
\label{eq:bpos_skalidis}
\end{equation}
where $\rho$, $\sigma_\mathrm{NT}$, and $\widehat{\sigma}_\mathrm{c}(\varphi)$ are the same as in Eq.\,\ref{eq:bpos}.

Similarly as in Eq.\,\ref{eq:bpos2}, after substituting all constants, the above equation becomes: 

\begin{equation}
 B_\mathrm{pos} \approx 1.8\cdot\sqrt{n_\mathrm{H_2}}\cdot\dfrac{\Delta\upsilon}{\sqrt{\widehat{\sigma}_\mathrm{c}(\varphi)}}\,[\rm \upmu G].
\label{eq:bpos2_skalidis}
\end{equation}

We favor using this approach rather than the classical DCF because it is physically motivated, considering the nature of the ISM in NGC~2024. The ST method takes into account the compressible nature of the gas, and \cite{orkisz_2017} found compressible, non-Alfvénic motions to dominate over solenoidal modes in NGC~2024. 

\subsection{Sliding window}
\label{subsec:slidding_window}

We present the dispersion of the mean direction of the magnetic field in Fig.\,\ref{fig:angle}. We produced this map by using a "sliding window" to remove the impact of gradients of the large-scale magnetic field \citep["unsharp-masking",][]{slidding_window, pattle_2017}. The "sliding window" technique is based on computing the mean and standard deviation of the magnetic field orientation within the window that is three times bigger than the beam and has $12\times12$ pixels. The size of the sliding window is made to ensure that we remove a large-scale magnetic field contribution without losing information on the small-scale perturbation of the magnetic field \citep{pattle_2017}. As shown in \cite{mardia99}, a suitable way to estimate the mean and standard deviation from the $12\times 12= 144$ directions $\varphi(l)$ of each window is to compute:
\begin{equation}
 z=\dfrac{1}{L} \sum_{l=1}^{L} e^{2i\varphi(l)},   
\end{equation}
where $L$ is the number of pixels within the sliding window. % (11). 
Defining $a$ and $b$ such that $z=a+i\cdot b$, one computes the circular mean using
\begin{equation}
    \widehat{m}_\mathrm{c}(\varphi) = \frac{1}{2} \arctan\dfrac{b}{a},
\label{eq:circ_mean}
\end{equation}
and the circular standard deviation using
\begin{equation}
    \widehat{\sigma}_\mathrm{c}(\varphi) = \sqrt{ \dfrac{1}{2} \left ( 1 - \left | z \right | \right ) }.
\label{eq:circ_std}
\end{equation}
Moreover, to consider the uncertainty associated with the different directions $\varphi(l)$, we use a weighted mean to compute $z$ with weight inversely proportional to the variance.
The error of the circular standard deviation is computed as follows:
\begin{equation}
    \Delta(\widehat{\sigma}_\mathrm{c}(\varphi)) = \dfrac{\widehat{\sigma}_\mathrm{c}(\varphi)}{\sqrt{2\cdot L - 2}}.
\end{equation}

We show maps of the magnetic field angle, the mean angle computed using Eq.\,\ref{eq:circ_mean} and $12\times12$ window, and the circular standard deviation (Eq.\,\ref{eq:circ_std}) in Fig.\,\ref{fig:angle}. As seen in the left panel of Fig.\,\ref{fig:dust_pol}, the magnetic field direction differs along the edges of the expanding shell and the filament. A similar behavior we observe in the mean angle of the POS magnetic field is shown in the middle panel of Fig.\,\ref{fig:angle}. The standard deviation of the angle that is caused by the small-scale turbulence also shows different values within the borders of the \HII\ region and the filament. In particular, the measured $\widehat{\sigma}_\mathrm{c}$ is higher along the filament (more than 10 degrees), whereas it is a few degrees at the edges of the bubble. This result indicates that the magnetic field is possibly stronger at the edges of a bubble than in the filament. This result could also be a consequence of a complex geometry of NGC~2024: near the edge of a bubble, we observe a coherent limb-brightened structure, whereas, near the center of NGC~2024, we find a superposition of several components.

\begin{figure*}
    \centering
    \includegraphics[width=\textwidth]{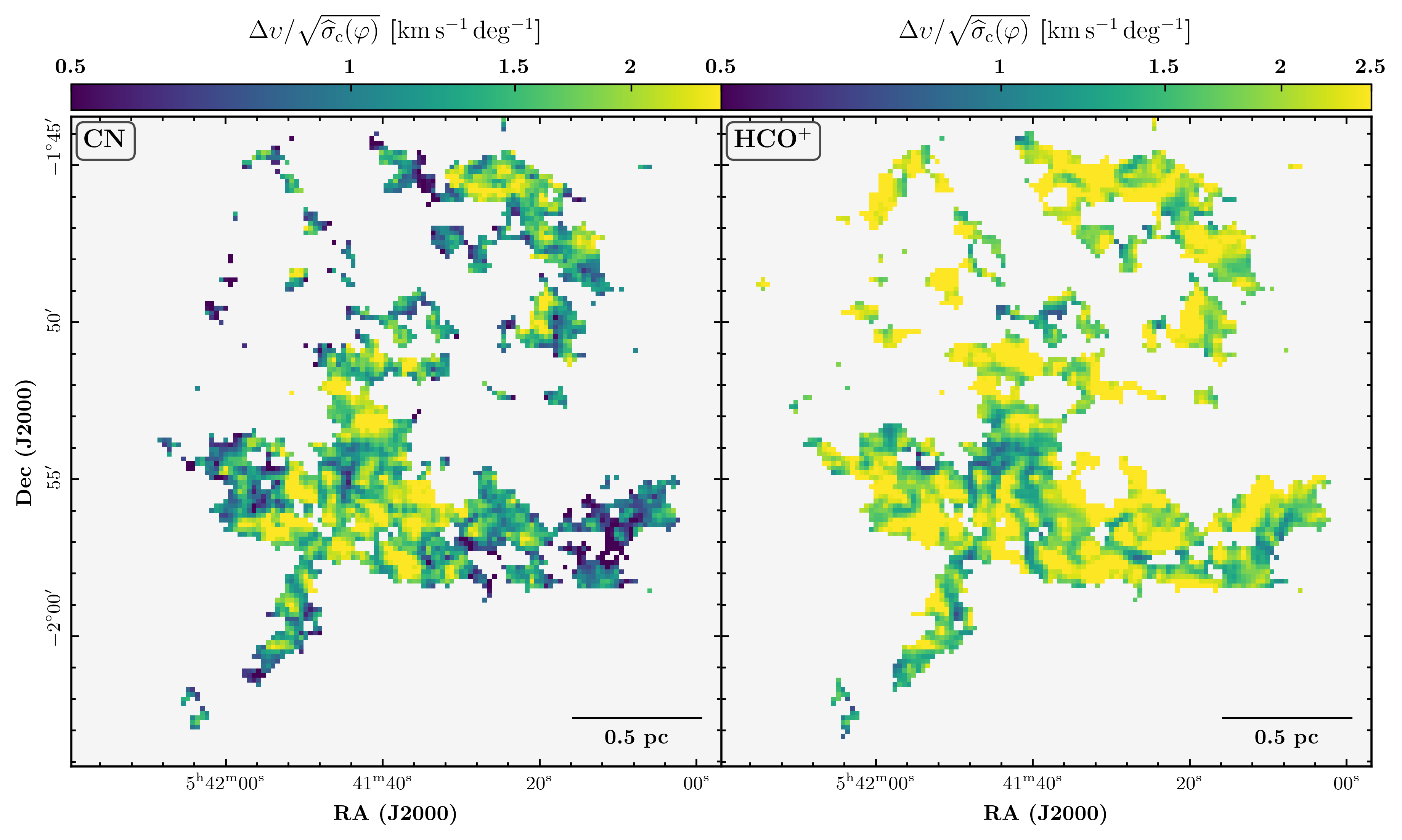}
    \caption{Ratio between measured line widths (right panels in Fig.\,\ref{fig:moments}) and the square root of the circular standard deviation of the magnetic field angle shown in the right panel of Fig.\,\ref{fig:angle}. The left panel shows the line width of CN divided by the square root of the rms of an angle, and the right panel shows the same as on the left panel, but here we use the \HCOp line width.}
    \label{fig:lw_vs_angle}
\end{figure*}

The ratio between the line width and the angle dispersion in Eq.\,\ref{eq:bpos2} is shown in Fig.\,\ref{fig:lw_vs_angle}. The two panels show the measured ratios between the line width of the CN line (left) and \HCOp\, emission (right) to the dispersion of the angle of the magnetic field. We used the line width information derived from the second moment map shown in the right panels in Fig.\,\ref{fig:moments}. This approach gives a robust estimate of the range of the magnetic field strength assuming a specific number density of the gas because the second moment map provides information on the line width of all emission along the line of sight without considering possible impacts (see Secs.\,\ref{sec:mom_maps} and \ref{sec:fwhm_etc}). 

By assuming a gas number density of, for instance, $10^{4}$\,cm$^{-3}$, using Eq.\,\ref{eq:bpos2}, the ratio of 0.5 km\,s$^{-1}$\,deg$^{-1}$ in Fig.\,\ref{fig:lw_vs_angle} corresponds to $90$\,\textmu G, whereas a ratio of 2 km\,s$^{-1}$\,deg$^{-1}$ corresponds to 360\,\textmu G. We note, however, that using the second order moment for getting information about line width does not take into account the opacity of the corresponding line, or a possible spectral complexity.

\subsection{Magnetic field strength in NGC~2024}
\label{subsec:bfield_strength}

We calculate the strength of magnetic field across the beam-averaged area at the western (Bubble West) and eastern side (Bubble East) of the bubble, the filament (Filament), and across the region where the dust filament overlaps with the expanding shell of the \HII\ region (Filament+Bubble) (Fig.\,\ref{fig:dust_pol}, also see Sec.\,\ref{sec:cn_analysis}). We measure the uncertainty of magnetic field strength derived using the DCF method (Sec.\,\ref{subsec:dcf}) as follows:

\begin{equation}
   \Delta B_\mathrm{POS} = B_\mathrm{POS}\cdot \sqrt{\dfrac{1}{4}\cdot\dfrac{ (\Delta n)^2}{n^2} + \dfrac{ (\Delta (\Delta\upsilon))^2 }{(\Delta\upsilon)^2} + \dfrac{(\Delta (\widehat{\sigma}_\mathrm{c}(\varphi)))^2}{ \widehat{\sigma}_\mathrm{c}(\varphi)^2 }  },
\label{eq:dbpos}
\end{equation}

\noindent where $\Delta n$, $\Delta(\Delta\upsilon)$, and $\Delta(\widehat{\sigma}_\mathrm{c}(\varphi))$ are measured uncertainties of $n$, $\Delta\upsilon$ and $\widehat{\sigma}_\mathrm{c}(\varphi)$ (Sec.\,\ref{subsec:slidding_window}) respectively. The uncertainty of the magnetic field strength derived from Eq.\,\ref{eq:bpos2} is similar to that given by Eq.\,\ref{eq:dbpos}, with an additional factor of $\frac{1}{4}$ in front of the last term in the equation.

We measure the uncertainties of $n$, $\sigma_\mathrm{NT}$ from the $\chi^2$ minimization (see the bottom panel in Fig.\,\ref{fig:radex_cn_west}-\ref{fig:radex_mix_hcop}). We select the range in parameter space within which the $\chi^2$ reaches 0.5 to derive corresponding $n$, and $\sigma_\mathrm{NT}$. Thus, we found $\Delta n$ to be $10^3$\,cm$^{-3}$ and $\Delta\sigma_\mathrm{NT} = 0.25$\,km/s for both CN and \HCOp\,.

The reported measurements of the angle dispersion and the POS magnetic field strength, including their uncertainties for both methods, are shown in Tab.\,\ref{tab:pixels}. We show the comparison between the magnetic field strengths derived using the DCF and the ST method, including the use of different methods to estimate the angle dispersion in Fig.\,\ref{fig:dcf_skalidis} in App.\,\ref{subsec:dcf_skalidis}. Overall, the POS magnetic field strength derived using the prescription from \cite{skalidis_2021a,skalidis_2021b} is lower than that derived using classical DCF. In the following, we comment on the magnetic field strengths derived using the Eq.\,\ref{eq:bpos2_skalidis} in Sec.\,\ref{subsec:skalidis}. The $B_\mathrm{POS}$ strength is in the range of $\sim20-90$\,\textmu G.
The magnetic field strength inferred from CN is overall smaller than that derived using \HCOp\,, mainly due to \HCOp\, showing slightly larger line widths than the CN in some regions, which directly impacts the strength of the magnetic field. Due to ambipolar diffusion effects \citep[][]{zweibel_2004,tritsis_2023}, it is expected that the emission lines of neutral molecular species (CN in our case) are wider than those of the ion molecular species (such as \HCOp) \citep{li-houde_2008,yin_2021}. However, in our case, we observe that \HCOp\, is broader than the CN spectral lines, even after corrections for the opacity broadening. This result has two implications. Firstly, our observations did not spectrally resolve \HCOp\, emission, which potentially results in larger line widths caused by the blending of velocity components. Secondly, it is possible that our measurements do not probe ambipolar diffusion scale \citep[$\sim10^{-3}$\,pc,][]{li-houde_2008} in NGC~2024. 

The strongest magnetic field, derived from CN and \HCOp\, is observed in the region at the intersection of the dusty filament and the edge of the expanding \HII\ region and at the eastern side of the bubble. This result is driven mainly by the large densities and the broader CN and \HCOp\, lines observed in these regions (see Sec.\,\ref{sec:radex_all}) found in this work. 
The magnetic field strength measured using both CN and \HCOp\, emission is stronger toward the eastern side of the bubble than on the western edge. We found lower line widths of CN and \HCOp\ at the western side of the bubble. The weakest magnetic field is derived toward the filament, although we note the $B_\mathrm{POS}$ derived from \HCOp\, in the western side of the bubble and filament to be comparable given the uncertainties.

\begin{table*}
\centering
\caption{Measured angle dispersion from HAWC+ data and results on the measured magnetic field strength and its uncertainty, using results from modeling the CN and \HCOp\, excitation (App.\,\ref{sec:radex_all}, Tab.\,\ref{tab:radex}). The POS magnetic field strength is derived using the classical DCF method (third and fourth column) and the ST method presented in \cite{skalidis_2021a, skalidis_2021b} (fifth and sixth column).}
%%%%%%%%%%%%%%%%%%%%%%%%%%%%%%%%%%%
\begingroup
\setlength{\tabcolsep}{8pt} % Default value: 6pt
\renewcommand{\arraystretch}{1.2} % Default value: 1
%%%%%%%%%%%%%%%%%%%%%%%%%%%%%%%%%%%
\resizebox{0.8\textwidth}{!}{ { \begin{tabular}{lccccc}
\hline
\smallskip
Region & $\sigma_{\theta}$ [deg] & $B_\mathrm{DCF, CN}$ [\textmu G] & $B_\mathrm{DCF, HCO^+}$ [\textmu G] & $B_\mathrm{ST, CN}$ [\textmu G] & $B_\mathrm{ST, HCO^+}$ [\textmu G] \\ \hline \hline
Bubble West & 5.5 $\pm$ 0.3 & 46 $\pm$ 23 & 90 $\pm$ 26 & 20 $\pm$ 10 & 40 $\pm$ 11 \\
Bubble East & 3.5 $\pm$ 0.2 & 235 $\pm$ 55 & 235 $\pm$ 55 & 82 $\pm$ 19 & 82 $\pm$ 19 \\
Filament & 6.1 $\pm$ 0.4 & 66 $\pm$ 28 & 102 $\pm$ 35 & 30 $\pm$ 13 & 47 $\pm$ 16 \\
Filament+Bubble & 6.3 $\pm$ 0.4 & 162 $\pm$ 45 & 185 $\pm$ 48 & 76 $\pm$ 21 & 87 $\pm$ 22 \\
\hline
\hline
\end{tabular} } }
\endgroup
\label{tab:pixels}
\end{table*}

%%%%%%%%%%%%%%%%%%%%%%%%%%%%%%%%%%%%%%%%%%%%%%%%%%%%%%%%%%%%%%%%%%%%%%%%%%%%%%%%%%%%%%%%%%%%%%%%%%%%%%%%%%%%%%%%%%%%%%%%%%%%%%%

\section{Discussion}
\label{sec:discussion}

\subsection{The magnetic field in NGC~2024}
\label{sec:disc_bfield}

In this section, we discuss our results on the direction of magnetic field and its morphology, presented in Sec.\,\ref{sec:mag_fields}. An overview of previous studies on the structure of magnetic field within NGC~2024 is reported in \cite{meyer_2008}. Information about magnetic field in NGC~2024 were based on the thermal dust continuum emission (linear dust polarization), at 100\,\textmu m \citep{hildebrand_1995, dotson_2000} and at 850\,\textmu m \citep{matthews_2002}, dichroic polarization of point sources \citep{kandori_2007}, and the Zeeman splitting of \HI\, and OH lines \citep{crutcher_1999}. 

All these studies found that the magnetic field shows a specific structure. In particular, the line of the sight (LOS) $B$-field strength in the central area of NGC~2024 dominated by the dusty filament \citep{crutcher_1999} weakens from the northeast to the southwest. \cite{matthews_2002} investigated (sub)millimeter dust polarization and modeled magnetic field in NGC~2024 with two components. The first component is related to the dense and dust obscured part, where magnetic field lines follow locations of the FIR sources (right panel in Fig.\,\ref{fig:fig1}). The second component of magnetic field is linked to gas affected by the stellar feedback. In our case, we observe regions where these two features are dominating: Band~D dust polarization traces the filament in the central and southern part of NGC~2024, whereas it is impacted by the stellar feedback at eastern and the western parts at ionization front (Eastern and Western Loop in the right panel in Fig.\,\ref{fig:fig1}).

We overlay dust polarization vectors at 100\,\textmu m from \cite{hildebrand_1995} and at 850\,\textmu m from \cite{matthews_2002} in the bottom row of Fig.\,\ref{fig:hawc+_sources} in App.\,\ref{sec:literature_comp}. The area where dust polarization measurements from \cite{hildebrand_1995} and \cite{matthews_2002} overlaps with our work is the central part of NGC~2024, where embedded FIR sources are located. We find the overall agreement in the direction of magnetic field inferred from dust polarization at 100, 154 and 850\,\textmu m. %However, 850\,\textmu m dust continuum does not trace the outflow associated with FIR\,5, in contrast to our Band~D data.

POS magnetic field lines reported in our work are impacted by the stellar feedback at the surfaces of molecular cloud. Similarly, \cite{crutcher_1999} reported that the LOS magnetic field is nearly zero east of the filament in NGC~2024, which indicates a possibility that total magnetic field lines are in the POS east of the filament in NGC~2024. %This result could possibly imply that the total magnetic field in NGC~2024 is being impacted by the stellar feedback.
However, we note that observations presented in \cite{crutcher_1999} do not cover the edges of \HII\ region, and that we require a study of LOS magnetic field component that covers larger field of view in NGC~2024. 

The magnetic field lines are mainly parallel to the filament in NGC~2024, which is in the agreement with results presented in other studies \citep[such as,][]{planck_2016a, santos_2016, pillai_2015, pattle_2017}. However, we observe a change in the magnetic field direction at the very southern part of the filament in Fig.\,\ref{fig:dust_pol}, particularly in the Band~D data, and also at the northern part of NGC~2024. Such changes in the direction of the POS magnetic field can indicate a few possibilites. Firstly, changes of the magnetic field often trace star formation \citep{pillai_2009, ward-thompson_2017} due to gravitational collapse of the gas that cause the magnetic field lines to have ''pinched" structure. Moreover, the variation of the magnetic field direction can be also a consequence of changes in the column density of the gas \citep{alina_2019}.

%%% \cite{doi_2021} studied a filament within NGC~1333, but this study did not determine the strength of magnetic field.
%%% add part of the filament

\subsection{The magnetic field strength in the PDR and filament}
\label{sec:disc_bpos_strength}

In this section, we discuss our findings on the measured magnetic field strength in several distinct regions: the edge of the expanding PDR and the filament, including the region where it is not possible to clearly separate these two environments. 

\subsubsection{The edge of the \HII\ region}
\label{sec:disc_hii}

In our work, we estimate the POS magnetic field strength at the border of the expanding \HII\ region to be $\sim20-40$\,\textmu G on the west side, and $\sim82$\,\textmu G on the east side. The good agreement between the CN and \HCOp\, $B_\mathrm{POS}$ strengths in these regions, particularly at the eastern side, could indicate that these two molecular lines are tracing similar gas. Nevertheless, the factor of almost 2 difference in the $B_\mathrm{POS}$ measured in the west and the east could be due to different gas densities at these sides of the bubble. As the eastern side of the bubble is denser than the western side \citep[see,][]{meyer_2008}, we expect that the gas here is less impacted by the incoming radiation. In addition, the opacity of the \HCOp\, emission is notably higher ($\tau > 4$) on the western side than on the eastern side ($\tau = 2$).

As previously presented in Sec.\,\ref{sec:Bfield_results}, the eastern edge of \HII\ region has a stronger magnetic field than the western side. This result can also be due the total magnetic field changing its direction with respect to the line of sight. A previous study showed that the line of sight (LOS) magnetic field changes its strength from being zero at the eastern part of NGC~2024, to 100\,\textmu G at the western side, which is indicative of the change of a direction of the overall magnetic field \citep{crutcher_2009}. Moreover, the molecular gas content located west from the center of NGC~2024 has a lower density than the gas located at the eastern side, suggesting that stellar radiation has stronger impact on the gas on the western side of the bubble \citep{crutcher_2009}. At the eastern edge of the bubble, due to higher gas densities (Tab.\,\ref{tab:radex}), stellar feedback did not sweep up gas as far as on the other side \citep{barnes_1989}. % A factor of $>2$ in the POS magnetic fields towards the eastern side of the bubble measured using the CN and \HCOp\, comes from significantly narrower \HCOp\, line. This result could stem from the fact that at side of the bubble, due to larger gas densities, stellar feedback did not sweep up gas as far as on the other side \citep{barnes_1989}. Therefore, the production of the CN could be possibly slower on the eastern side. This does not mean, however, that the CN is not adequate tracer for the UV-illuminated gas.

The values of $B_\mathrm{POS}$ calculated in our work are generally lower than those reported in studies of other PDRs. For example, the magnetic field strength measured from the Zeeman splitting of \HI\, and OH in M~17 is $\sim750\,$\textmu G and $\sim250\,$\textmu G respectively \citep{brogan_2001}, and $\sim (1000-1700)$\,\textmu G using dust polarization data \citep{hoang_2022}. The magnetic field strength of the PDR in the Horsehead nebula \citep[SMM1,][]{hwang_2023} is a few tens of \textmu G, and comparable to our results in NGC~2024. That study used \CeiO\, to derive the line width and dust column density and effective radius to estimate the gas volume density. Although the edges of the bubble in our work and the Horsehead nebula have comparable densities (a few $\times$ $10^3$\,cm$^{-3}$), measured line widths are different. It is worth pointing out that \CeiO\, and CN and \HCOp\, are tracing different gas \citep{philipp_2006,bron_2018}. \CeiO\, traces more compact structures than the CN and \HCOp, and it gets destructed by the stellar feedback, contrary to CN and \HCOp, whose emission becomes enhanced in these regions \citep{bron_2018}. % Moreover, \cite{hwang_2023} found significantly narrower line widths of \CeiO\, than those of CN and \HCOp\, (measured in our work). 
In addition, \cite{hwang_2023} used a modified DCF method that measures the ratio of the ordered and turbulent component of the $B$-field \citep{hildebrand_2009} to derive the magnetic field strength. % Another possible explanation of the observed difference is that the NGC~2024 \HII\, region is younger than the Horsehead nebula and still embedded in its parental molecular cloud.
%is powered by a single O star, whereas NGC~2024 is energized by a deeply embedded massive stellar cluster.
% {\color{blue} DCL: I would think that C18O traces different gas compared to HCO+/CN. See Fig. 1 of Philipp et al. https://www.aanda.org/articles/aa/pdf/2006/28/aa3533-05.pdf}

Other PDR regions impacted by super stellar clusters have stronger magnetic field to those reported in our work. For example, \cite{pattle_2018} reported strong magnetic fields of a few hundred \textmu G in the Pillars of M~16. This work used the DCF method to derive the magnetic field strength. Gas density of the Pillars are somewhat higher than those in our work $5\cdot10^4$\,cm$^{-3}$ \citep{ryutov_2005}. Similarly, the line width measurements used in \cite{pattle_2018} are taken from the Gaussian fitting of several molecular lines from and these are wider than those reported in our work \citep[see Tab.\,3 in][reported line widths in the range from 1.2 to 2.2\,km/s]{white_1999}. \cite{guerra_2021} showed the POS magnetic field strength across the Orion Bar PDR in OMC-1 is of a few hundreds of \textmu G, also using the DCF method. 

The main difference in magnetic field strengths computed in our work and found within the literature comes from the methods used to derive the POS $B$-field. The DCF method generally overestimates the magnetic field strength. On the other hand, simulations have shown that the POS $B$-field is proportional to inverse square root of the angle dispersion. However, in this case, the prefactor is smaller by a factor of 5, resulting in generally lower $B_\mathrm{POS}$. We show how $B_\mathrm{POS}$ varies with the method we selected in Fig.\,\ref{fig:dcf_skalidis}.
% this is the fillamentary work: \cite{arzoumanian_2021} within bubbles in NGC~6334. 

%%%%%%%%%%%%%%%%%%%%%%%%%%%%%%%%%%%%%%%%%%%%%%%%%%%%%%%%%%%%%%%%%%%%%%%%%%%%%%%%%

\subsubsection{Filament}
\label{sec:disc_fil}

The filament going across NGC~2024 is super-critical \citep{orkisz_2019}, which means it is gravitationally unstable and a potential site for star formation. We measure the lowest values of the POS magnetic field ($\sim30-50$\,\textmu G) in this region, suggesting that the magnetic field cannot dominate gas dynamics. We return to this point in Sec.\,\ref{sec:disc_bcrit}. These low values in the magnetic field strength come from the largest dispersion angles and narrow spectral lines computed for this region (Tab.\,\ref{tab:radex} and \ref{tab:pixels}). Large dispersion angles suggest that the turbulence does not impact the magnetic field, which is also supported by the narrow CN and \HCOp\, lines. Since the turbulence level in this part appears low, this could lead to star formation in the filament, previously confirmed at its southern part \citep{hwang_2023}.

Overall, our measurement of the magnetic field strength in the filament is lower than in other filamentary structures. For example, \cite{pattle_2017} measured a POS magnetic field strength of a few mG in the Orion A filament. Similarly, as discussed in the previous section, it is important to highlight that \cite{pattle_2017} used \CeiO\, emission to measure the turbulence and number density and different method to derive the $B_\mathrm{POS}$. Similarly, \cite{ching_2018} investigated magnetic field toward the DRL1 filament and found $B_\mathrm{POS} = 600\,$\textmu G.

\subsection{Magnetic field support in NGC~2024}
\label{sec:disc_bcrit}

\begin{table*}[t!]
\centering
\caption{Estimated mass-to-flux ratio $\mu_{\Phi}$, Alfvénic Mach number $\mathcal(M)_\mathrm{A}$ and plasma-beta $\beta$ using $B_\mathrm{POS}$ reported in Tab.\,\ref{tab:pixels}. These parameters indicate role of radiation, gravitational and magnetic field and their impact on the gas in four regions in NGC~2024. }
%%%%%%%%%%%%%%%%%%%%%%%%%%%%%%%%%%%
\begingroup
\setlength{\tabcolsep}{8pt} % Default value: 6pt
\renewcommand{\arraystretch}{1.2} % Default value: 1
%%%%%%%%%%%%%%%%%%%%%%%%%%%%%%%%%%%
\resizebox{0.8\textwidth}{!}{{ \begin{tabular}{lcccccc}
\hline
Region & $\mu_{\Phi}(\mathrm{CN})$ & $\mu_{\Phi}(\mathrm{HCO^+})$ & $\mathcal{M}_\mathrm{A}(\mathrm{CN})$ & $\mathcal{M}_\mathrm{A}(\mathrm{HCO^+})$ & $\beta_B(\mathrm{CN})$ & $\beta_B(\mathrm{HCO^+})$ \\ \hline \hline
% Bubble West & 0.10 & 0.09 & 1.56 & 1.06 & 0.206 & 0.078 \\
% Bubble East & 0.18 & 0.16 & 1.23 & 1.60 & 0.179 & 0.247 \\
% Filament & 2.25 & 1.45 & 1.69 & 1.18 & 0.389 & 0.079 \\
% Filament+Bubble & 0.90 & 0.74 & 1.72 & 1.09 & 0.201 & 0.055 \\
% \hline
Bubble West & 0.38 & 0.19 & 1.78 & 2.03 & 1.051 & 0.345 \\
Bubble East & 0.09 & 0.09 & 1.42 & 1.43 & 0.083 & 0.084 \\
Filament & 2.50 & 1.61 & 1.88 & 1.31 & 0.483 & 0.098 \\
Filament+Bubble & 1.00 & 0.88 & 1.91 & 1.21 & 0.247 & 0.077 \\
\hline
\hline
\end{tabular} } }
\endgroup
\label{tab:parameters}
\end{table*}

Next, we investigate the role of the magnetic field at the edges of the bubble, in the filament, and the overlap region (bubble and the filament) in NGC~2024. To do so, we compute several parameters that describe gas stability in relation to local environmental conditions: radiation, magnetic and turbulent field. Since we do not have information on the line of sight component of magnetic field, we cannot derive the total magnetic field strength. Therefore, the following quantities need to be treated as upper limits, as they are proportional to the magnetic field strength as $B^{-n}$.

We make an estimate on the mass-to-flux ratio of gas in NGC~2024 $\left ( \frac{M}{\Phi} \right )$ and compare it to the critical mass-to-flux ratio. The critical mass-to-flux ratio \mbox{$\left ( \frac{M}{\Phi} \right )_\mathrm{crit}$} depends on the assumed geometry. For instance, for a uniform disk, \mbox{$\left ( \frac{M}{\Phi} \right )_\mathrm{crit,cyl}= \dfrac{1}{2\pi\sqrt{G}}$} \citep{nakano_1978, joos_2012, hanawa_2019}. In addition, for a spherical geometry, the critical mass-to-flux ratio will be expressed as \mbox{$\left ( \frac{M}{\Phi} \right )_\mathrm{crit,sph}= \dfrac{1}{3\pi\sqrt{\frac{G}{5}}}$} \citep{mouschovias_1976}. The $\mu_{\Phi}$ parameter is the ratio between estimated mass-to-flux and $\left ( \frac{M}{\Phi} \right )_\mathrm{crit}$ and can be computed as:

\begin{equation}
    \left ( \mu_{\Phi} \right ) _\mathrm{cyl} = \dfrac{\left ( M / \Phi \right ) }{ \left ( M / \Phi \right )_\mathrm{crit} } = 7.6\cdot10^{-21} \dfrac{N}{B},
\label{eq:mu-phi}
\end{equation}

\noindent in the case of cylindrical geometry. $N$ is the column density of molecular gas taken from \cite{lombardi_2014} in units of cm$^{-2}$, and $B$ is measured POS magnetic field strength (Eq.\,\ref{eq:bpos2_skalidis}) in \textmu G. In the case of spherical geometry, using the expression for $\left ( \frac{M}{\Phi} \right )_\mathrm{crit}$ and the above equation:

\begin{equation}
    \left ( \mu_{\Phi} \right ) _\mathrm{sph} = 0.67\cdot \left ( \mu_{\Phi} \right ) _\mathrm{cyl}.
\end{equation}

Therefore, the $\mu_{\Phi}$ will vary by a factor of $0.67$ that comes from the assumed geometry. It is necessary to point out that there is not a critical mass-to-flux ratio for lateral contraction of a filament when it threaded by magnetic field parallel to its axis of symmetry \citep{mouschovias_1991b}. In our work, we use Eq.\,\ref{eq:mu-phi} and assume the cylindrical geometry, but we note that this factor can vary depending on the assumed geometry of a system. Moreover, in case of the filament, the interpretation and physical meaning of mass-to-flux ratio in not straightforward, and it should be threated with causion.

Next, we compute the Alfvénic Mach number, $\mathcal{M}_\mathrm{A}$:

\begin{equation}
    \mathcal{M}_\mathrm{A} = \dfrac{\sqrt{3}\cdot\sigma_{\upsilon, \mathrm{NT}}}{\upsilon_\mathrm{A}}.
\label{eq:alf_mach}
\end{equation}

\noindent where $\sigma_{\upsilon, NT}$ is the non-thermal velocity dispersion reported in Tab.\,\ref{tab:radex}. The $\sqrt{3}$ factor comes from the assumption of isotropic turbulence \citep{crutcher_2004, stewart_2022}. $\upsilon_\mathrm{A}$ is the Alfvén velocity defined as

\begin{equation}
    \upsilon_\mathrm{A} = \dfrac{B}{\sqrt{4\pi\rho}},
\end{equation}

\noindent where $B$ is the total magnetic field strength in units of G, and $\rho$ is the gas mass volume density in units of g\,cm$^{-3}$. Eq.\,\ref{eq:alf_mach} can also be represented as a ratio between turbulent gas and magnetic energies. Therefore, the Alfvénic Mach number provides information about a dominant driver of gas flows.

Finally, we compute the plasma-beta parameter, $\beta_B$, that gives information about the ratio of thermal and magnetic pressure:

\begin{equation}
    \beta_B = \dfrac{n k_\mathrm{B}T}{ B^2 / (8\pi)},
\label{eq:beta}
\end{equation}

\noindent where $n$ is the number density, $T$ is the gas temperature (Tab.\,\ref{tab:radex}), and $B$ magnetic field strength (Tab.\,\ref{tab:pixels}). 

We report our results in Tab.\,\ref{tab:parameters} for values of magnetic field strengths computed from CN and \HCOp\, respectively using approach from \cite{skalidis_2021a,skalidis_2021b}. We estimate $\mu_{\Phi} < 1$ in each region. This result could imply the presence of magnetically supported gas \citep[][]{pattle_2019}. In general, it is worth noting that we see a difference of one and two orders of magnitude in the $\mu_{\Phi}$ between the edges of the bubble and the filament and the overlap region.

We find $M_\mathrm{A}$ greater than 1 in all regions, which suggests that the magnetic field does not govern the gas motions. Moreover, we measure a plasma-beta parameter lower than 1 in almost all environments, which agrees with the lower values of $\mu_{\Phi}$, particularly at the edges of the bubble. A plasma beta derived from CN measurements of magnetic field at the western side of the bubble is slightly higher than one. Low values of plasma-beta suggest that the magnetic energy dominates the thermal motions of the gas.

The results reported in Tab.\,\ref{tab:parameters} should be taken as upper limits considering we calculated them using the POS component of the magnetic field. In particular, the reported mass-to-flux ratio must be taken with caution. The complex geometry of NGC~2024, large uncertainties of reported POS magnetic field strength, and non availability of the LOS component of the $B$-field have a big impact on values reported in Tab.\,\ref{tab:parameters}. Our results imply that the edges of the bubble show different properties than the filamentary structure and the overlap region in NGC~2024. These regions are indeed different in terms of the impact of stellar feedback. However, further quantification of the gravitational stability of the gas in these regions requires systematic analysis of the magnetic field. This also includes both POS and LOS components in these regions and constraints on the geometry of NGC~2024, which is beyond the scope of this study.

The properties of the gas we study in our work vary with the environment and this gas shows different characteristics. Our results thus highlight the non-negligible role of the magnetic field in NGC~2024 in regions impacted by the stellar feedback. The edges of the bubble in NGC~2024 are magnetically dominated and gas in these regions is magnetically supported against gravitational collapse. However, the high values of the Alfvénic Mach number suggest that these regions are highly turbulent. The gravitational stability of gas located at the shell of expanding \HII\, region is not always the case. For example, the expansion of the \HII\, region can also trigger star formation, as seen in Galactic \HII\, regions RCW\,82 \citep{pomares_2009} and RCW\,120 \citep{figueira_2017}. Since the \HII\, region in NGC~2024 is relatively young, it is possible that the stellar feedback has not yet been strong enough to trigger star formation. Nevertheless, it is crucial to point out that our result is based on CN and \HCOp\, measurements, which trace the UV-illuminated and UV-shielded gas, which do not necessarily trace the star-forming gas.

The environment in which we observe gas coming from the edge of the bubble and the filament is also magnetically supported, which could imply that the gas in this region mainly comes from the ionization front. This result can be explained by the presence of compressive motions in NGC~2024 \citep{orkisz_2017} that possibly originate from expansion of the \HII\ region.

The role of the magnetic field is somewhat different in the filament. High $\mu_{\Phi}>1$ implies gravitational instability that could lead to star formation, as observed at the southern part of this filament \citep{hwang_2023}. Measured Alfvénic Mach numbers are also higher than 1 in the filament. $\mathcal{M}_\mathrm{A}$ measured from CN and \HCOp\, is also higher than 1.

Nevertheless, we should note the following. Firstly, in this work we do not investigate the central region of NGC~2024, where the protostellar candidates and ongoing star formation is observed. Therefore, our results focus only on the gas impacted by the stellar feedback and one located in the front of the bubble and a filament. We also note that due to relatively high uncertainties (around $30\%$) of measured magnetic field strengths presented in Tab.\,\ref{tab:parameters}, these uncertainties impact our measurements, particularly they could change our presentation of the physical conditions in the filament. Given the uncertainties, it is possible that the filament is in the transition zone between being fully magnetically supported and gravitationally unstable.

%%%%%%%%%%%%%%%%%%%%%%%%%%%%%%%%%%%%%%%%%%%%%%%%%%%%%%%%%%%%%%%%%%%%%%%%%%%%%%%%%%%%%%%%%%%%%%%%%%%%%%%%%%%%%%%%%%%%%%%%%%%%%%%

\section{Conclusions}
\label{sec:summary}

We present new SOFIA HAWC+ dust polarization measurements across the NGC~2024 \HII\ region and associated molecular cloud in the Orion B molecular cloud. We combined these measurements with molecular data observations from the ORION-B Large Program on the IRAM 30-meter telescope. We summarize our findings obtained using these observations as follows:

   \begin{enumerate}
      \item Our results focus on a subset of environments found in NGC~2024, particularly on the shell near the edge of the \HII\ region and the filament in the front of NGC~2024, which are not locations containing protostellar cores (Fig.\,\ref{fig:fig1} and sites of active star formation, located in the center of NGC~2024.
      \item We investigate the magnetic field morphology traced by dust polarization from HAWC+ observations using Band~D  at $154$\,\textmu m and Band~E at $216$\,\textmu m. The direction of the magnetic field derived from these two bands shows a good agreement.
      \item We use HAWC+ Band~D at $154$\,\textmu m to characterize the geometry of the magnetic field across NGC~2024. We find that the structure of the magnetic field is ordered and follows the morphology of the expanding \HII\ region and the direction of the filament. 
      \item Using the CN\Jone\, and \HCOp\Jone\, molecular emission obtained using the IRAM 30-meter telescope, we characterize physical conditions (turbulence and gas density) in four specific regions in NGC~2024: the edges of the expanding \HII\, shell located to the east and the west, the filament to the south, and the environment in the northern part of NGC~2024. In our analysis, we include collisions with electrons and the hyperfine structure of CN\Jone, which are essential aspects of our calculations.
      %\item 
      Both CN\Jone\, and \HCOp\Jone\, emission lines are optically thick across NGC~2024, which broadens the observed line widths. The gas number density derived from CN\Jone\, is comparable or somewhat higher than that obtained from analyzing the excitation of \HCOp\Jone.
      \item We derived the POS magnetic field strength using the ST (and classical DCF) method, with values ranging from 20 to 90\,\textmu G (50 - 240\,\textmu G) in the environments mentioned above.
      The strongest magnetic field is found in the region composed of the dusty filament and the edge of the expanding \HII\ region, located in the northern part of NGC~2024. The high magnetic field strength derived in this area is driven by the largest gas densities and line widths.
      \item Magnetic field strengths derived using CN\Jone\, and \HCOp\Jone\, are comparable within the uncertainties at the edges of the bubble, especially in the eastern side of the bubble. The magnetic field strength calculated from \HCOp\Jone\, is generally higher than that inferred from CN\Jone. We note that $B_\mathrm{POS}$ measured from CN\Jone\, and \HCOp\Jone\, is larger on the eastern side of the bubble than in the western edge. This observed difference may result from changes in the magnetic field direction indicated in previous studies. In addition, the western side of the bubble is more impacted by stellar radiation due to its lower density.
      \item By analyzing the mass-to-flux ratio, Alfvénic Mach number, and plasma-beta parameter, we find that the edges of the bubble and the filament show different properties.
      Gas impacted by the stellar feedback and traced by the CN\Jone\ and \HCOp\Jone\, emission seems better supported against gravitational collapse by the magnetic field than the gas in the filament, which represents a location where star formation can take place. However, we find the Alfvénic Mach number higher than one in all regions, which suggests that the magnetic field does not control the gas motions. Our results should be treated as upper limits given the use of the POS component of the $B$ field. In addition, the estimated mass-to-flux ratio has a large uncertainty due to the specific geometry of NGC~2024. We also note that these regions of NGC~2024 could be in the transition phase regarding gravitational stability.
      % \item {\color{blue} JP: We need to add something about the relation with the SFE in NGC2024.}
   \end{enumerate}

Our research emphasizes the importance of utilizing dust polarization measurements to characterize the structure of the magnetic field and the need to combine these measurements with molecular data to accurately infer the magnetic field's strength. Our results demonstrate that the magnetic field plays a critical role and that its contribution and other factors cannot be ignored. 
We also demonstrate the significant impact that different methods used to derive the POS magnetic field can have on our results and how it affects the interpretation of the gravitational stability of the gas based on the corresponding parameters. Moreover, by removing the contribution of large-scale gradients of the magnetic field direction using techniques such as the sliding window, we can consider only the local magnetic field. 
In addition, we highlight the necessity of conducting a careful analysis of the line radiative transfer and using appropriate rate coefficients for inelastic collisions with H$_2$ and electrons, as well as line opacity corrections.

%%%%%%%%%%%%%%%%%%%%%%%%%%%%%%%%%%%%%%%%%%%%%%%%%%%%%%%%%%%%%%%%%%%%%%%%%%%%%%%%%%%%%%%%%%%%%%%%%%%%%%%%%%%%%%%%%%%%%%%%%%%%%%%

% \input{Sections_referee/02-Data}

% \input{Sections_referee/03-Results}

% \input{Sections_referee/04-Discussion}

% \input{Sections_referee/05-Summary}

\begin{acknowledgements}
We thank the anonymous referee for useful comments that have improved the manuscript.

This research has made use of spectroscopic and collisional data from the EMAA database (\href{https://emaa.osug.fr}{emaa.osug.fr} and \href{https://dx.doi.org/10.17178/EMAA}{EMAA reference}). EMAA is supported by the Observatoire des Sciences de l’Univers de Grenoble (OSUG).

This work was supported by the French Agence Nationale de la Recherche through the DAOISM grant ANR-21-CE31-0010. This research was carried out in part at the Jet Propulsion Laboratory, which is operated by the California Institute of Technology under a contract with the National Aeronautics and Space Administration (80NM0018D0004).

MGSM and JRG thank the Spanish MICINN for funding support under grant PID2019-106110GB-I00.

Based in part on observations made with the NASA/DLR Stratospheric Observatory for Infrared Astronomy (SOFIA). SOFIA is jointly operated by the Universities Space Research Association, Inc. (USRA), under NASA contract NNA17BF53C, and the Deutsches SOFIA Institut (DSI) under DLR contract 50 OK 2002 to the University of Stuttgart. Financial support for SC was provided by NASA through award \#08\_0186 issued by USRA.
\end{acknowledgements}

\bibliographystyle{aa}
\bibliography{references}

%\clearpage 

\appendix

\section{HAWC+ dust continuum and polarization}

We present a comprehensive analysis of the flux that is filtered out in our dust polarization measurements using SOFIA HAWC+ in Sec. \ref{sec:sofia_flux}. Then, we show HAWC+ Band~E dust continuum data, and provide comparison with Band~D in Sec.\,\ref{app:nir_pol}. The comparison between the dust polarization shown in this work with the literature data is described in Sec.\,\ref{sec:literature_comp}. 
%Finally, we describe the fitting analysis we apply to our CN and \HCOp\ data in Sec. \ref{sec:fit_cn} and \ref{sec:fit_hcop}, respectively.

\subsection{Filtering of low-level extended flux in HAWC+ images}
\label{sec:sofia_flux}

\begin{figure}[!hbt]
    \centering
    \includegraphics[width = 0.49\textwidth]{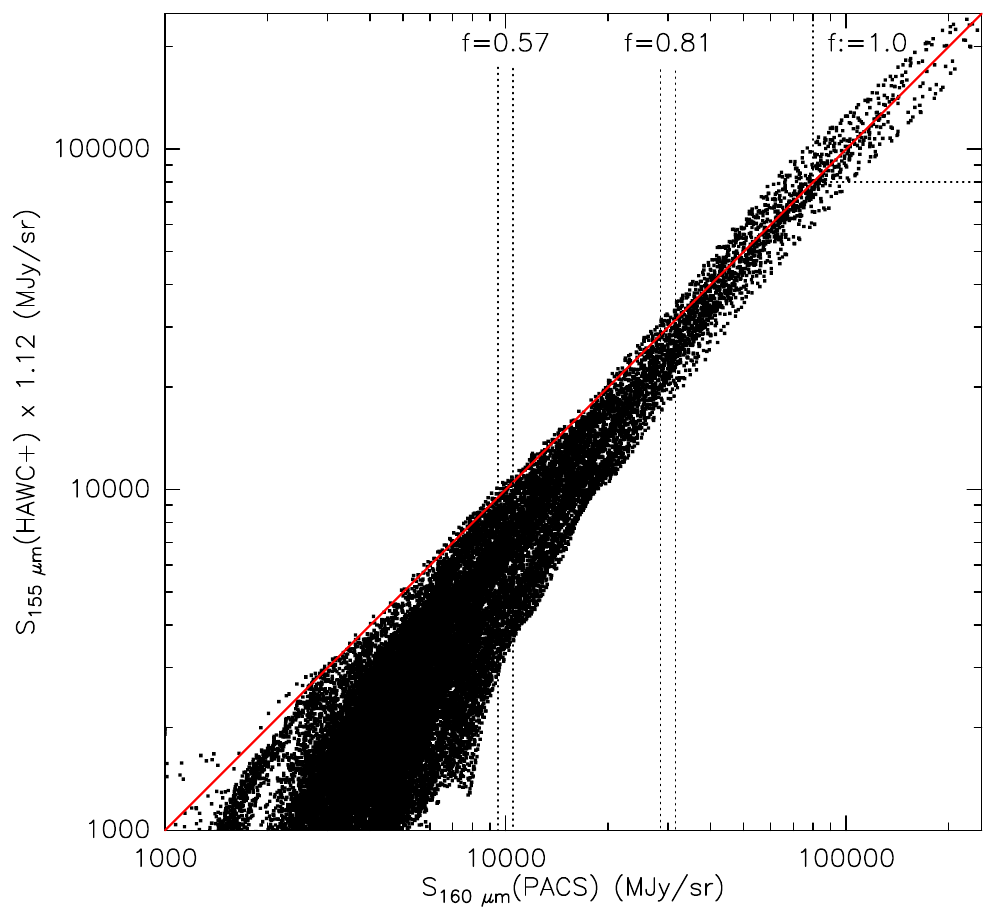}
    \caption{Pixel-by-pixel comparison of the HAWC+ Band~D and PACS 160 \textmu m fluxes. The PACS data have been convolved to the spatial resolution of the HAWC+ image. Only high S/N pixels included in the polarization analysis are shown. As described in the text, the HAWC+ fluxes were scaled by a factor of 1.12.} \label{fig:pacs}
\end{figure}

Dust continuum images obtained with ground-based or airborne far-infrared total power continuum cameras suffer from some degree of correlated atmospheric noise, which cannot be completely separated from the extended low-level emission of the source. The HAWC+ instrument data reduction pipeline attempts to remove correlated components from the time stream using the algorithms described in \cite{Kovacs2008}. To estimate the magnitude of the resulting spatial filtering at different flux levels in our HAWC+ Stokes $I$ images, we compared the Band D image to that obtained using the 160~\textmu m channel of the PACS instrument on \emph{Herschel} (OBSID 1342206080). Only high S/N pixels are included in the dust polarization analysis (see Sec.\,\ref{subsec:hawc+}). Fig.~\ref{fig:pacs} shows a pixel-by-pixel comparison of the HAWC+ and PACS fluxes at the spatial resolution of the HAWC+ Band D data. Given the difference in the filter passbands and overall calibration uncertainties of the two instruments, we scaled the HAWC+ data by a factor of 1.12 to enforce a unity flux ratio at pixels with fluxes above 80,000 MJy\,sr$^{-1}$ (marked by the dotted line in the upper-right corner of the figure). 

A good linear correlation between the two data sets is seen for points with fluxes above about 60,000 MJy\,sr$^{-1}$. However, at lower flux levels, the HAWC+ points consistently fall below the slope-one line marked in red. To quantify this effect, we computed average HAWC+/PACS flux ratios in two narrow ($\pm 5\%$) intervals centered at PACS fluxes of 30,000 and 10,000 MJy\,sr$^{-1}$, marked with vertical dotted lines in Fig.~\ref{fig:pacs}. The average HAWC+/PACS flux ratios in the two intervals are 0.81 and 0.57, respectively. 

The effect of the filtering of extended flux in the Stokes $Q$ and $U$ images is impossible to estimate, given the absence of space-based polarization data at a sufficient angular resolution. However, the filtering likely affects the polarization fraction more strongly than the polarization angle. Consequently, we do not use the polarization fraction in our analysis. Using the PACS image as a prior in the data reduction pipeline would improve the fidelity of the HAWC+ Stokes $I$ images. However, such a future modification to the pipeline is unlikely. 

\begin{figure*}
    \centering
    \includegraphics[width = \textwidth]{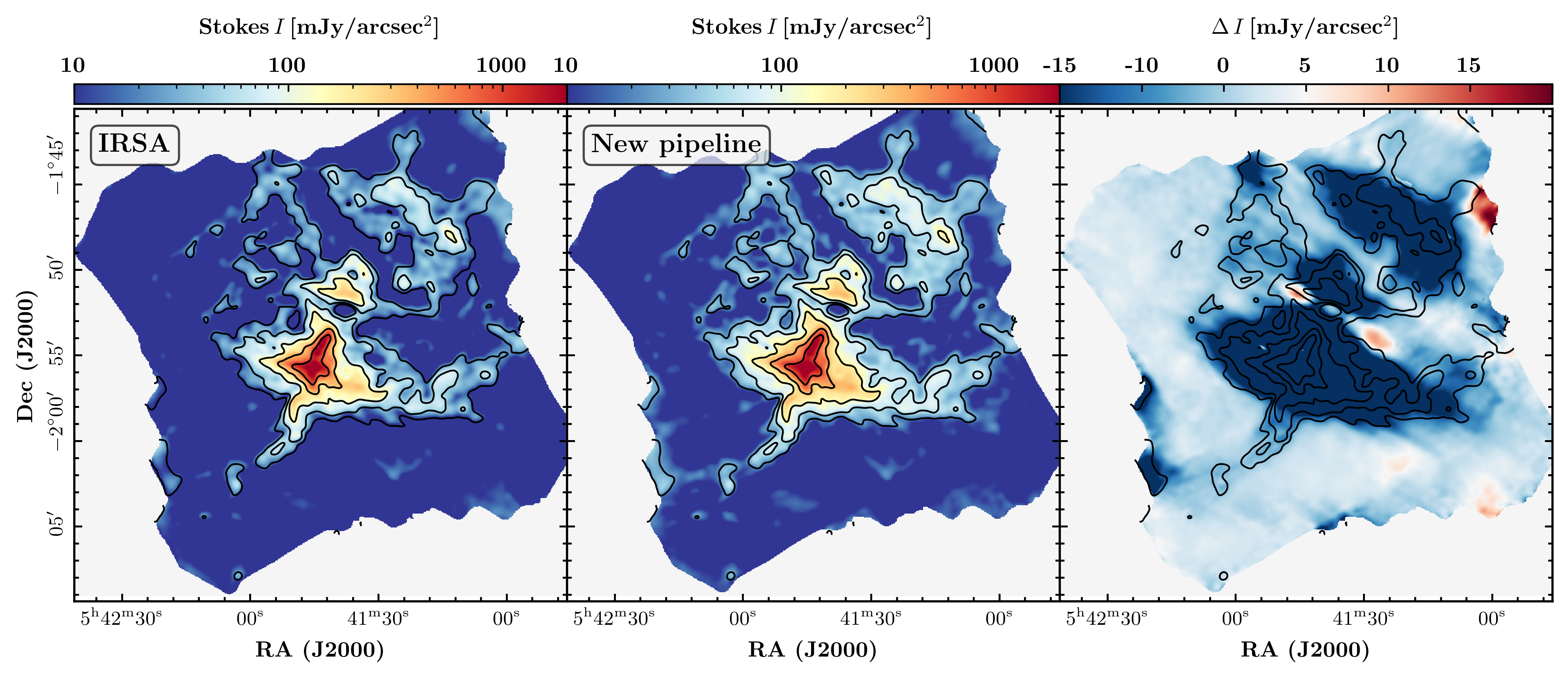}
    \includegraphics[width = \textwidth]{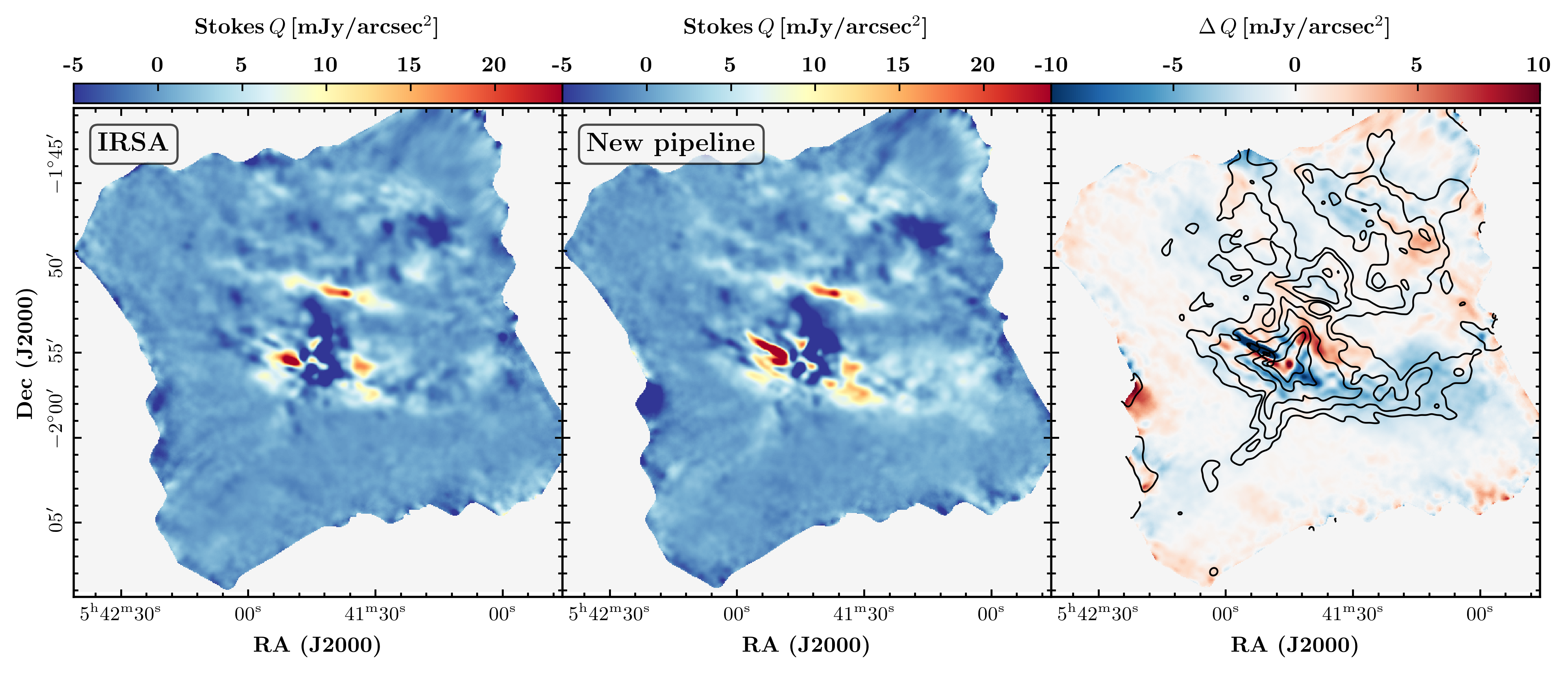}
    \includegraphics[width = \textwidth]{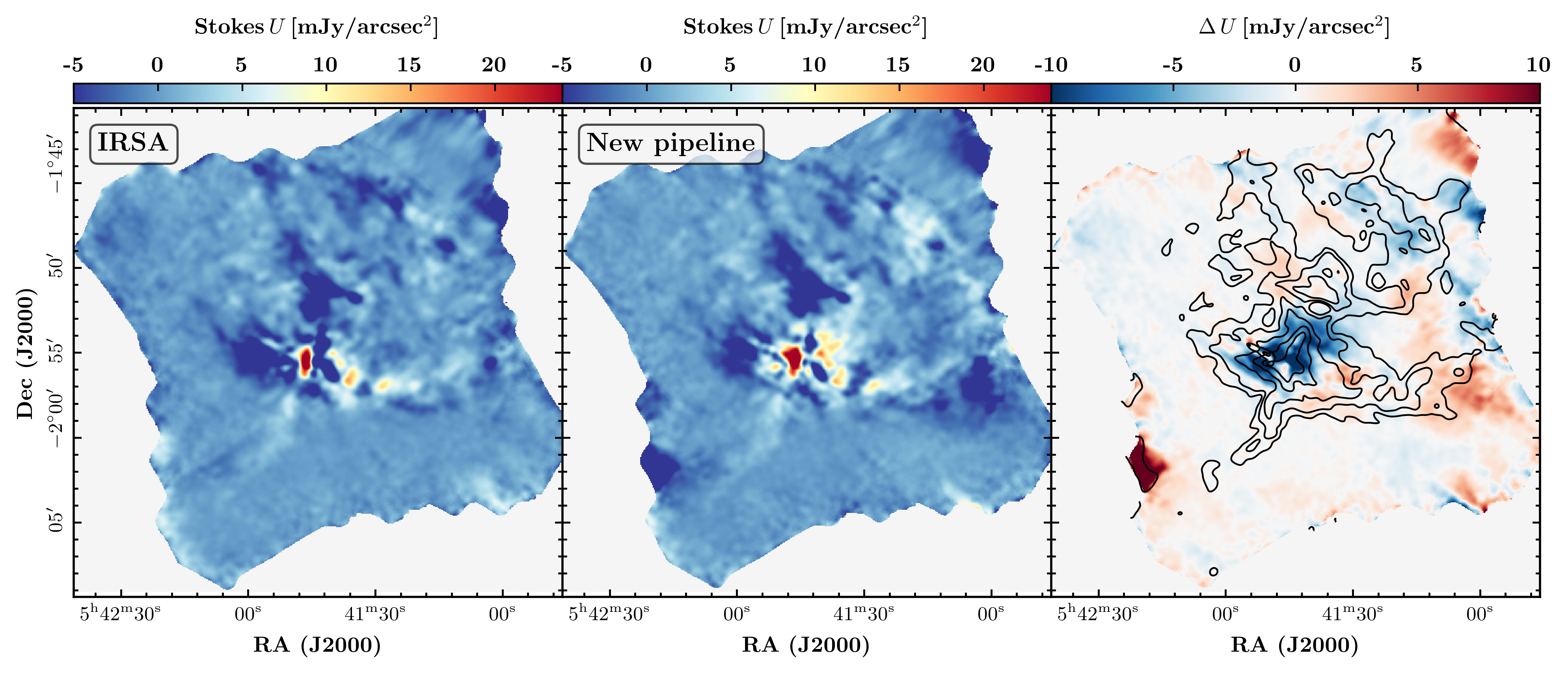}
    \caption{Comparison between Stokes parameters, $I$, $Q$, and $U$ (each row) for two setups of data reduction (the one used in this work is shown in the left panel, and the other one in the middle panel; see Sec.\,\ref{subsec:hawc+}). The right panels show the difference between the left and middle panels (computed for each Stokes parameter).}
    \label{fig:data_comp}
\end{figure*}

\begin{figure*}
    \centering
    \includegraphics[width = \textwidth]{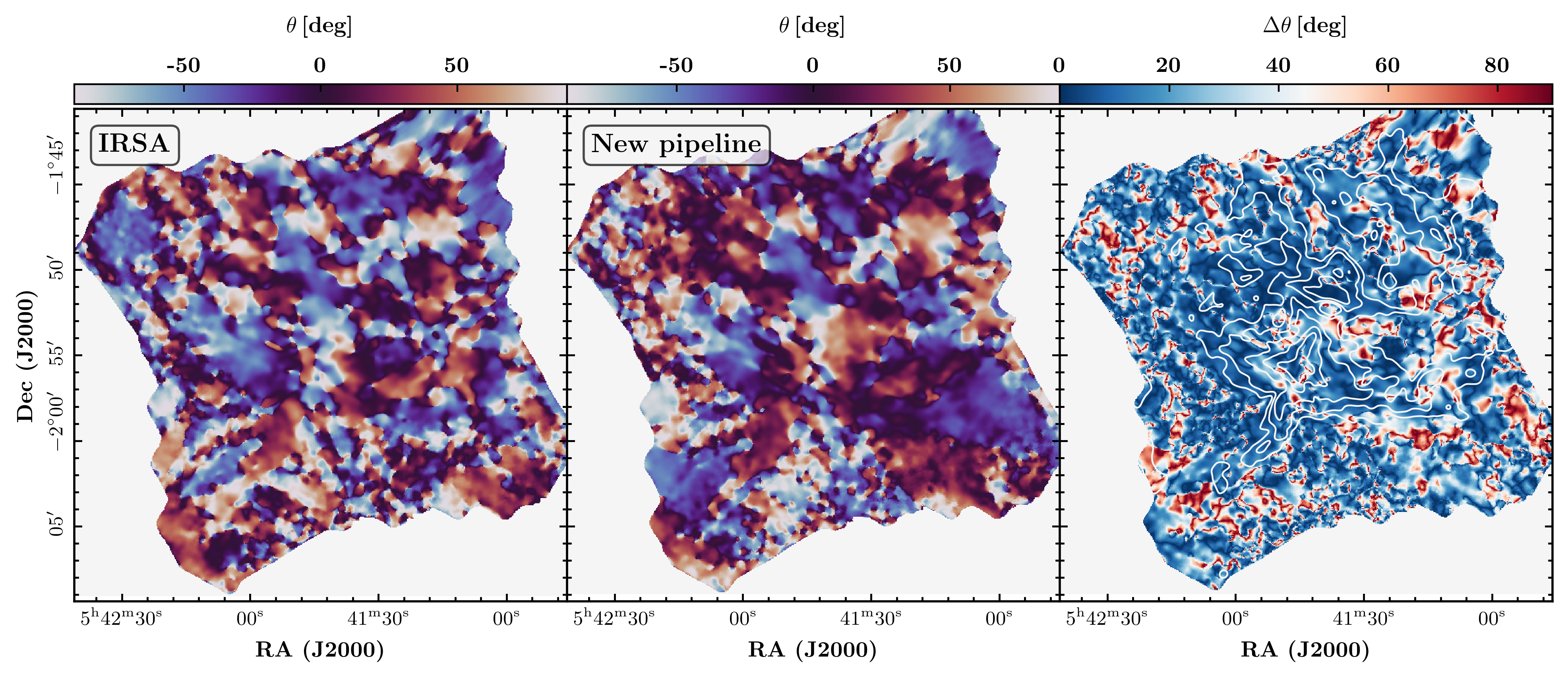}
    \caption{Same as in Fig.\,\ref{fig:data_comp}, but for the polarization angles (see Eq.\,\ref{eq:theta} in Sec.\,\ref{subsec:hawc+}).}
    \label{fig:data_comp2}
\end{figure*}

\subsection{SOFIA HAWC+ Band~E dust continuum}
\label{app:nir_pol}

\begin{figure*}%[t!]
    \centering
    \includegraphics[width = 0.583\textwidth]{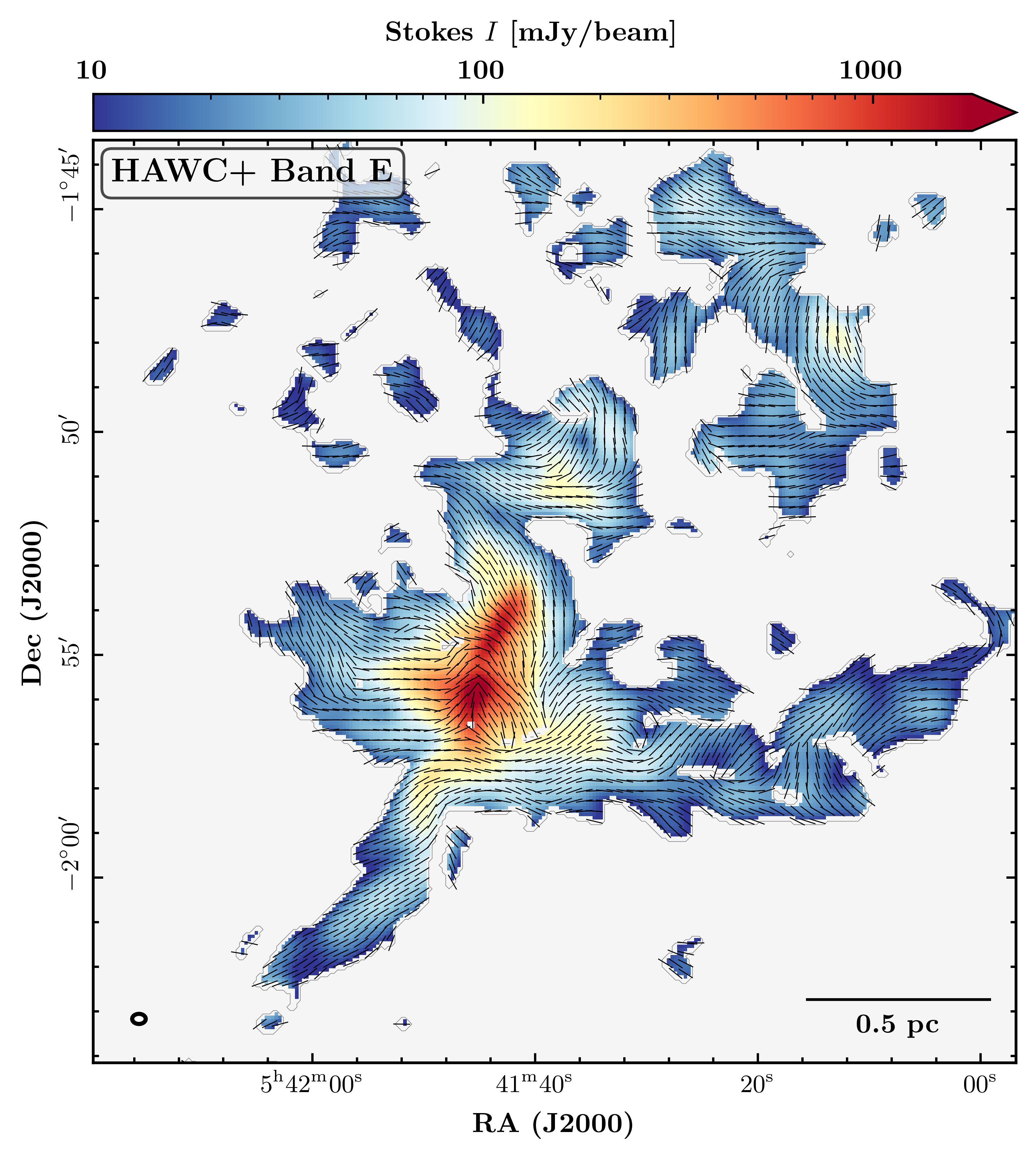}
    \caption{SOFIA HAWC+ $214\,$\textmu m (Band~E) dust continuum map at $18.2^{\prime\prime}$ angular resolution corresponding to linear scales of $\sim 0.037$\,pc. The map is masked the same as the Band~D map (Fig.\,\ref{fig:dust_pol}, see Sec.\,\ref{subsec:hawc+}). Black lines in both panels show the orientation of the magnetic field for every fifth pixel.} \label{fig:hawc+_e}
\end{figure*}

We show SOFIA HAWC+ Band~E dust continuum map in Fig.\,\ref{fig:hawc+_e} at its native resolution of $18.7''$. This map is processed and S/N masked the same way as Band~D data, as described in Sec.\,\ref{subsec:hawc+}. 

\begin{figure*}[h!]  
    \centering
    \includegraphics[width = \textwidth]{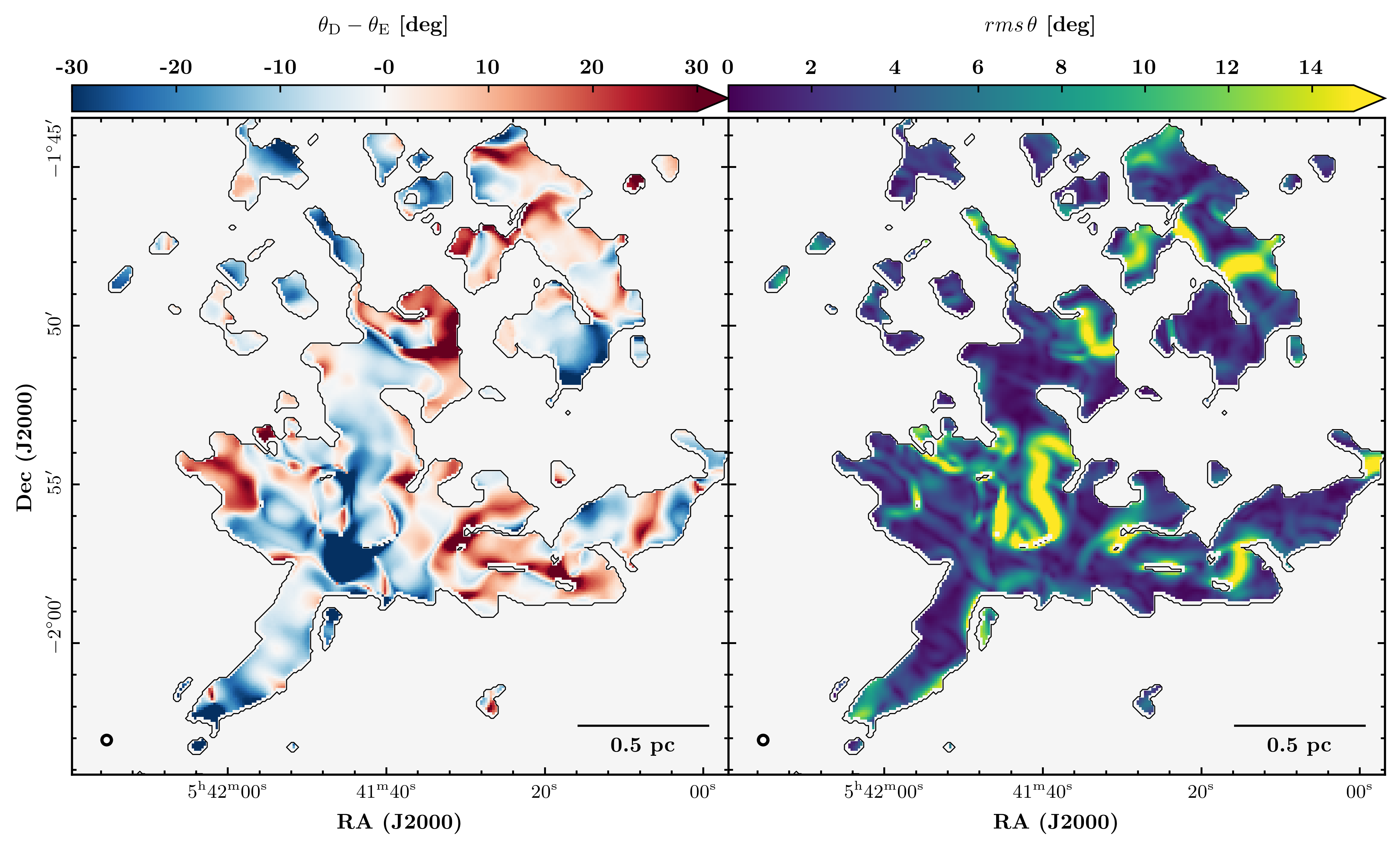}
    \caption{Difference in polarization angles measured from Band~D and E data is shown in the left panel. The right panel shows the rms of the angle differences.} \label{fig:hawc+_comparison}
\end{figure*}

To compare dust polarization angles derived from Bands D and E observations, we first convolve Band~D to match the lower spatial resolution of the Band E data. Next, we compute a difference between the two angles, $\theta_{D}-\theta_{E}$. Finally, we measure the rms of this difference, as described in Sec.\,\ref{subsec:slidding_window}, Eq.\,\ref{eq:circ_std}. We show a map of $\theta_{D}-\theta_{E}$ in the left panel of Fig.\,\ref{fig:hawc+_comparison}. The right panel in Fig.\,\ref{fig:hawc+_comparison} shows the corresponding rms of the map shown in the left panel.

\subsection{Comparison with the literature data}
\label{sec:literature_comp}

\begin{figure*}  
    \centering
    \includegraphics[width = \textwidth]{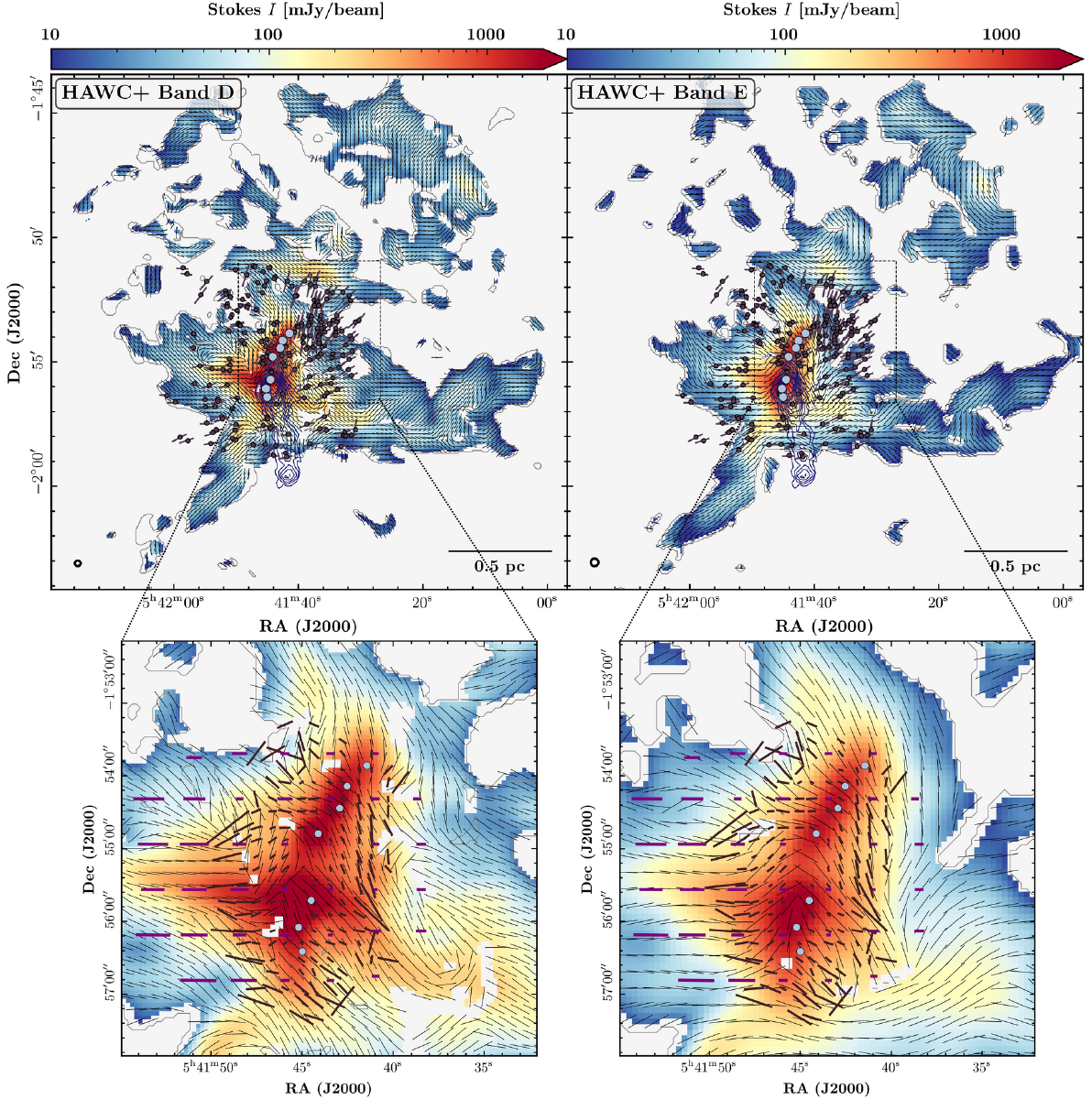}
    \caption{SOFIA HAWC+ dust continuum Band~D and Band~E measurements (left and right panels in the top row, respectively), but with overlayed positions of protostars and NIR polarization \citep[dark purple circles and lines, ][]{kandori_2007}. Dark blue contours represent the outflow observed in the \HCOp\, emission. Light blue points indicate the positions of FIR sources. Black dashed rectangles show the central area and the zoom-in panels on the bottom row. Here, we show magnetic field lines from the FIR dust polarization at 100\,\textmu m \citep{dotson_2000} in dark magenta and from (sub)millimeter dust polarization at 850\,\textmu m \citep{matthews_2002} in black.} 
    \label{fig:hawc+_sources}
\end{figure*}

In Fig.\,\ref{fig:hawc+_sources}, we show a comparison to our SOFIA HAWC+ data with other polarization measurements from the literature. The background in all panels is Stokes $I$ map of Bands D (left panels) and E (right panels), whereas thin black lines show the direction of the magnetic field. We overlay NIR polarization measurements \citep{kandori_2007} as thick black lines and the position of the outflow indicated by the dark blue contours on the top panels of Fig.\,\ref{fig:hawc+_sources}. The bottom panels show the zoom-in of the inner region in NGC~2024 (indicated by the black dashed rectangle in the top panels) and 100\,\textmu m \citep[brown lines,][]{dotson_2000} and magnetic field direction inferred from the 850\,\textmu m dust continuum observations in purple \citep{matthews_2002}.

\section{Spectral line fitting}
\label{sec:line_fitting}
% Finally, we describe the fitting analysis we apply to our CN and \HCOp\ data in Sec. \ref{sec:fit_cn} and \ref{sec:fit_hcop}, respectively.

We fit the CN and \HCOp\, emission lines using the \texttt{CUBE} software, part of \texttt{GILDAS/CLASS}, currently under development by IRAM in Grenoble. We describe the CN fitting procedure in Sec.\,\ref{sec:fit_cn}. Additionally, we use the Semi-automated multi-COmponent Universal Spectral-line fitting Engine \citep[\texttt{SCOUSE},][]{henshaw_2016, henshaw_2019} for fitting the \HCOp\, emission prior to using \texttt{CUBE}, which is further explained in Sec.\,\ref{sec:fit_hcop}.

\subsection{Hyperfine structure of the CN line}
\label{sec:fit_cn}

The CN molecule has a hyperfine structure, and the important parameters of each component of the multiplet studied in this work are shown in Tab.\,\ref{tab:hfs} in Sec.\,\ref{sec:cn_analysis}. Prior to fitting the hyperfine structure, we assume the following. The first assumption is that the components of a multiplet do not overlap with each other. Second, we assume the same excitation temperature for all hyperfine components in the multiplet and the same line width. 

Four parameters describe the hyperfine structure model: $p_1, p_2, p_3$, and $p_4$. The first parameter is the antenna temperature multiplied by the optical depth of the CN emission:

\begin{equation}
    p_1=T_\mathrm{ant}\cdot\tau
\label{eq:p1}
\end{equation}

The second parameter is the centroid velocity of the main hyperfine component (the $F=5/2-3/2$ transition):

\begin{equation}
    p_2 = \upsilon_{0, F=5/2-3/2}.
\label{eq:p2}
\end{equation}

The next parameter is the full width at half maximum (FWHM) of the hyperfine components:

\begin{equation}
    p_3 = FWHM,
\label{eq:p3}
\end{equation}

and the last parameter is the opacity of all components of the studied multiplet:

\begin{equation}
    p_4 = \tau.
\label{eq:p4}
\end{equation}

We assume that the optical depth of each component can be described using a Gaussian function of velocity:

\begin{equation}
    \tau_\mathrm{i} (\upsilon) = \tau_\mathrm{i} \cdot \exp{ \left [ -4\ln{2} \left ( \dfrac{\upsilon - \upsilon_\mathrm{0,i}}{p_3} \right )^2 \right ]}.
\label{eq:taui}
\end{equation}

The opacity of the multiplet is then calculated as a sum of opacities of all hyperfine components:

\begin{equation}
    \tau = \sum_{\mathrm{i}=1}^{N} \tau_i
\label{eq:tau}
\end{equation}

\noindent where $\tau_\mathrm{i}$ is the opacity of the $i$-th component.

Finally, the antenna temperature, $T_\mathrm{ant}$ is thus derived as the ratio between the first and the fourth fitting parameters, $p_1$, and $p_4$: 

\begin{equation}
    T_\mathrm{ant} = \dfrac{p_1}{p_4}\left ( 1 - e^{-\tau(\upsilon)} \right ).
\label{eq:tant}
\end{equation}

To get the excitation temperature, we mask all pixels having the opacity higher than 10 and lower than 0.2 to avoid degeneracies. The excitation temperature is derived assuming the Local Thermodynamical Equilibrium (LTE) case:

\begin{equation}
    T_\mathrm{ex} = \dfrac{h\nu}{k} \left \{ \ln{ \left [ 1 + \dfrac{h\nu}{kT_\mathrm{ant}} \left ( 1 - e^{-\tau_{\nu}} \right ) \right ] } \right \}^{-1}.
\end{equation}

We fit the CN emission in \texttt{CUBE} that uses an optimized version of the minimization method taken from the MINUIT system of CERN. To do so, we load the data into \texttt{CUBE} and run the commands \texttt{$\backslash$fit/minimize} and add \texttt{$\backslash$hfs} to specify we want to fit the hyperfine structure. Therefore, we also provide a file containing information about the components of the hyperfine multiplet, shown in Tab.\,\ref{tab:hfs}. In the case of the hyperfine fitting, \texttt{CUBE} fits four parameters, $p_1, p_2, p_3,$ and $p_4$, described in Eq.\,\ref{eq:p1}, \ref{eq:p2}, \ref{eq:p3}, and \ref{eq:p4} respectively.

\begin{figure*}[t!]
    \centering
    \includegraphics[width=\textwidth]{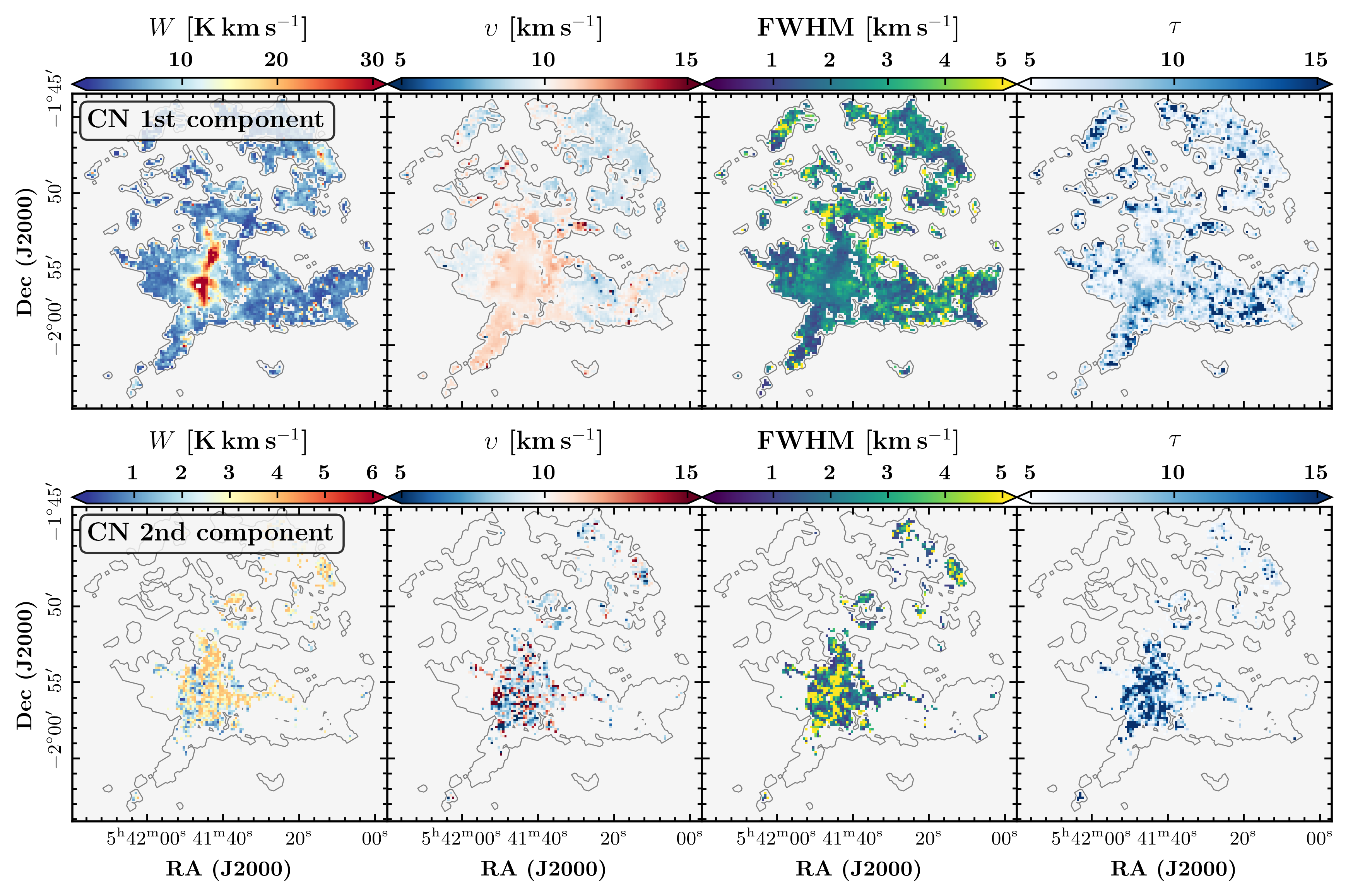}
    \caption{Fitting parameters for the hyperfine CN emission in NGC~2024 measured for both velocity components (top and bottom row each). The first parameter, $p_1$ is shown in the left panel, followed by the centroid velocity, FWHM and the final parameter, the opacity ($p_4$). }
    \label{fig:cn_fit}
\end{figure*}

Before fitting, we define a two-dimensional mask based on the S/N of the CN integrated intensity. We select the threshold of 10 and fit two CN components in the area that goes inside the mask and one component outside the mask. By adding this step, we provide \texttt{CUBE} with additional information about the area where the CN emission is bright enough to observe double components in its spectrum. \texttt{CUBE} requires the initial values of the free parameters for each component we want to fit. We specify our initial guess for the centroid velocities of the components to be 10\,km/s and 7\,km/s. However, we do not specify initial conditions for the rest of the fitting parameters and let \texttt{CUBE} find the best possible values. We show the results of the CN emission fitting for both components in Fig.\,\ref{fig:cn_fit}.

\begin{figure*}[!hbt]
    \centering
    \includegraphics[width=\textwidth]{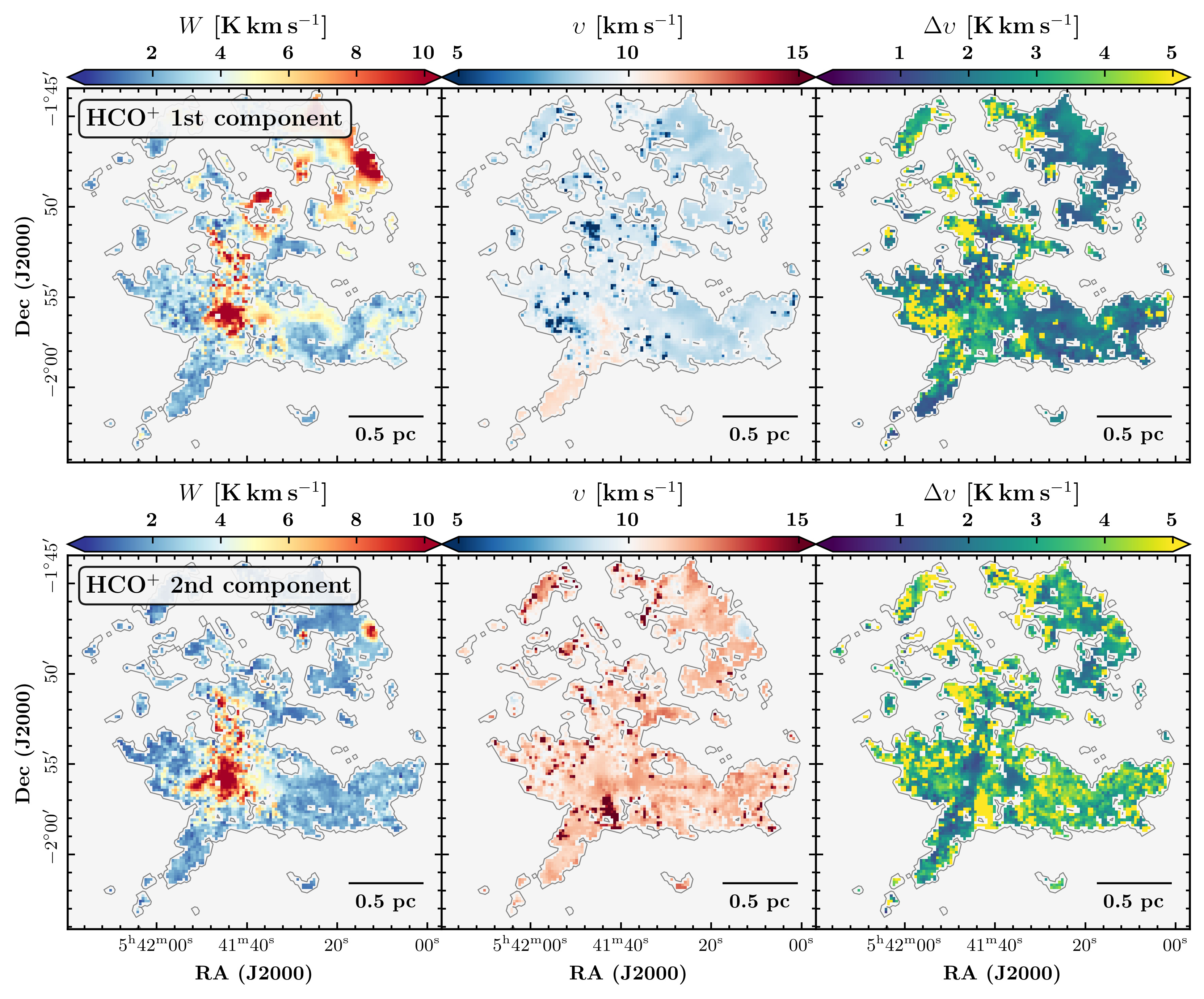}
    \caption{Fitting parameters for the \HCOp\, emission assuming a Gaussian line profile. Each row shows parameters for each fitted component. The results in the first row represent the brightest \HCOp\, component. The second row shows the fainter component and the outflow observed close to the center of NGC~2024.}
    \label{fig:hcop_fit}
\end{figure*}

\subsection{Gaussian lines of the \HCOp\, emission}
\label{sec:fit_hcop}

The \HCOp\, emission line can be described using the Gaussian function. Similarly, as in the case of the CN, by inspecting the \HCOp\, data cube, we notice the presence of two velocity components across the large portion of the map. In addition, we observe a region in which \HCOp\, spectra contain extended line wings, suggestive of the molecular outflow also seen in the CO emission, whose presence is also known from previous studies \citep[for instance, see,][]{richer_1992}. Different velocity components we observe correspond to the primary and intermediate velocity layers seen in the CO emission and its isotopologues across NGC~2024 \citep{gaudel_2023}.

Before fitting the Gaussian line to the \HCOp\ spectra, similarly as in the case of CN, we want to locate regions where \HCOp\, the emission has more than one peak. Therefore, we decompose the \HCOp\, emission using \texttt{SCOUSEPY}. For the full description of the fitting procedures in \texttt{SCOUSEPY}, we refer to work by \cite{henshaw_2016,henshaw_2019}. In the first step, \texttt{SCOUSEPY} divides the spectral cube into spectral averaging areas (SAAs). The spectrum of each SAA is the average spectrum of all pixels found within the SAA. Then, by assuming the shape of the spectral line, \texttt{CUBE} decomposes the spectrum of each SAA, identifies a number of velocity components, and suggests a model in a second step. In this step, the user can modify a model suggested by \texttt{SCOUSEPY}. Next, in the third step, \texttt{SCOUSEPY} fits emission in each pixel based on the fitting model of each SAA. The final step allows the user to check the fitting result within each pixel and, similarly to the second step, modify the model if needed.

We fit a Gaussian line profile to the \HCOp\, emission in \texttt{SCOUSEPY}. The output parameters describing a Gaussian function are each component's peak temperature, centroid velocity, and FWHM. Additionally, \texttt{SCOUSEPY} computes the rms, S/N and the residuals. Based on these results we derive from \texttt{SCOUSEPY}, we define two two-dimensional masks. The first mask contains a region where \texttt{SCOUSEPY} identifies three velocity components. The second mask is a region within which \texttt{SCOUSEPY} finds two velocity components. We use these masks as the input to \texttt{CUBE} to mark regions where we observe three (first mask), two (second mask), and one velocity component (outside these two masks).

Similarly, as for the CN emission, we use the command \texttt{$\backslash$fit/minimize} in \texttt{CUBE} and add \texttt{$\backslash$gaussian} to specify the shape of the spectral line. In \texttt{CUBE}, the Gaussian line is described using three parameters: the area (in K\,km/s), the centroid velocity, and the FWHM. We also specify initial guesses for fitting parameters. In the last step, we select the brightest component to be the first, and then combine results of the second (fainter) component and the outflowing feature into one map. We show the results of this fitting procedure in Fig.\,\ref{fig:hcop_fit}.

\section{Radiative transfer modeling of CN and \HCOp\, emission}
\label{sec:radex_all}

\begin{figure*}
    \centering
    \includegraphics[width = \textwidth]{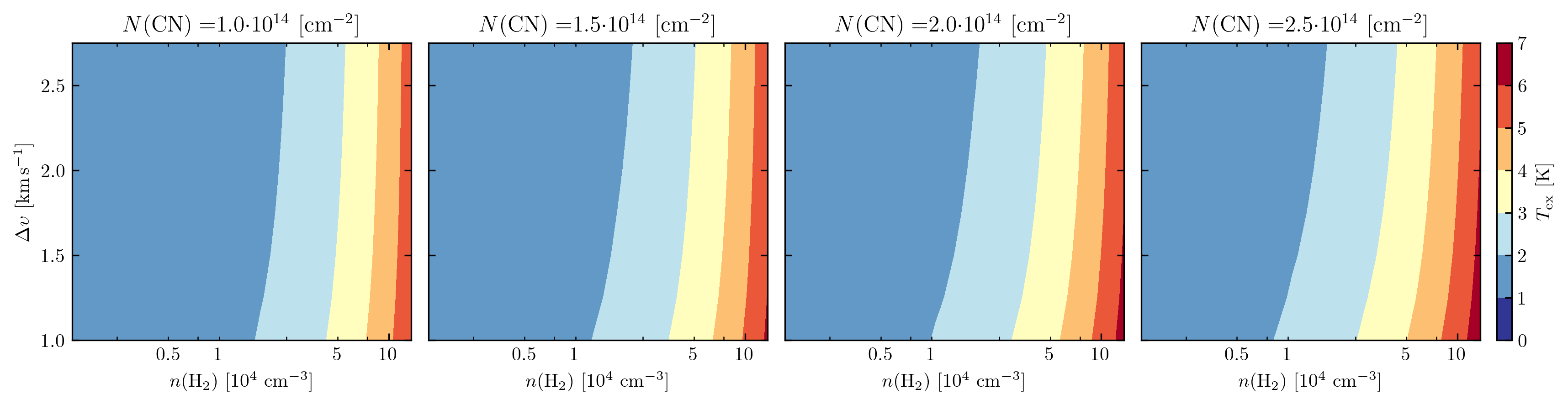}
    \includegraphics[width = \textwidth]{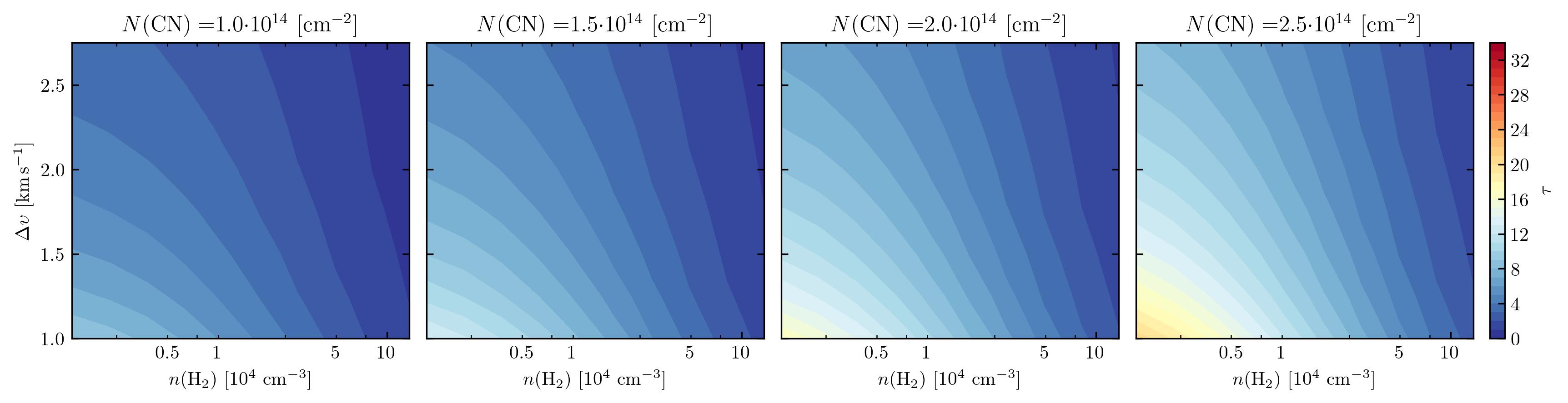}
    \includegraphics[width = \textwidth]{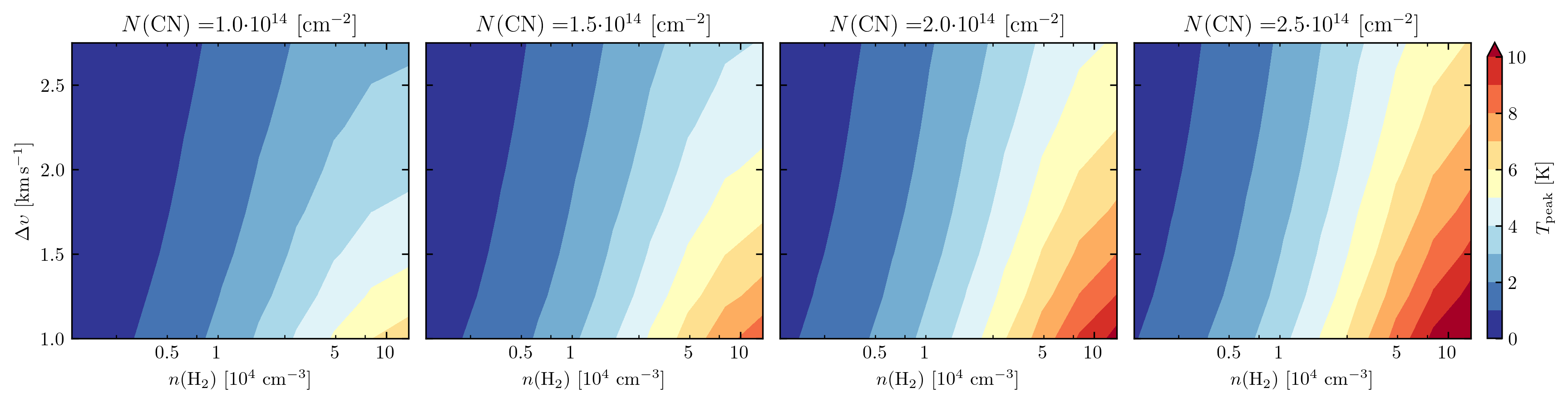}
    \includegraphics[width = \textwidth]{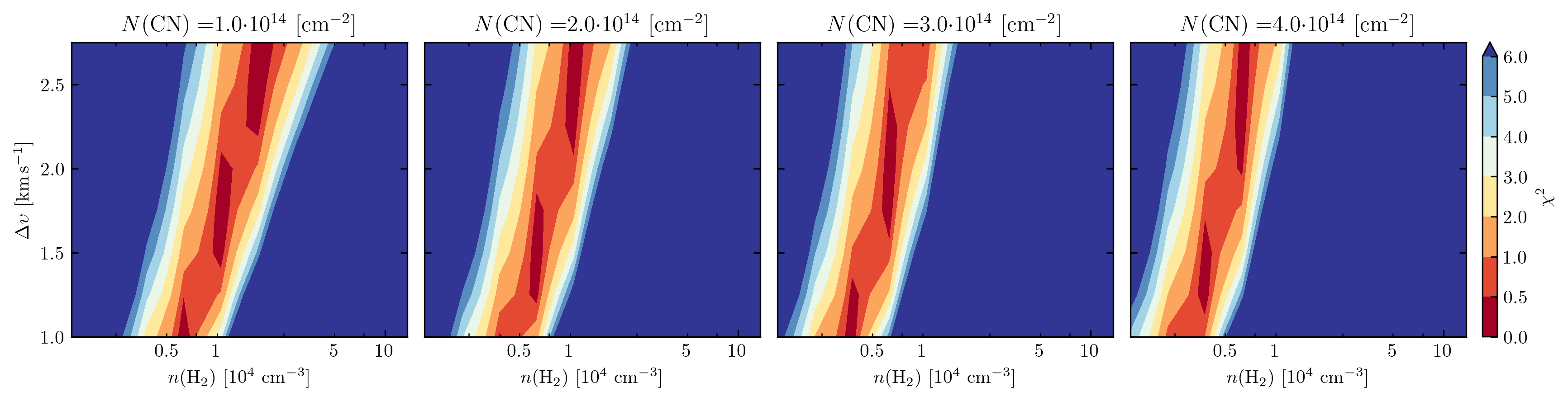}
    \caption{Parameter space used for the radiative transfer modeling of CN in the region at the western side of the bubble. The first row shows results for the excitation temperature. The color bar shows values of excitation temperature, $T_\mathrm{ex}$ for the grid of volume densities (x-axis), line widths (y-axis) and column densities (each panel) for fixed electron fraction. The second row shows results for opacity, $\tau$, and the third row shows the peak temperature, $T_\mathrm{peak}$. The last row shows the $\chi^2$ minimization of the modeled peak temperature and the observed value computed from the CN\Jone\ spectra.}
    \label{fig:radex_cn_west}
\end{figure*}

\begin{figure*}
    \centering
    \includegraphics[width = \textwidth]{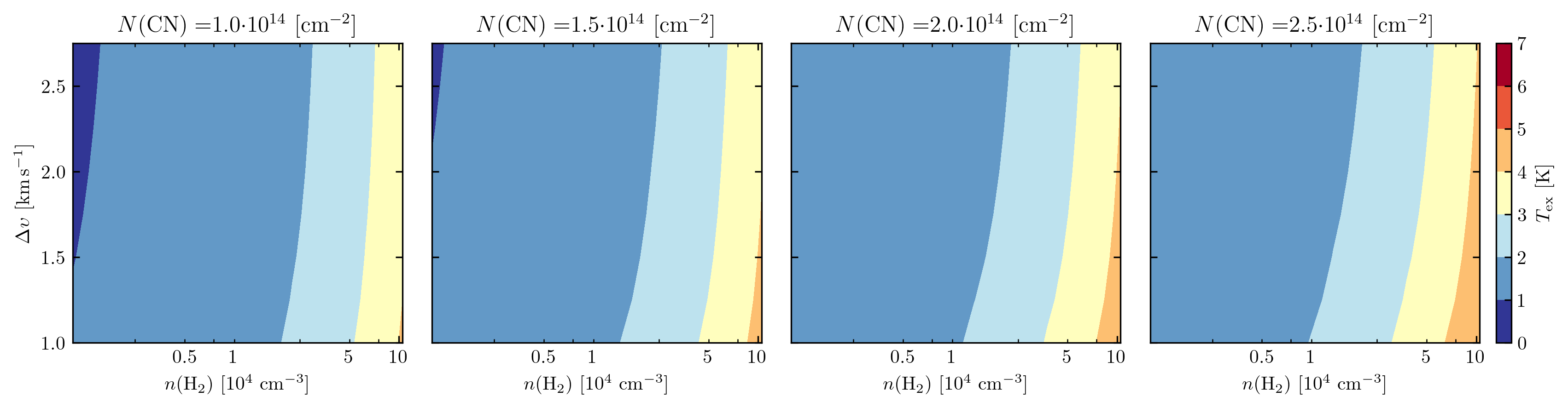}
    \includegraphics[width = \textwidth]{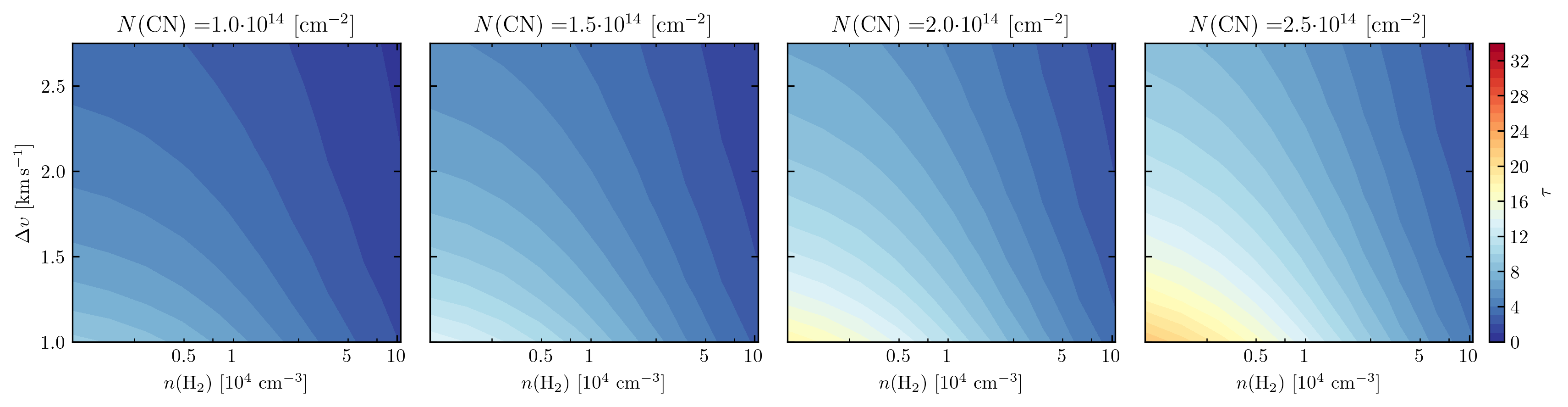}
    \includegraphics[width = \textwidth]{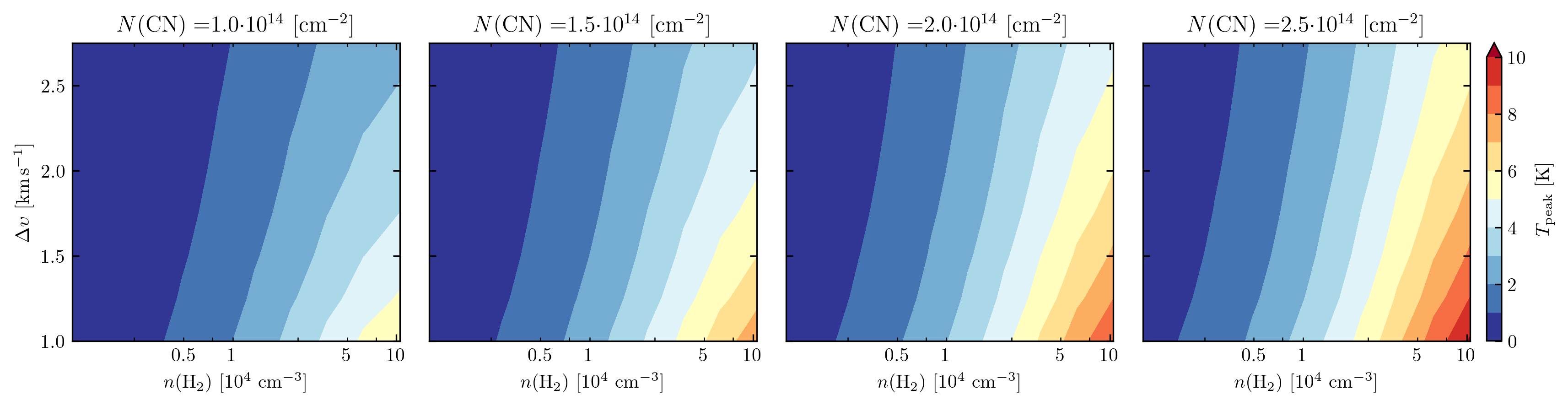}
    \includegraphics[width = \textwidth]{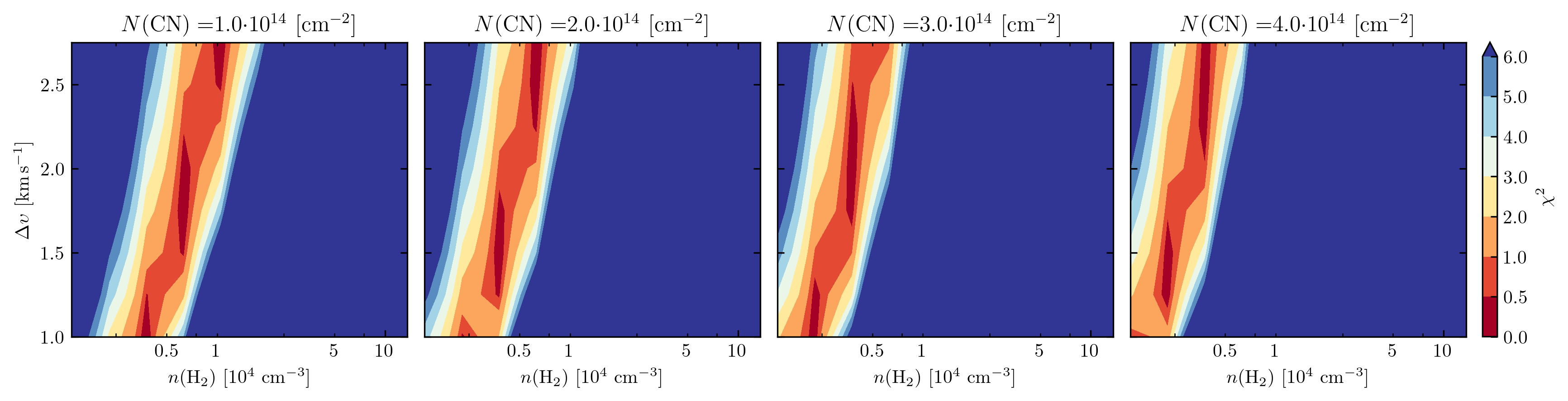}
    \caption{Same as in Fig.\,\ref{fig:radex_cn_west}, but for the eastern part of the bubble.}
    \label{fig:radex_cn_east}
\end{figure*}

\begin{figure*}
    \centering
    \includegraphics[width = \textwidth]{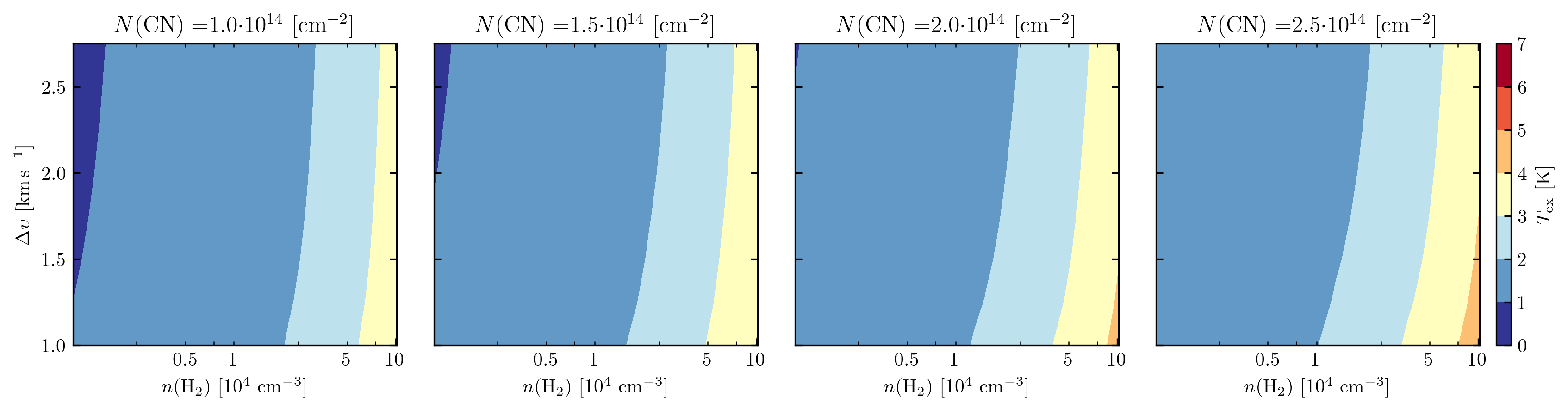}
    \includegraphics[width = \textwidth]{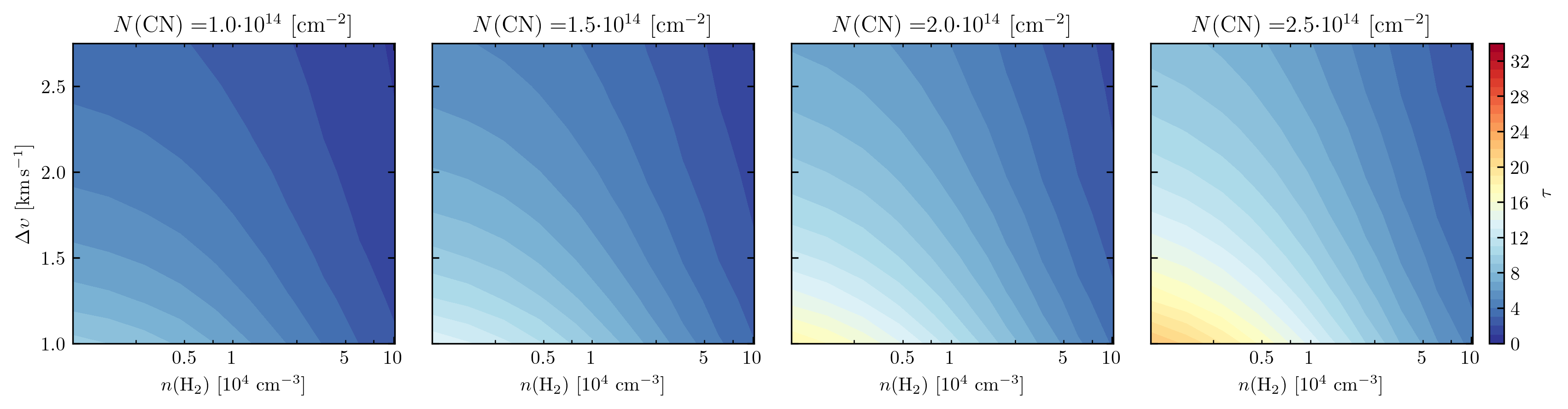}
    \includegraphics[width = \textwidth]{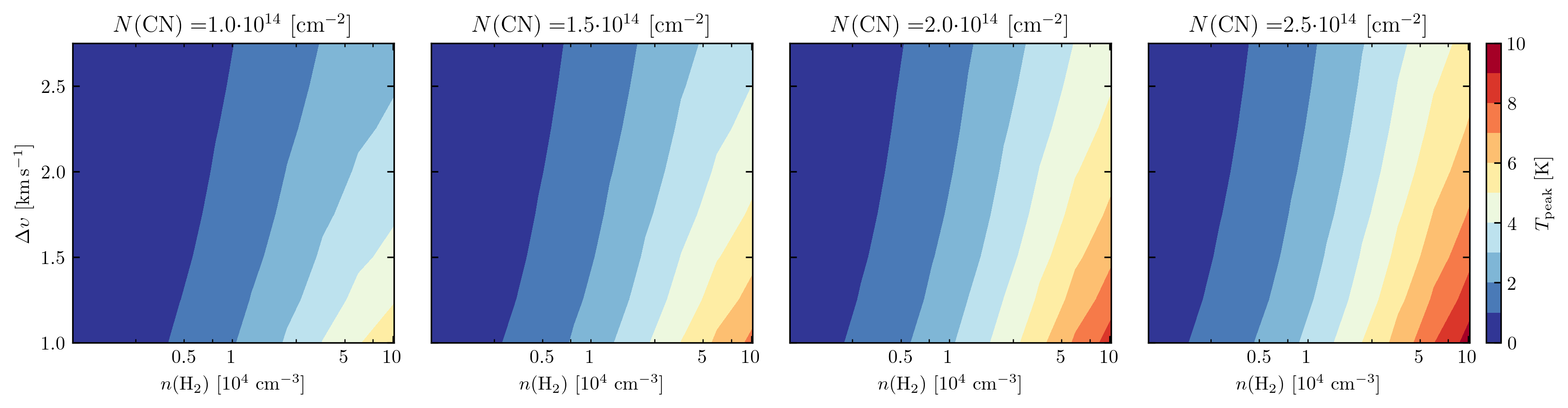}
    \includegraphics[width = \textwidth]{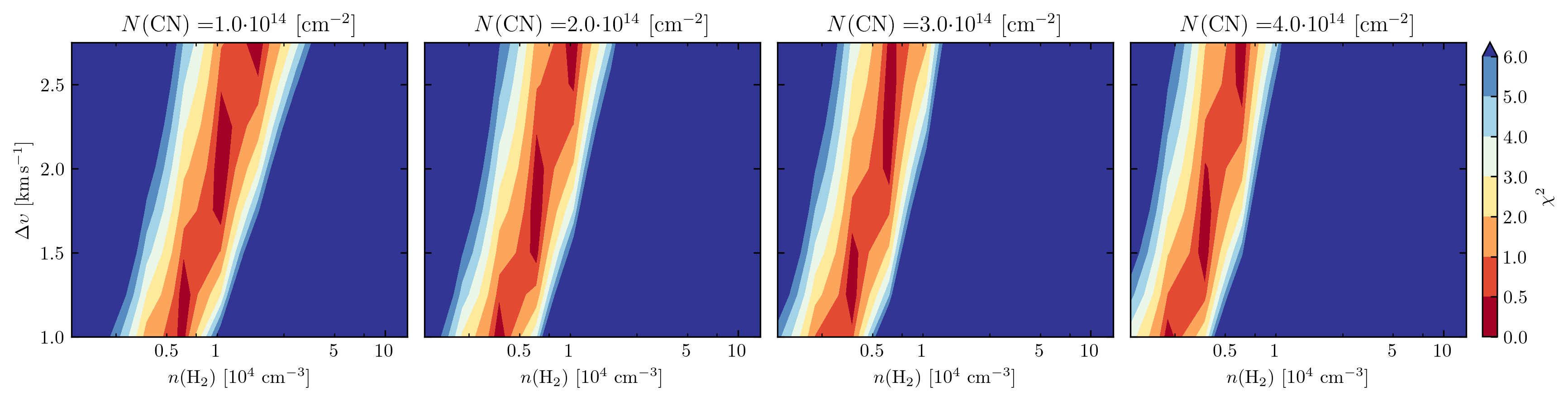}
    \caption{Same as in Fig.\,\ref{fig:radex_cn_west}, but for the filament.}
    \label{fig:radex_cn_fil}
\end{figure*}

\begin{figure*}
    \centering
    \includegraphics[width = \textwidth]{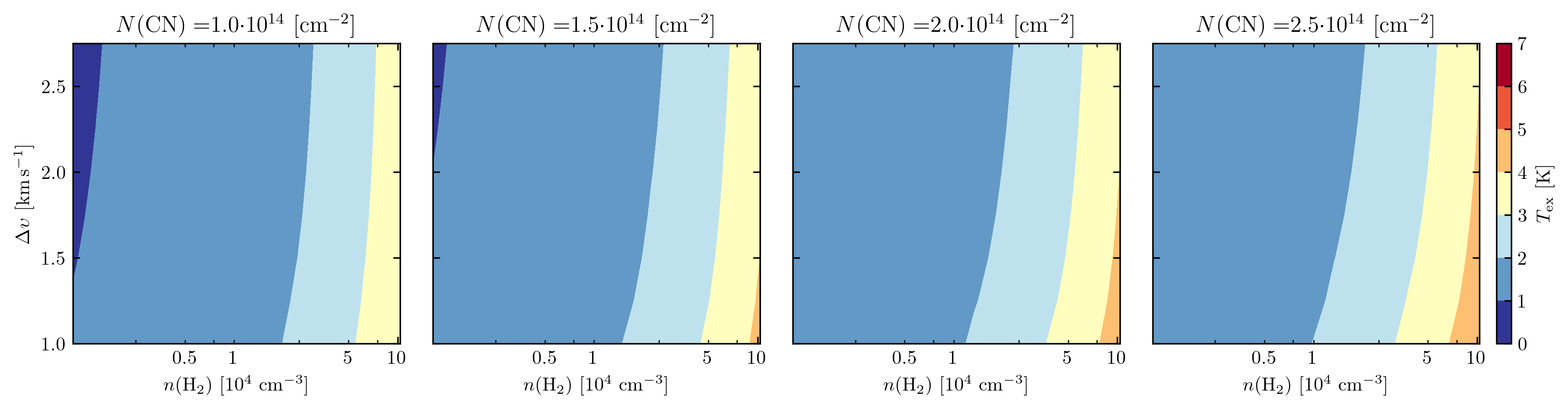}
    \includegraphics[width = \textwidth]{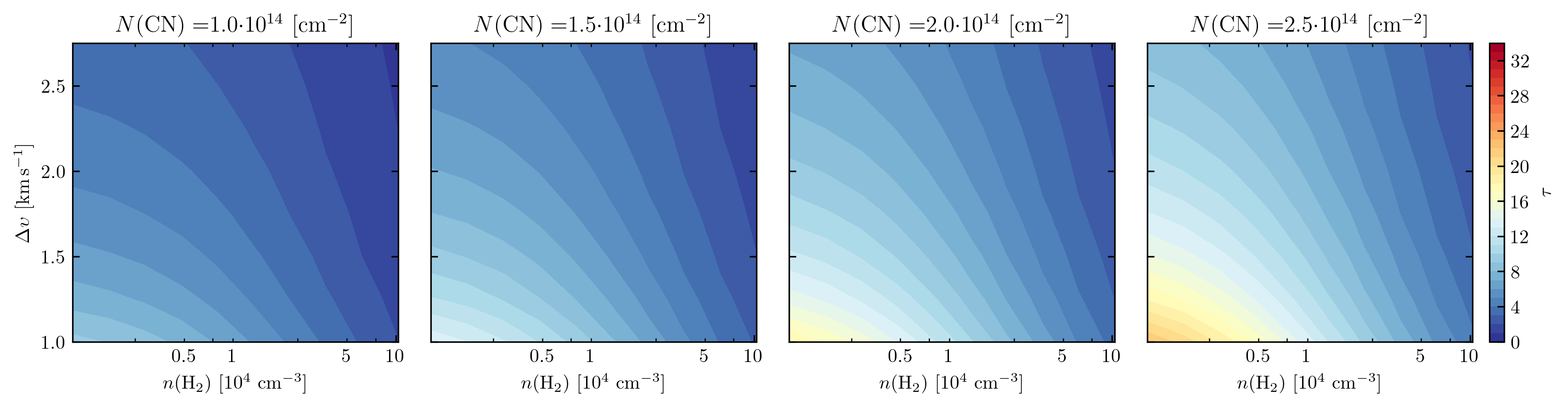}
    \includegraphics[width = \textwidth]{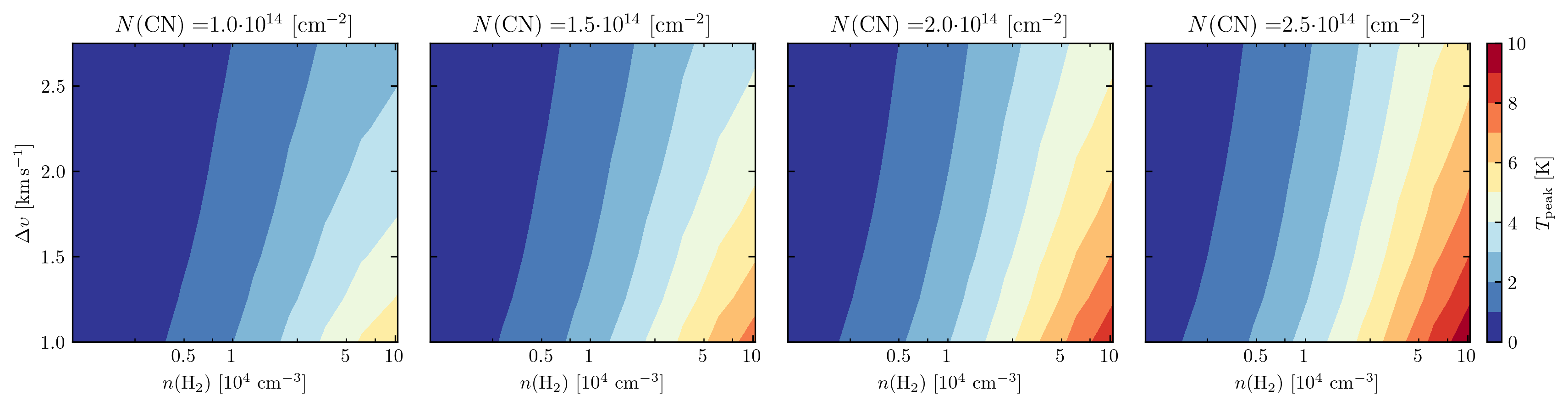}
    \includegraphics[width = \textwidth]{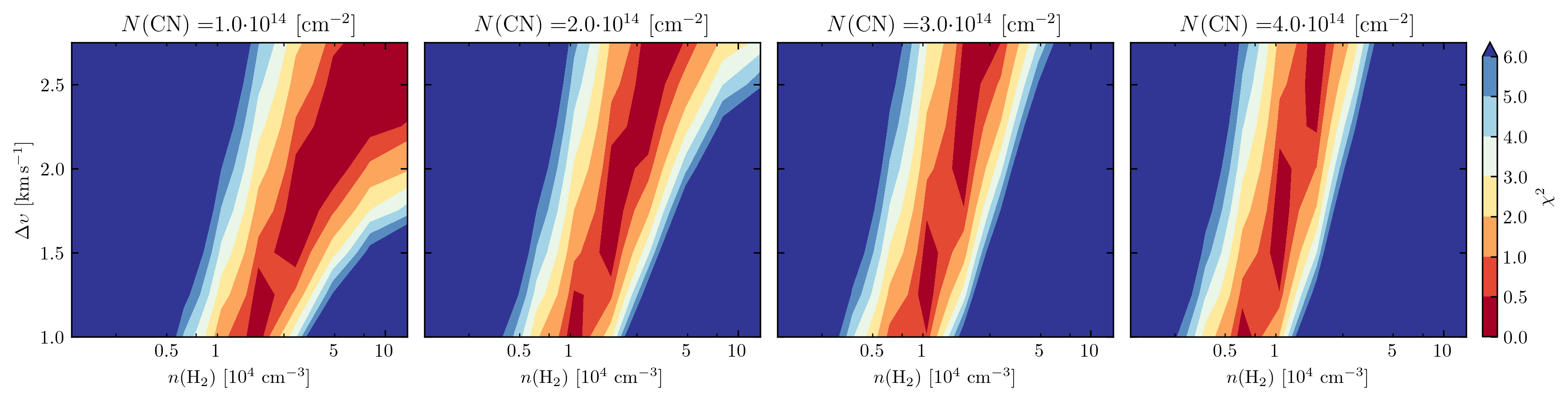}
    \caption{Same as in Fig.\,\ref{fig:radex_cn_west}, but for the overlap region.}
    \label{fig:radex_cn_mix}
\end{figure*}

\begin{figure*}
    \centering
    \includegraphics[width = \textwidth]{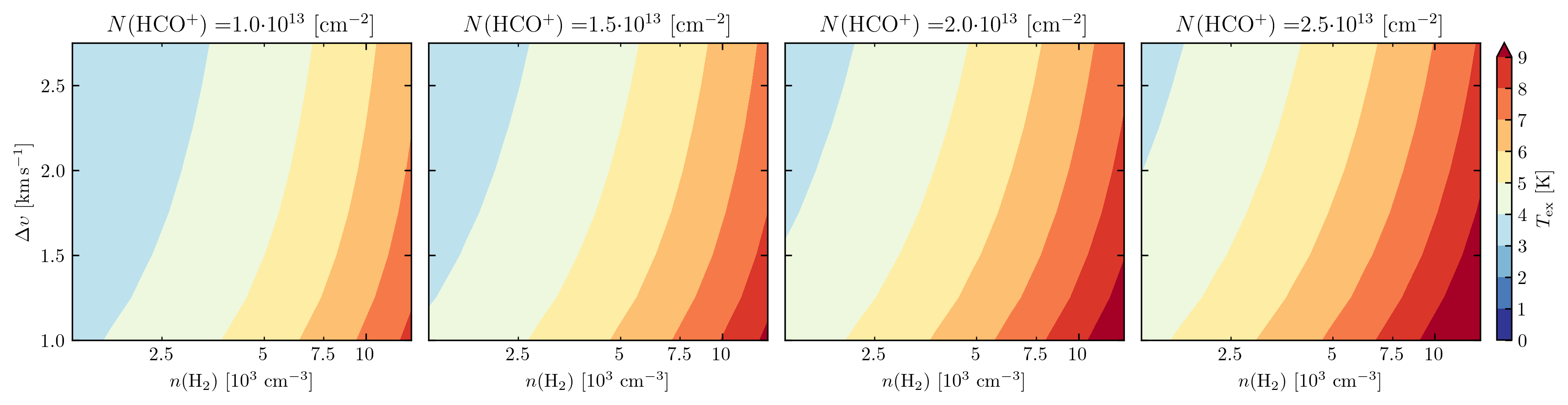}
    \includegraphics[width = \textwidth]{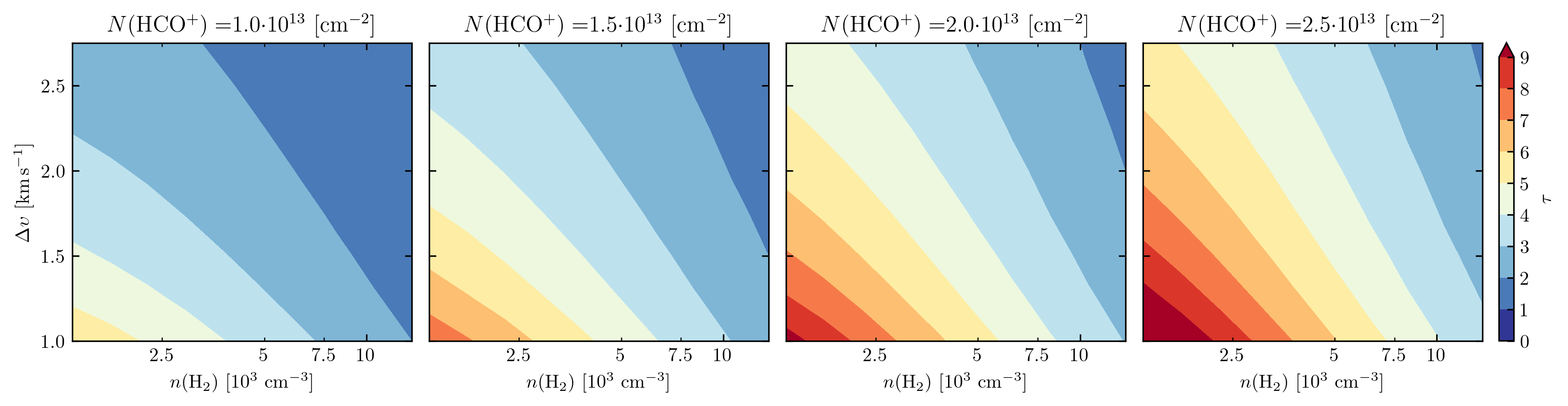}
    \includegraphics[width = \textwidth]{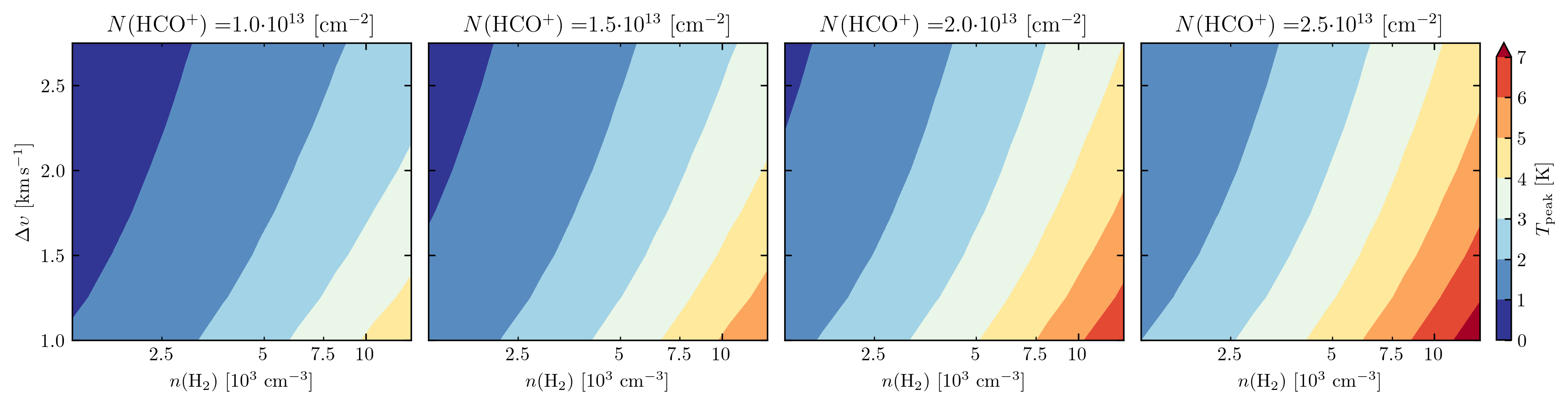}
    \includegraphics[width = \textwidth]{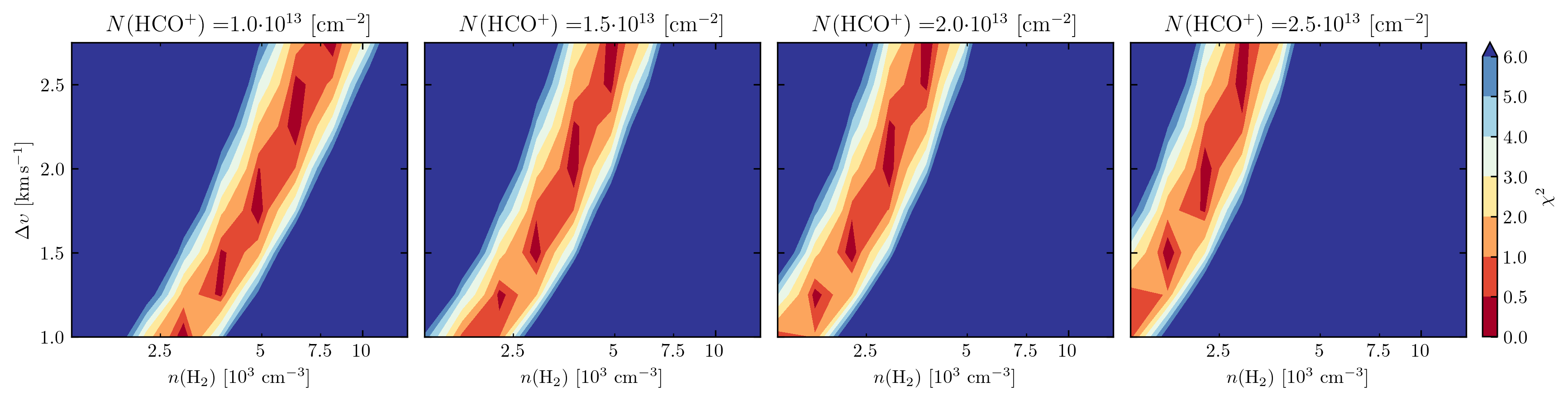}
    \caption{Same as in Fig.\,\ref{fig:radex_cn_west}, but for \HCOp\Jone.}
    \label{fig:radex_west_hcop}
\end{figure*}

\begin{figure*}
    \centering
    \includegraphics[width = \textwidth]{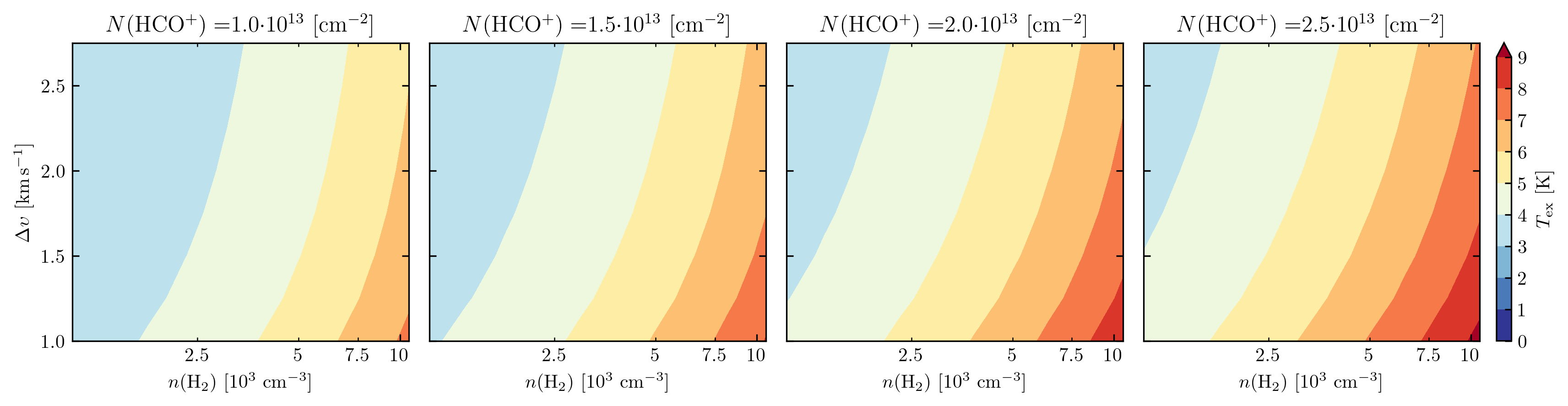}
    \includegraphics[width = \textwidth]{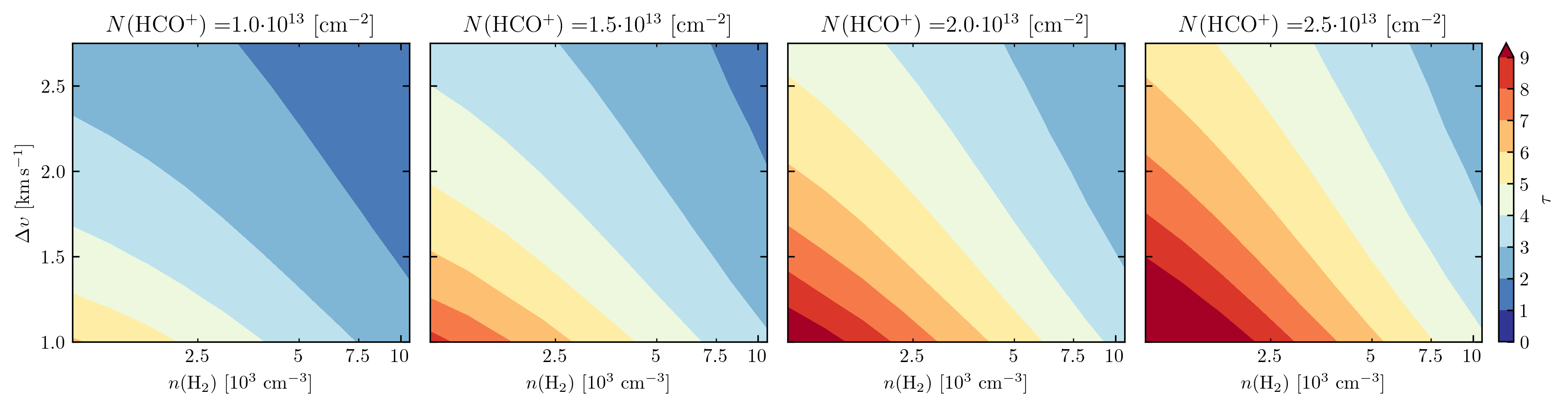}
    \includegraphics[width = \textwidth]{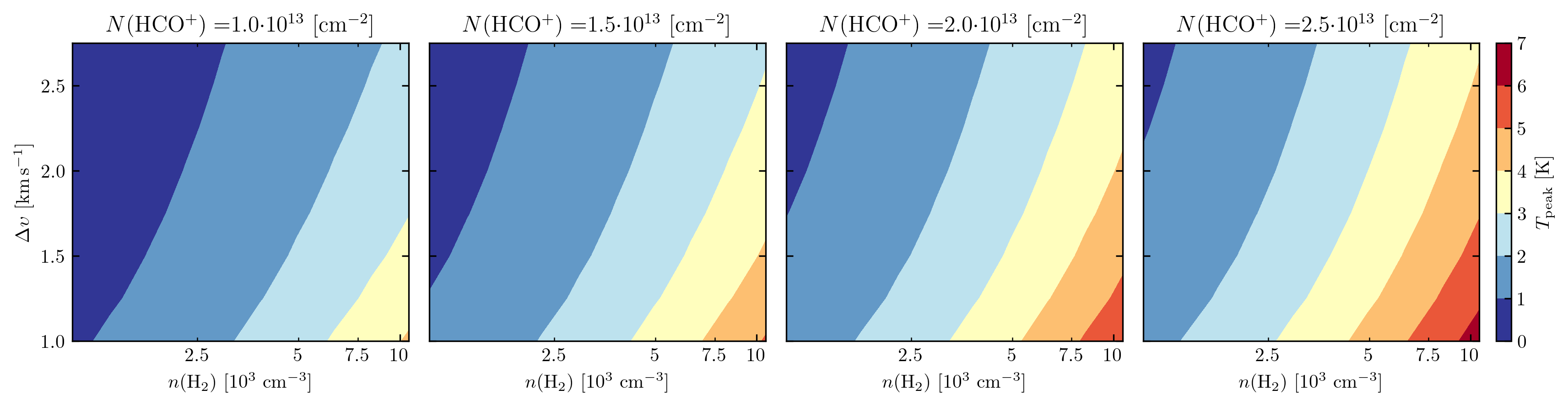}
    \includegraphics[width = \textwidth]{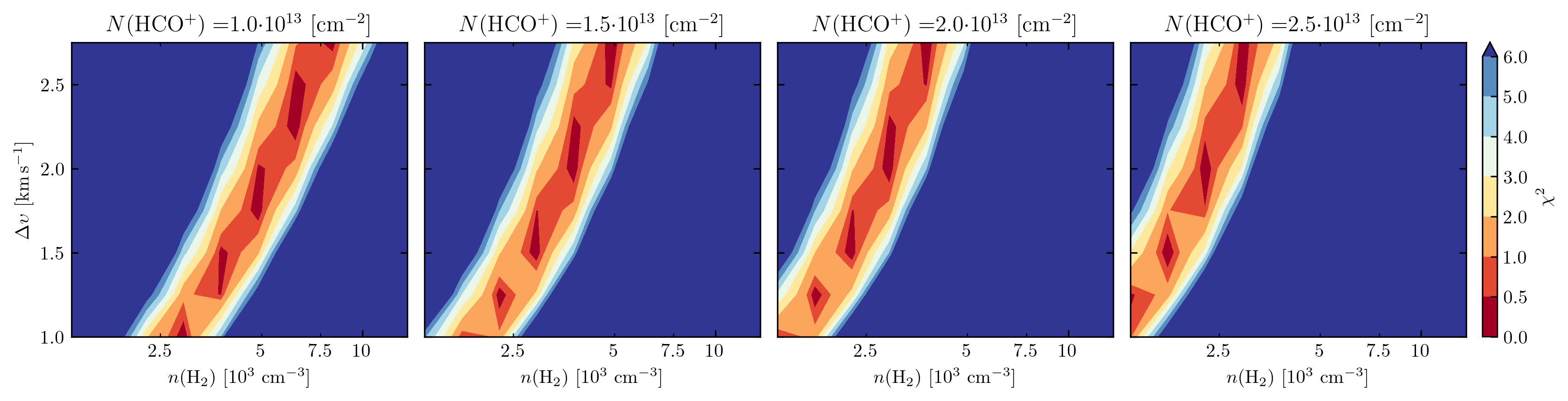}
    \caption{Same as in Fig.\,\ref{fig:radex_west_hcop}, but for the eastern side of the bubble.}
    \label{fig:radex_east_hcop}
\end{figure*}

\begin{figure*}
    \centering
    \includegraphics[width = \textwidth]{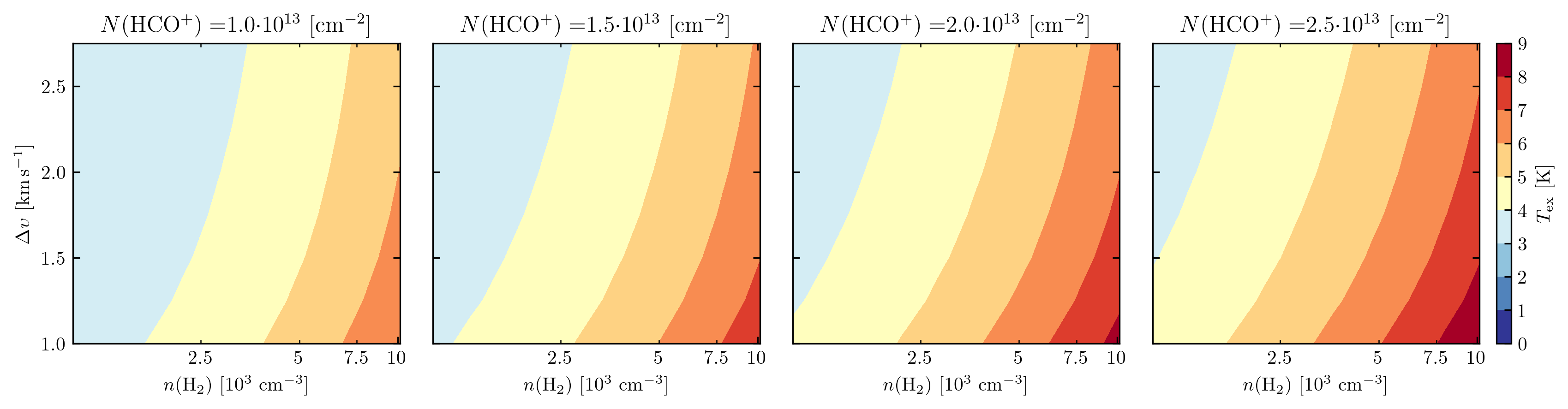}
    \includegraphics[width = \textwidth]{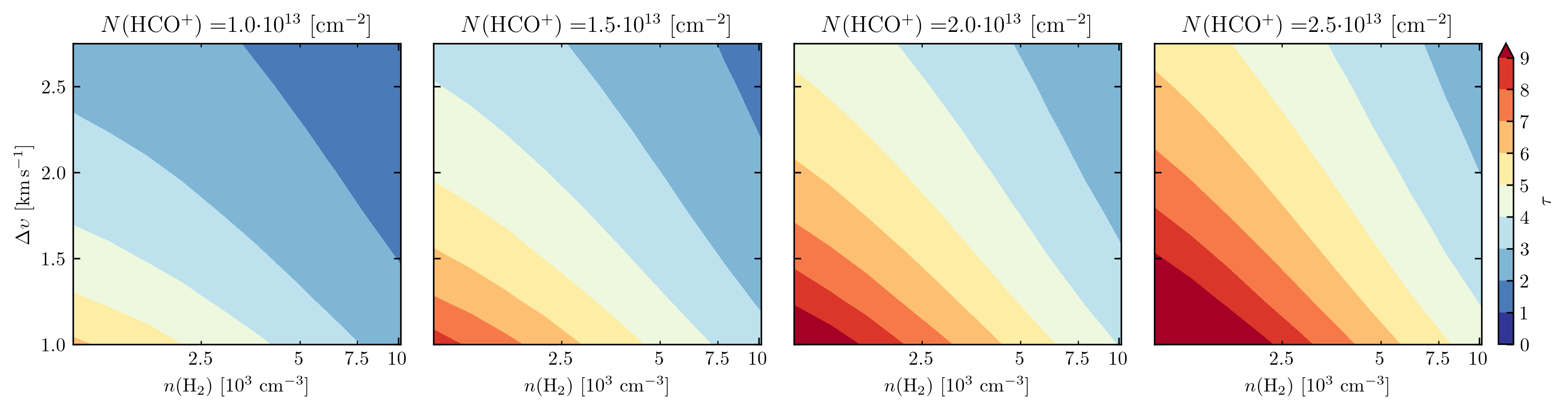}
    \includegraphics[width = \textwidth]{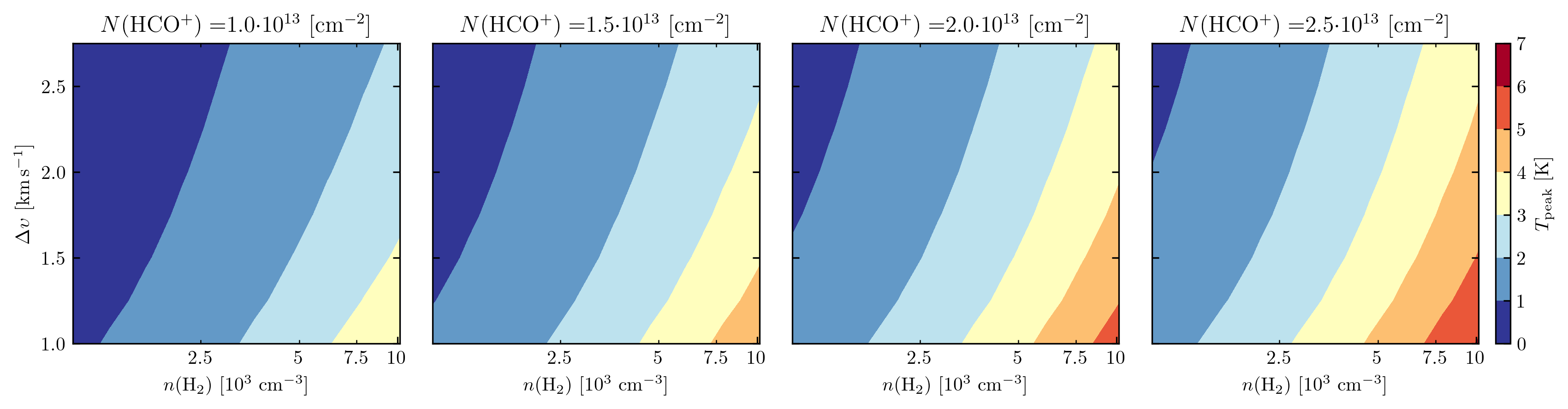}
    \includegraphics[width = \textwidth]{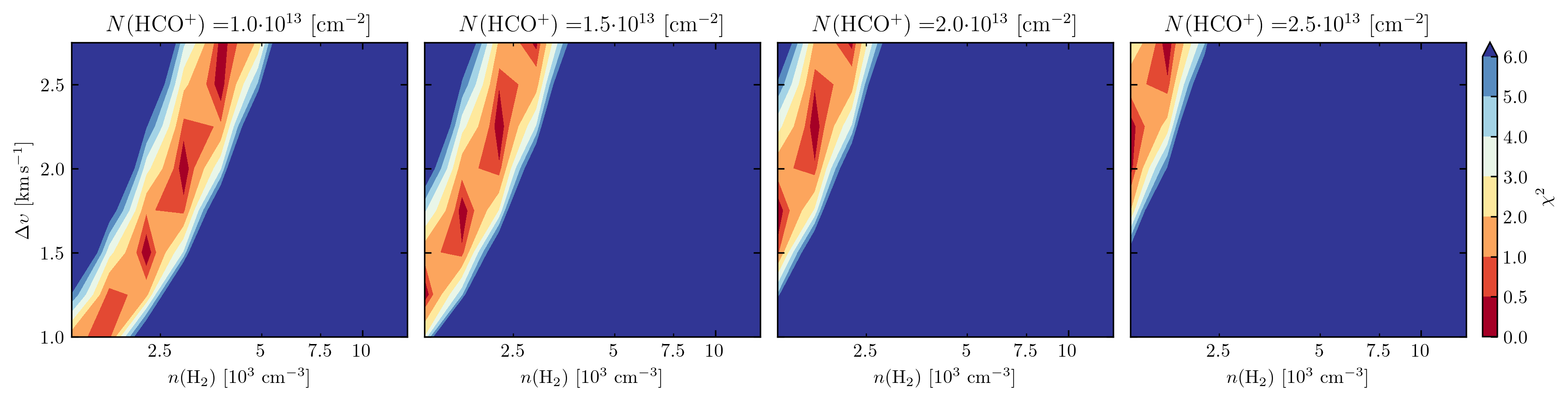}
    \caption{Same as in Fig.\,\ref{fig:radex_west_hcop}, but for the filament.}
    \label{fig:radex_fil_hcop}
\end{figure*}

\begin{figure*}
    \centering
    \includegraphics[width = \textwidth]{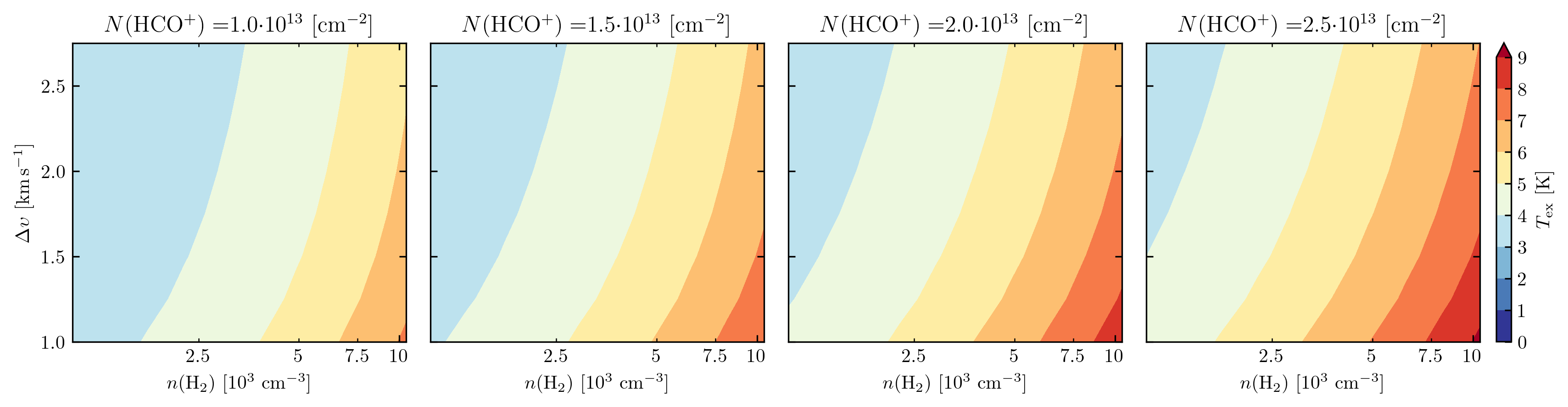}
    \includegraphics[width = \textwidth]{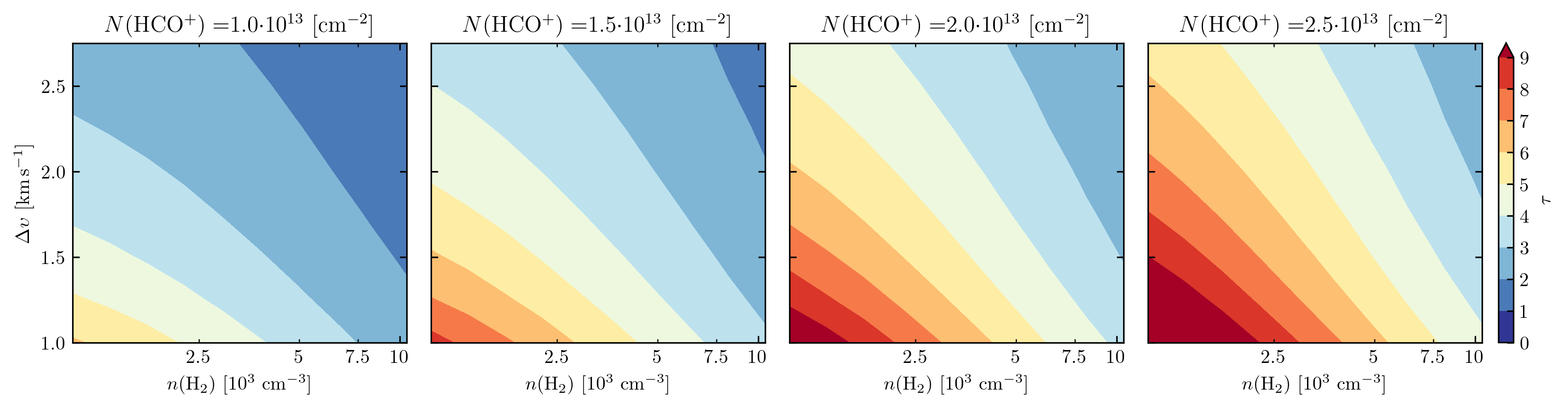}
    \includegraphics[width = \textwidth]{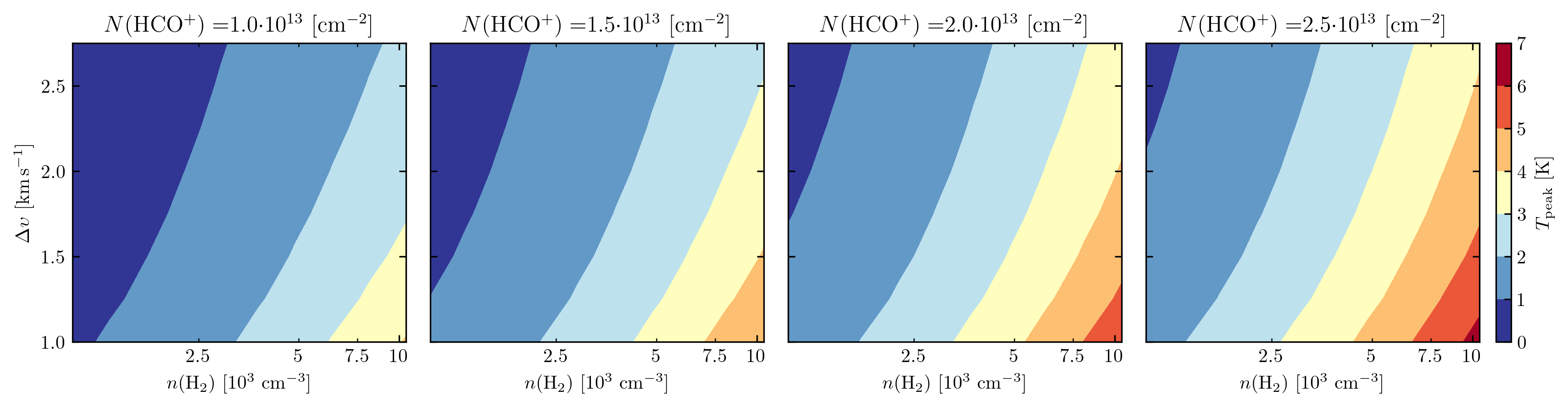}
    \includegraphics[width = \textwidth]{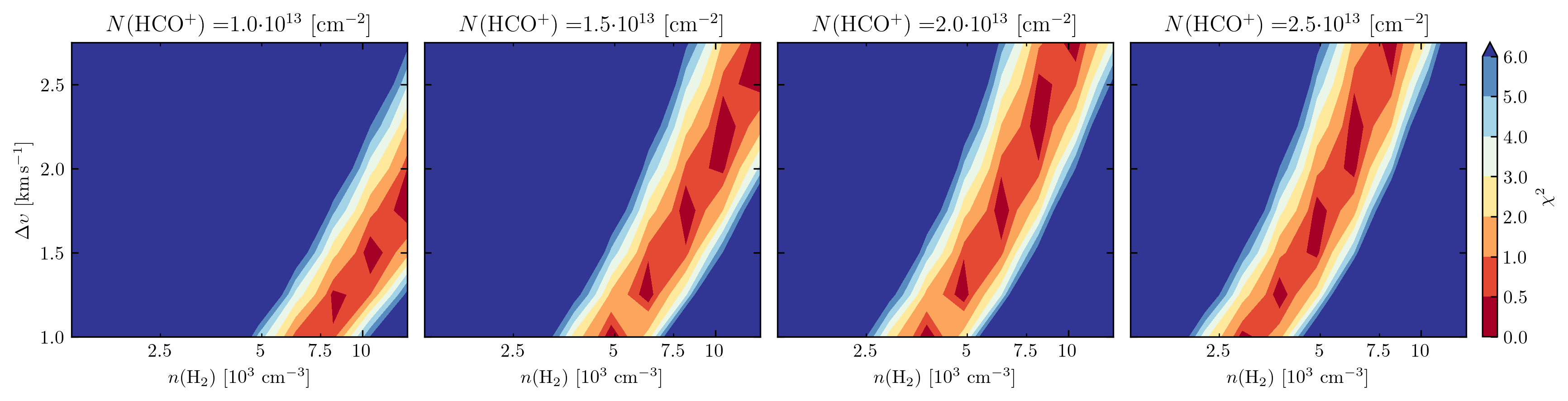}
    \caption{Same as in Fig.\,\ref{fig:radex_west_hcop}, but for the overlap region.}
    \label{fig:radex_mix_hcop}
\end{figure*}

Here, we show results from the radiative transfer modeling of CN and \HCOp\, excitation, presented in Sec.\,\ref{subsec:radex}. For each region we study in this work (edges of the bubble, filament, and the overlap region), we show how our output parameters from \texttt{RADEX} (excitation temperature, opacity, and the peak temperature) vary as a function of input parameters (line width, column and volume density) in Fig.\,\ref{fig:radex_cn_west}, \ref{fig:radex_cn_east}, \ref{fig:radex_cn_fil} and \ref{fig:radex_cn_mix} for CN emission. In the case of \HCOp\, emission, we show these results in Fig.\,\ref{fig:radex_west_hcop}, \ref{fig:radex_east_hcop}, \ref{fig:radex_fil_hcop}and \ref{fig:radex_mix_hcop}. We also include a $\chi^2$ as a function of the input parameters at the bottom row in each figure.

\begin{figure*}
    \centering
    \includegraphics[width = 0.95\textwidth]{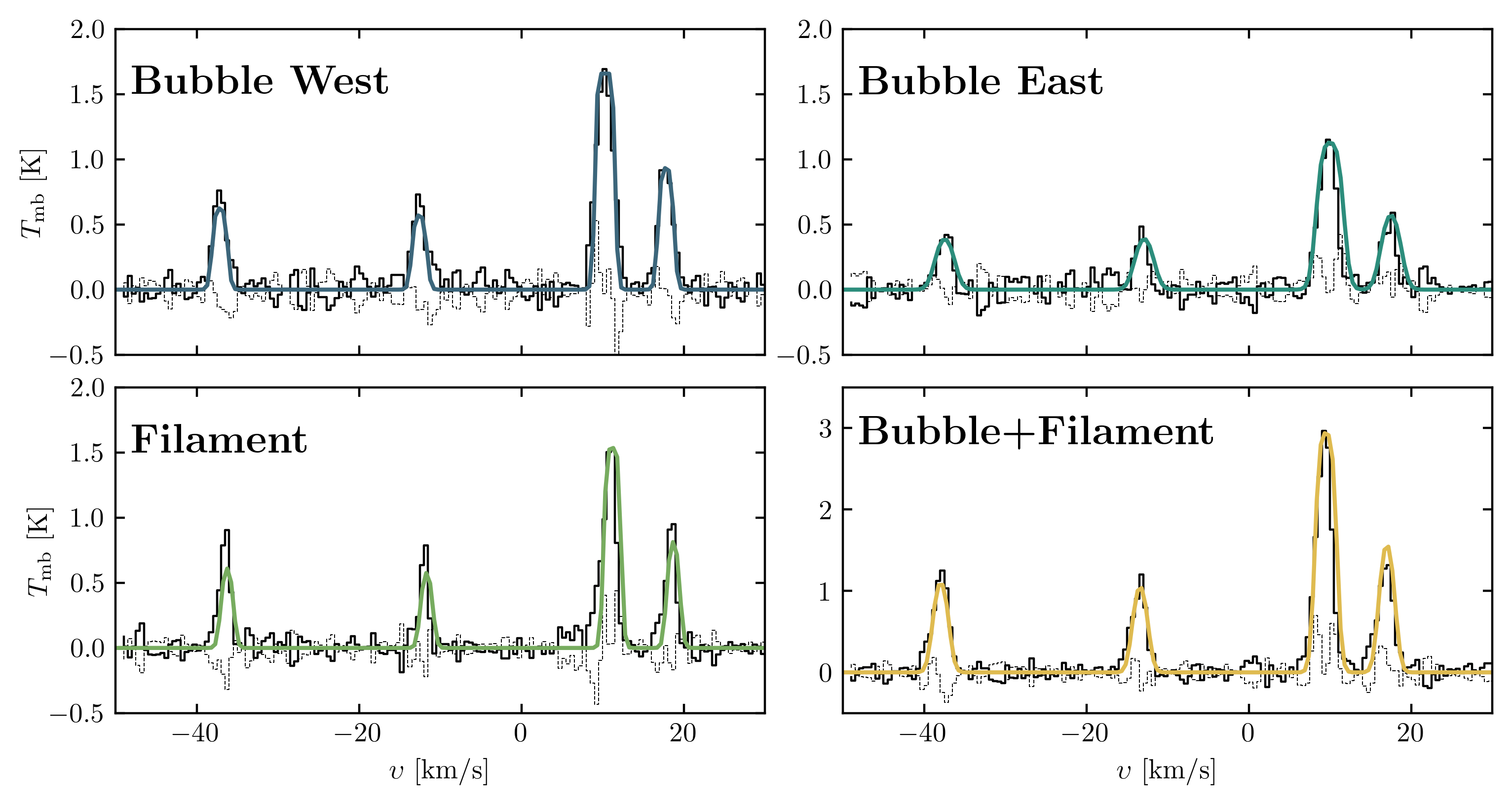}
    \includegraphics[width = 0.95\textwidth]{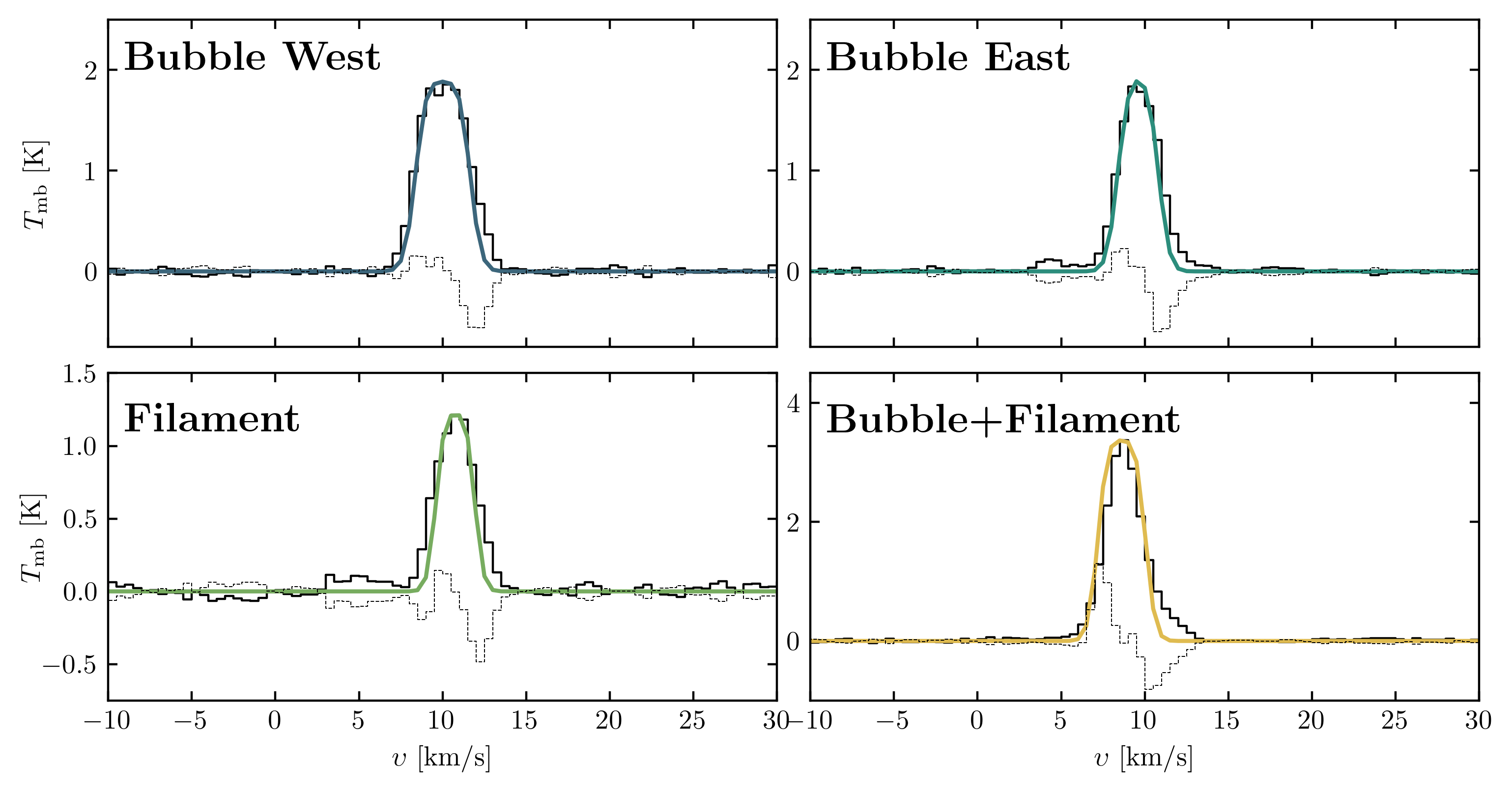}
    \caption{Beam averaged spectrum of CN and \HCOp\, taken within four different regions shown in the left panel of Fig.\,\ref{fig:fig1} and in Fig.\,\ref{fig:dust_pol}. Colored line represent the corresponding model spectrum infrerred from the non-LTE RADEX modeling for physical conditions presented in Sec.\,\ref{sec:radex_all} and shown in Tab.\,\ref{tab:pixels}. Dashed lines show the corresponding residuals.}
    \label{fig:cn_hcop}
\end{figure*}

Based on the results presented in Sec.\,\ref{subsec:radex}, we computed model spectrum of CN and \HCOp\ for edges of the bubble, filament and the overlap area. We show the beam-averaged spectrum of CN and \HCOp\, of each region studied in this work and their model in Fig.\,\ref{fig:cn_hcop}.

%%%%%%%%%%%%%%%%%%%%%%%%%%%%%%%%%%%%%%%%%%%%%%%%%%%%%%%%%%%%%%%%%%%%%%%%%%%%%%%%%%%%%%%

\section{Measuring magnetic field strength in NGC~2024}
\label{app:B_field}

In this part, we provide additional information on the sliding window technique in Sec.\,\ref{app:slidding_window}, an alternative approach to measuring the angle dispersion using the histogram analysis in Sec.\,\ref{sec:histogram}, and results of magnetic field strength computed using different methods in Sec.\,\ref{subsec:dcf_skalidis}. 

\subsection{Sliding window}
\label{app:slidding_window}

\begin{figure*}
    \centering
    \includegraphics[width=0.7\textwidth]{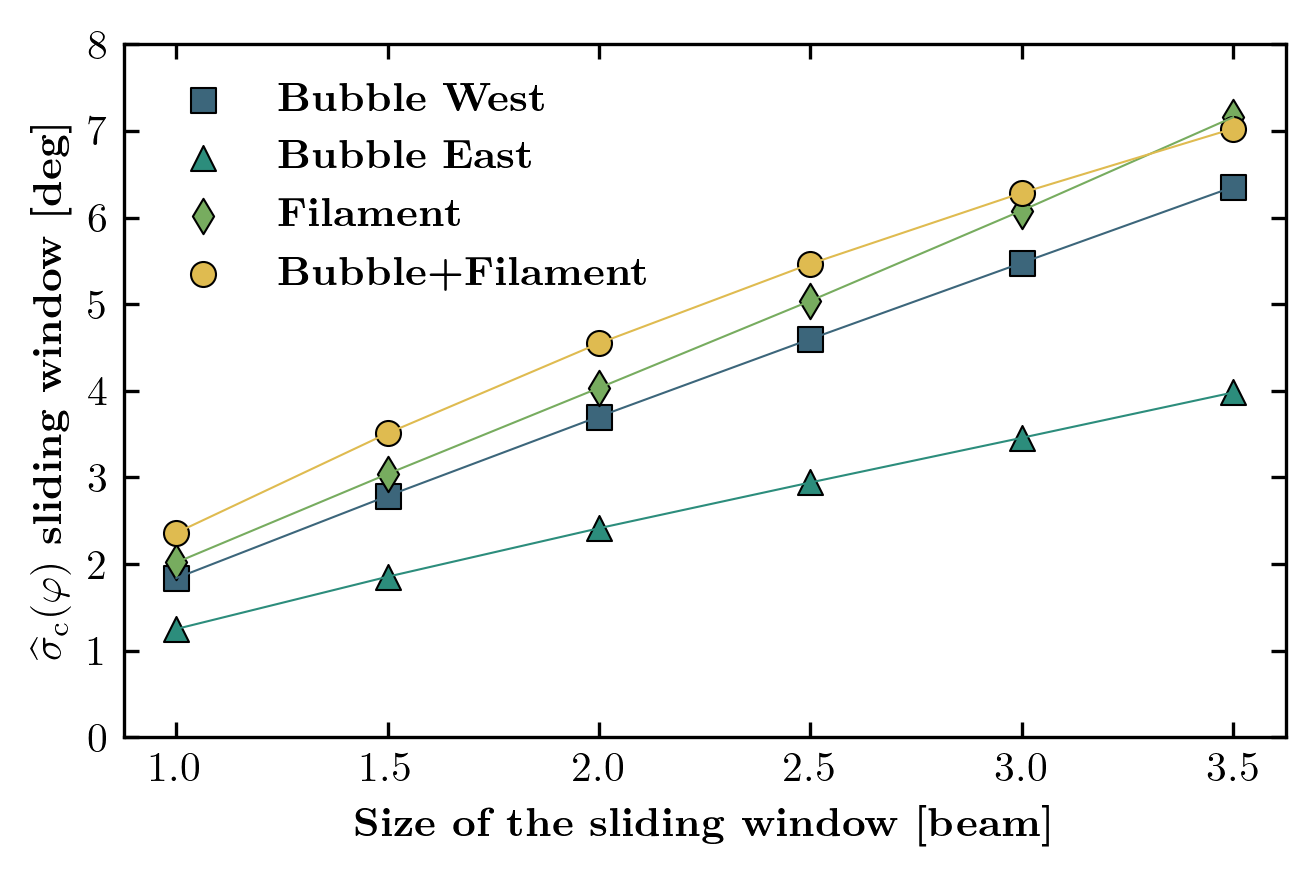}
    \caption{Angle rms as a function of the size of the sliding window.}
    \label{fig:sliding_window}
\end{figure*}

We investigate the change of the angle dispersion measured using the sliding window (Sec.\,\ref{subsec:slidding_window}) by varying its size. We used the size of the sliding window from 1 to 3.5 beam sizes, with a step of 0.5, which corresponds to sizes from $4\times4$ to $14\times14$ pixels, respectively. We show these results in Fig.\,\ref{fig:sliding_window}. The beam-sized sliding window is too small to remove any large-scale contribution of the magnetic field. We find that the rms of the angle changes by a factor of around 3 as we increase the size of the slidding window.

\subsection{Histogram analysis}
\label{sec:histogram}
\begin{figure*}[t!]
    \centering
    \includegraphics[width = 0.49\textwidth]{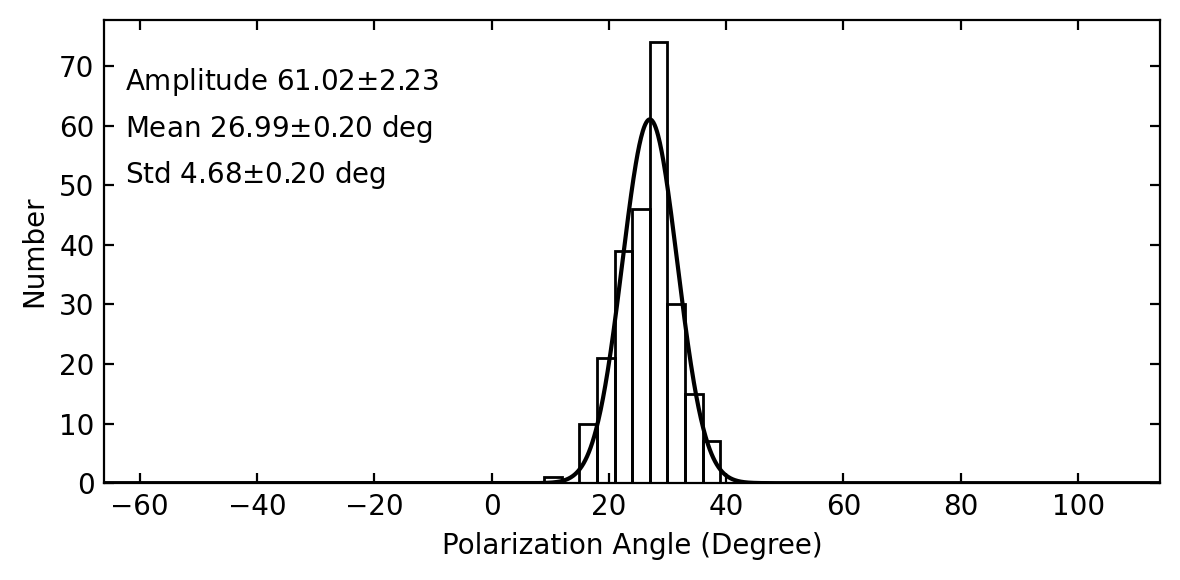}
    \includegraphics[width = 0.49\textwidth]{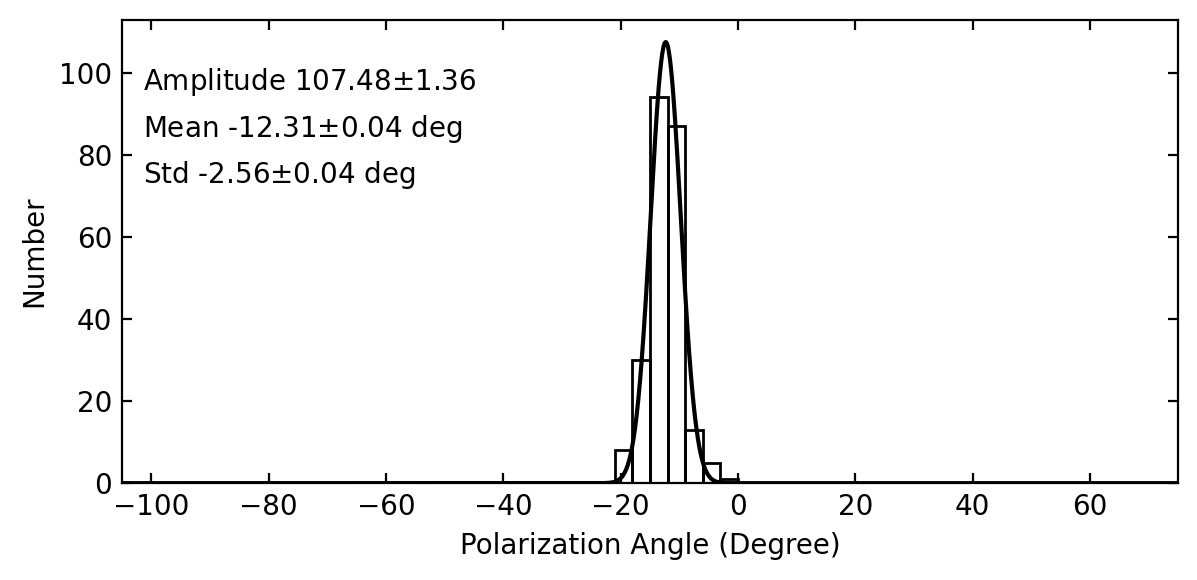}
    \includegraphics[width = 0.49\textwidth]{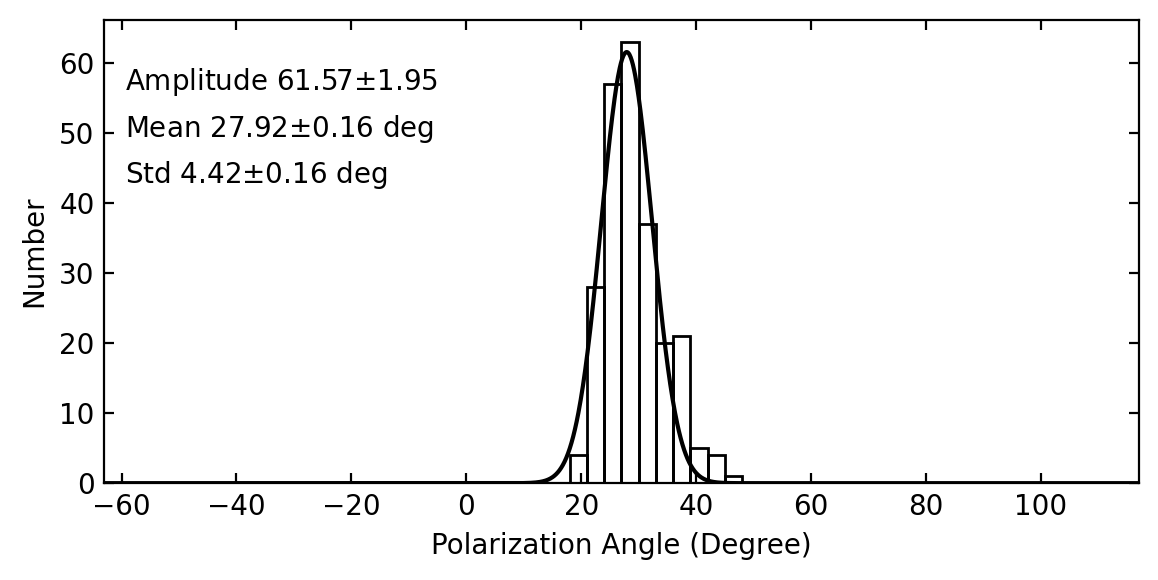}
    \includegraphics[width = 0.49\textwidth]{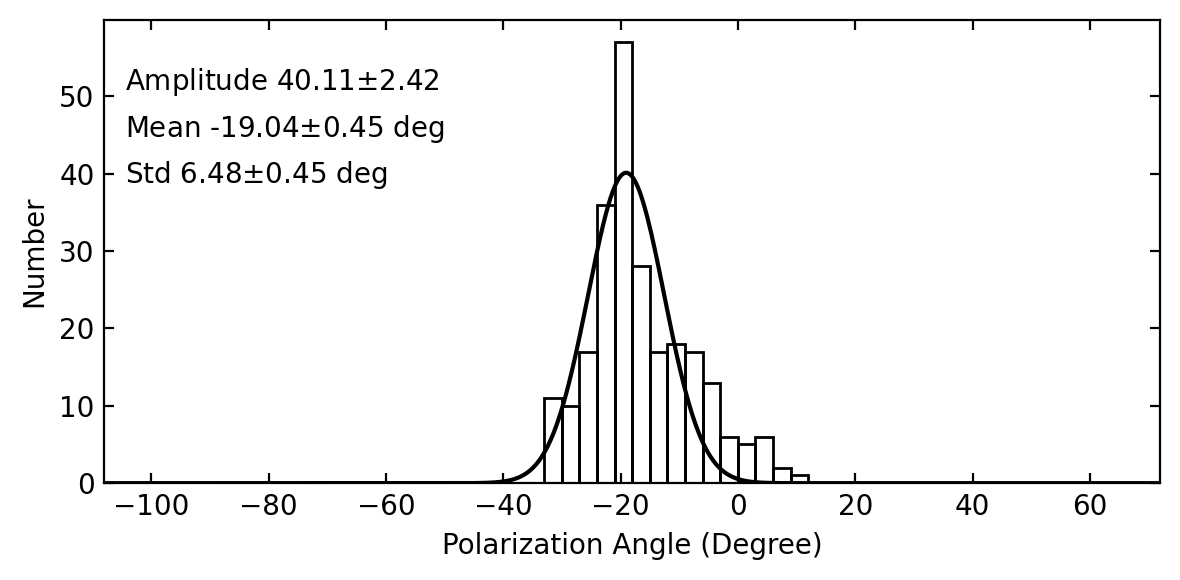}
\caption{Top left: Bubble on the west. Top right: Bubble on the eastern side of NGC~2024. Bottom left: Filament. Bottom right: Mixture of bubble and dusty filament.}
\label{fig:histograms}
\end{figure*}

\begin{figure*}
    \centering
    \includegraphics[width=0.6\textwidth]{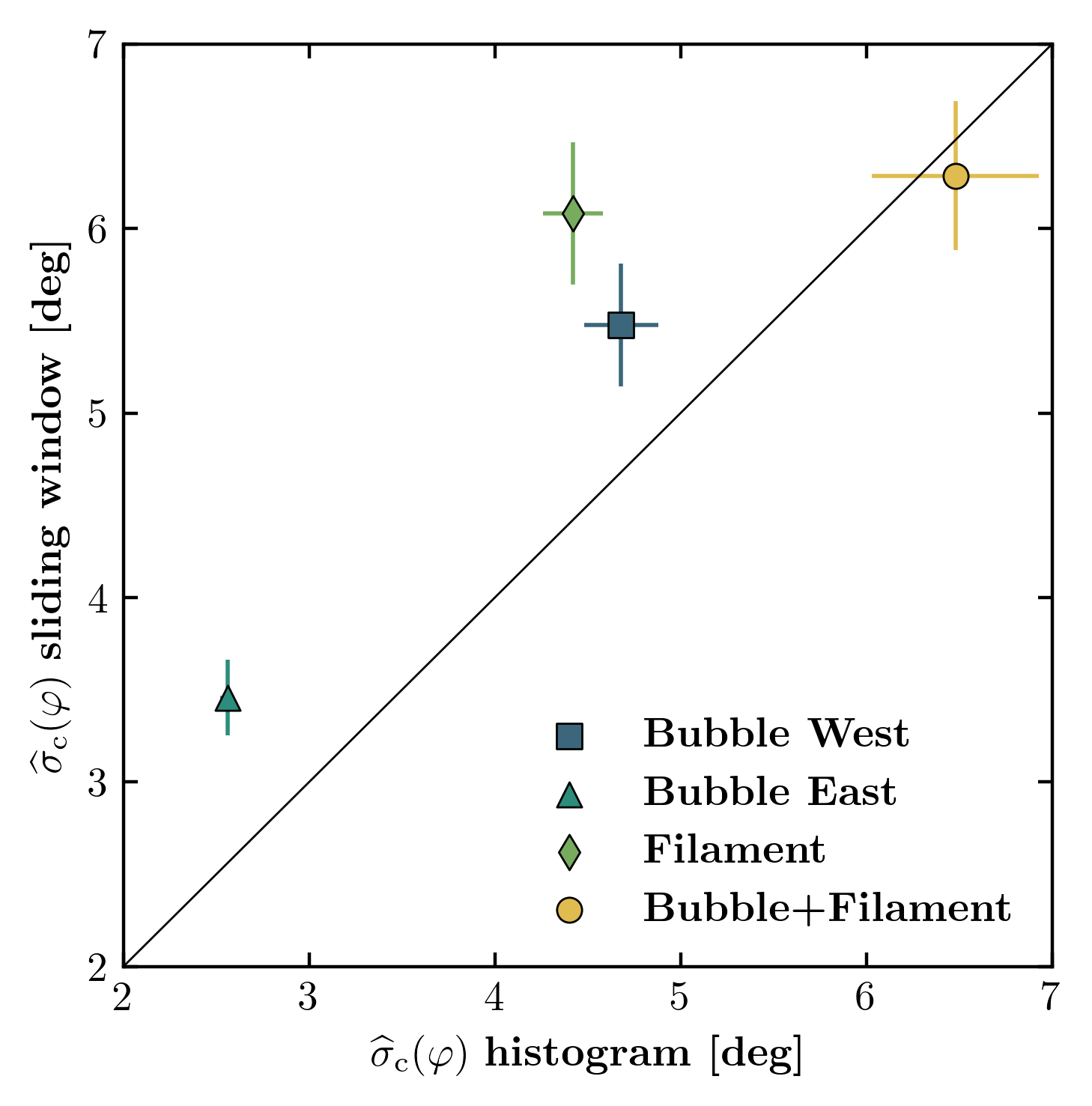}
    \caption{Comparison of the angle dispersion using two different approaches, sliding window (y-axis, see Sec.\,\ref{sec:Bfield_results}) and histogram analysis (x-axis).}
    \label{fig:b_hist_win}
\end{figure*}

% We take a beam-averaged value as shown in \textcolor{red}{Fig.\,\ref{fig:dust_pol}}. 
In this section, we describe an alternative approach to compute $\widehat{\sigma}_\mathrm{c}(\varphi)$. We consider all points within each circularly shaped region and create a histogram showing the distribution of magnetic field angles. Next, we compute the circular mean and fit the Gaussian function to the distribution of angles. We show the histograms for each beam-averaged area we used in this work, including the fitting parameters of each distribution in Fig.\,\ref{fig:histograms}. In most cases, the histogram corresponds to a single Gaussian. In Fig.\,\ref{fig:b_hist_win}, we show comparison between $\widehat{\sigma}_\mathrm{c}(\varphi)$ from sliding window (y-axis) and histogram analysis (x-axis). The derived standard deviations from the sliding window (Sec.\,\ref{sec:Bfield_results}) are higher than those derived from the histogram analysis.

\begin{table*} %[t!]
\centering
\caption{Results on the magnetic field strength from histogram analysis.}
%%%%%%%%%%%%%%%%%%%%%%%%%%%%%%%%%%%
\begingroup
\setlength{\tabcolsep}{8pt} % Default value: 6pt
\renewcommand{\arraystretch}{1.2} % Default value: 1
%%%%%%%%%%%%%%%%%%%%%%%%%%%%%%%%%%%
\resizebox{0.8\textwidth}{!}{ {\begin{tabular}{lccccc}
\hline
Region & $\widehat{\sigma}_\mathrm{c}(\varphi)$ [deg] & $B_\mathrm{DCF, CN}$ [\textmu G] & $B_\mathrm{DCF, HCO^+}$ [\textmu G] & $B_\mathrm{ST, CN}$ [\textmu G] & $B_\mathrm{ST, HCO^+}$ [\textmu G] \\ 
\hline \hline
%Bubble West & 4.7 $\pm$ 0.2 & 176 $\pm$ 43 & 194 $\pm$ 53 & 71 $\pm$ 17 & 79 $\pm$ 21 \\
%Bubble East & 2.56 $\pm$ 0.04 & 145 $\pm$ 56 & 160 $\pm$ 52 & 43 $\pm$ 17 & 48 $\pm$ 16 \\
%Filament & 4.4 $\pm$ 0.2 & 91 $\pm$ 38 & 141 $\pm$ 48 & 36 $\pm$ 15 & 55 $\pm$ 19 \\
%Filament+Bubble & 6.5 $\pm$ 0.5 & 157 $\pm$ 44 & 191 $\pm$ 48 & 75 $\pm$ 21 & 91 $\pm$ 23 \\
%\hline
Bubble West & 4.7 $\pm$ 0.2 & 53 $\pm$ 27 & 106 $\pm$ 30 & 22 $\pm$ 11 & 43 $\pm$ 12 \\
Bubble East & 2.56 $\pm$ 0.04 & 317 $\pm$ 72 & 318 $\pm$ 72 & 95 $\pm$ 22 & 95 $\pm$ 22 \\
Filament & 4.4 $\pm$ 0.2 & 91 $\pm$ 38 & 141 $\pm$ 48 & 36 $\pm$ 15 & 55 $\pm$ 19 \\
Filament+Bubble & 6.5 $\pm$ 0.5 & 158 $\pm$ 44 & 180 $\pm$ 47 & 75 $\pm$ 21 & 86 $\pm$ 22 \\
\hline
\hline 
\end{tabular} } }
\endgroup
\label{tab:pixels_hist}
\end{table*}

\subsection{$B_\mathrm{POS}$ using different approaches}
\label{subsec:dcf_skalidis}

\begin{figure*}
    \centering
    \includegraphics[width=0.7\textwidth]{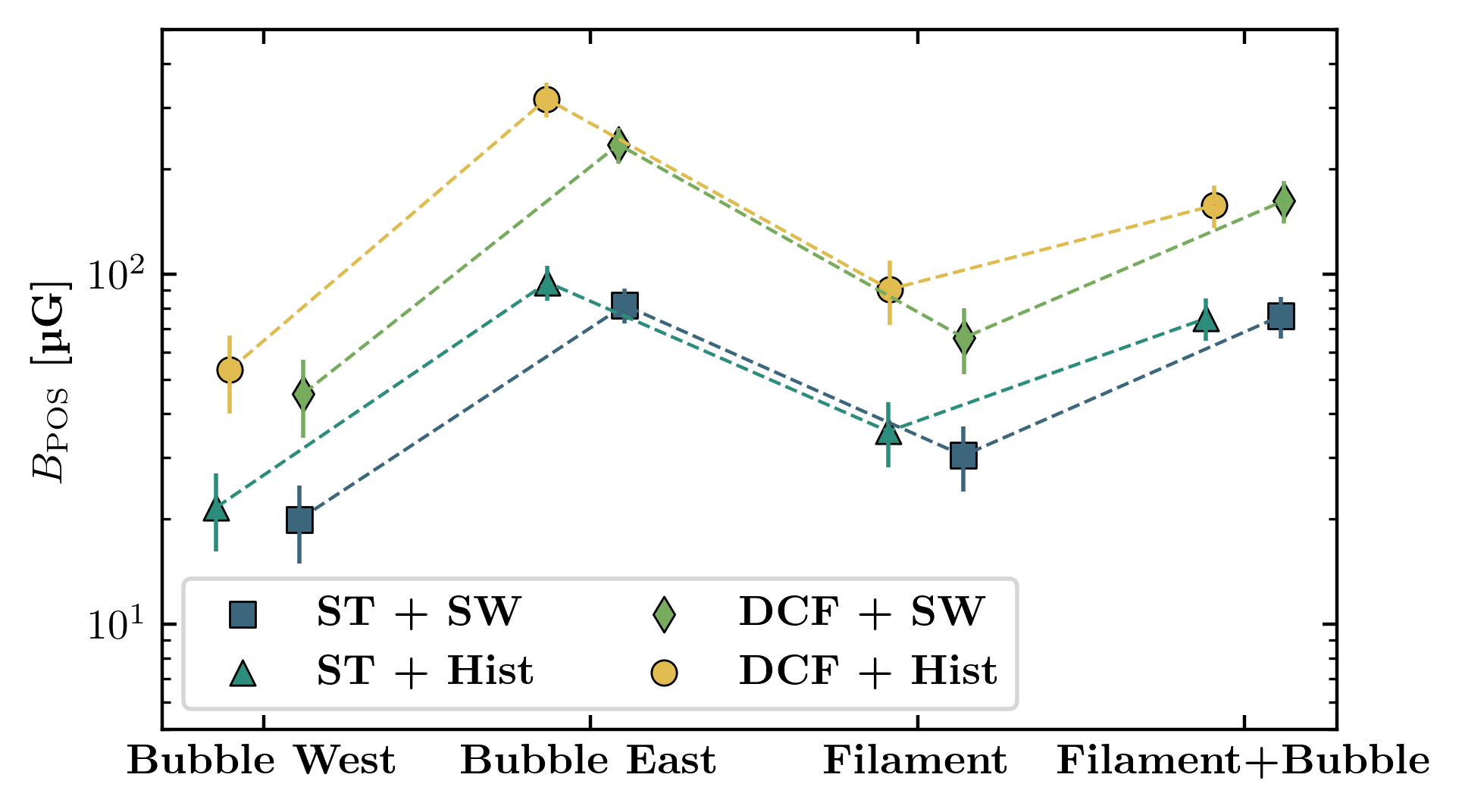}
    \caption{Magnetic field strength derived using different methods (Sec.\,\ref{subsec:dcf}, \ref{subsec:dcf_skalidis}) for computing the magnetic field strength and the angle dispersion (Sec.\,\ref{subsec:slidding_window}, \ref{sec:histogram}) as a function of the environment we analyze in this work.}
    \label{fig:dcf_skalidis}
\end{figure*}

We compare results on POS $B$-field strengths using several methods in Fig.\,\ref{fig:dcf_skalidis}. As described in Sec.\,\ref{sec:Bfield_results}, results derived from the classical DCF are higher than those derived using ST, even when using the histogram analysis instead of the sliding window approach.

% \input{Sections_referee/Appendix_C}

% WARNING
%-------------------------------------------------------------------
% Please note that we have included the references to the file aa.dem in
% order to compile it, but we ask you to:
%
% - use BibTeX with the regular commands:
%   \bibliographystyle{aa} % style aa.bst
%   \bibliography{Yourfile} % your references Yourfile.bib
%
% - join the .bib files when you upload your source files
%-------------------------------------------------------------------

\end{document}